\documentclass[a4paper,11pt]{article}
\usepackage{graphicx,graphics,xcolor}
\usepackage{epsfig}
\usepackage{amsmath}
\usepackage{amssymb}
\usepackage{mathtools}
\usepackage[top=1cm, bottom=1.5cm, left=1.2cm, right=1.2cm]{geometry}
\usepackage{microtype}
\usepackage{lmodern}
\usepackage[utf8]{inputenc}
\usepackage{float}
\usepackage{hyperref}
\usepackage{bera}
\newcommand{\ds}{\displaystyle }

\newcommand{\beq}{\begin{equation} }
\newcommand{\eeq}{\end{equation}}
\begin{document}

\title{Stochastic compartment model with mortality and its application to epidemic spreading in complex networks}
\author{
T\'eo Granger$^1$, Thomas M. Michelitsch$^1$\footnote{Corresponding, \, E-mail:\, thomas.michelitsch@sorbonne-universite.fr}, Michael Bestehorn$^2$, \\ Alejandro P. Riascos$^3$, Bernard A. Collet$^1$\\[3ex] 
$^1$ Sorbonne Universit\'e \\ Institut Jean le Rond d'Alembert 
CNRS UMR 7190 \\ 4 Place Jussieu, 75252 Paris Cedex 05, France
\\ \\[1ex]
$^2$
Brandenburgische Technische Universit\"at Cottbus-Senftenberg \\
Institut f\"ur Physik \\ Erich-Weinert-Stra{\ss}e 1, 03046 Cottbus, Germany
\\ \\[1ex]
$^3$
$^d$ Departamento de Física, Universidad Nacional de Colombia \\
Bogot\'a, Colombia  
}
\maketitle
\begin{abstract}
We study epidemic spreading in complex networks by a multiple random walker approach. Each walker performs an independent simple Markovian random walk on a complex undirected (ergodic) random graph where we focus on Barab\'asi-Albert (BA), Erd\"os-R\'enyi (ER) and Watts-Strogatz (WS) types. 
Both, walkers and nodes can be either susceptible (S) or infected and infectious (I) representing their states of health. 
Susceptible nodes may be infected by visits of infected walkers, and susceptible walkers may be infected by visiting infected nodes. No direct transmission of the disease among walkers (or among nodes) is possible. This model mimics a large class of diseases such as Dengue and Malaria with transmission of the disease via vectors (mosquitos). 
Infected walkers may die during the time span of their infection introducing an additional compartment D of dead walkers. Infected nodes never die and always recover from their infection after a random finite time. This assumption is based on the observation that infectious vectors (mosquitos) are not ill and do not die from the infection. The infectious time spans of nodes and walkers, and the survival times of infected walkers, are represented by independent random variables. 
We derive stochastic evolution equations for the mean-field compartmental populations with mortality of walkers and delayed transitions among the compartments. 
From linear stability analysis, we derive the basic reproduction numbers $R_M, R_0$ with and without mortality, respectively, and prove that $R_M < R_0$. For $R_M,R_0>1$ the healthy state is unstable whereas for zero mortality a stable endemic equilibrium exists (independent of the initial conditions) which we obtained explicitly.
We observe that the solutions of the random walk simulations in the considered networks agree well with the mean-field solutions for strongly connected graph topologies, whereas less well for weakly connected structures and for diseases with high mortality.
Our model has applications beyond epidemic dynamics, for instance in the kinetics of chemical reactions, 
the propagation of contaminants, wood fires, among many others.
 \\[2ex]
{\it Keyworkds. Epidemic spreading, compartment model with mortality, memory effects, random walks, random graphs}
\end{abstract}
\newpage
\tableofcontents
\section{Introduction}
\label{Intro}
Sudden or recurrent emergence of epidemics has been an everlasting thread to humanity. 
Highly infectious and fatal diseases such as 
pestilence, typhus, cholera, and leprosy were among the main causes of death in medieval times in Europe and 
until 20th century a major scourge of humanity \cite{public_health}. This permanent challenge has naturally driven interest in protective measures and predictive models.

The systematic mathematical study of epidemic spreading began only a century ago with the seminal work of Kermack and McKendrick \cite{KermackMcKendrick1927}. They were the first to introduce what we call nowadays a ``compartment model''. In their so called SIR-model the individuals are categorized in compartments
susceptible - S, infected - I, recovered (immune) - R characterizing their states of their health. 
Whereas standard SIR type models are able to capture main features of a certain class of infectious diseases such as 
mumps, measles and rubella, they fail to describe persistent oscillatory behaviors and spontaneous outbursts which are observed in many epidemics.

A large amount of work still is devoted to compartmental models
\cite{LiuHeathcote1987,Li-etal1999} where an impressive field has emerged \cite{Anderson1992,Martcheva2015} and the interest was again considerably enhanced by the context of COVID-19 pandemics \cite{Harris2023}.
Beside purely macroscopic models, the study of epidemic dynamics in complex networks 
has attracted considerable attention \cite{Satoras-Vespignani-etal2015,Pastor-SatorrasVespignani2001,OkabeShudo2021}. In these works the importance of the graph topology for spreading phenomena has been highlighted.
In particular, Pastor-Satorras and Vespignani showed that for a wide range of scale free networks no critical threshold for epidemic spreading exists \cite{Pastor-SatorrasVespignani2001}.
The topological features crucial for epidemic spreading include the small world property (short average network distances\footnote{The network `distance' of two nodes is the number of edges of the shortest path connecting them.}) and a high clustering coefficient measuring the existence of redundant paths between pairs of nodes \cite{NewmanWattsStrogats2002,Eraso-Hernandez-etal2023}.
 
Further interesting directions are represented by combinations of network science and stochastic compartmental models
\cite{Barrat-etal2008,Barabasi2016,BarabasiAlbert1999,RiascosSanders2021,fractional_book_MiRia2019}. Such models 
include Markovian and non-Markovian approaches  \cite{vanKampen1981,Ross1996,VanMiegem2014,BesMi-etal2021,BesMi2023} where non-Markovianity is introduced by non-exponentially distributed sojourn 
times in the compartments \cite{BesMiRias2022,Granger-et-al2023}.
In these works explicit formulae for the endemic equilibrium in terms of mean compartmental sojourn times and the basic reproduction number are derived and numerically validated in random walk simulations. 
A further non-Markovian model appeared recently \cite{BasnakovSandev-etal2020} where non-Markovianity comes into play by introducing an "age of infection" allowing individuals to recover when
their infection period exceeds a certain threshold, generalizing the initial idea of Kermack and McKendrick. 

Other recent works emphasize the importance of spatial heterogeneity effects of infection patterns
in epidemic spreading phenomena \cite{ZhuShenWang2023} and the role of local clusters to generate periodic epidemic outbursts 
\cite{Gostiaux-etal2023} and see \cite{Peyard2023} for a review of these effects. A cluster model to explain periodic behavior was introduced a long time ago \cite{Soper1929}.
The role of the complex interplay of retardation (delayed compartmental transitions) and fluctuations for oscillatory behavior has been investigated in one of our recent works \cite{BesMi2023}.

The aim of the present paper is to study the spreading of a certain class of diseases in a population of individuals (random walkers) moving
on complex graphs aiming to mimic human mobility patterns in complex environments such as cities, street, and transportation networks. 
Essential elements in our model are the account for the mortality of infected individuals (random walkers) and an indirect transmission pathway via vectors (detailed below).

The present paper is organized as follows. In Section \ref{SIS-motality} 
we establish a stochastic mean field model for the evolution of the compartmental populations.
The special case of zero mortality is considered in Section \ref{zero_mortality} 
where we obtain an explicit formula for the endemic equilibrium (stationary constant compartmental populations for infinite time). In this way we identify a crucial parameter controlling the stability of the healthy state having the interpretation of the basic reproduction number $R_0$ (Section \ref{stability_healthy}) where the healthy state is stable for $R_0 <1$ and unstable for $R_0 >1$. A detailed proof of the stability of the endemic state for $R_0>1$ is provided in Appendix \ref{endemic_stability}. In Section \ref{stability_healthy_mortality} we analyze stability of the healthy state with mortality and derive the basic reproduction number $R_M$ and prove that $R_M < R_0$, i.e. mortality reduces the basic reproduction number.
In Section \ref{simulations} we test robustness of our mean field model under
"complex real world conditions" by
implementing its assumptions into multiple random random walkers simulations on
Barabási-Albert (BA), Erdös-Rényi (ER) and Watts-Strogatz (WS) type graphs
(\cite{BarabasiAlbert1999,Barabasi2016,Newman2010} and Appendix \ref{complex_graphs}).
These graph types have different complexity and connectivity features with impact on the spreading. 

\section{Compartmental model with mortality}
\label{SIS-motality}
The goal of this section is to develop a mean field model for a certain class of diseases with indirect infection transmission via vectors which includes
Dengue, Malaria (transmission by mosquitos)
or Pestilence (transmission by fleas) and others \cite{Whitehead2007,Satoras-Vespignani-etal2015}.
To that end we consider a population of $Z$ random walkers navigating independently on a connected (ergodic) graph. 
Each walker performs independent steps from one to another connected node on the network (specified subsequently).
We assume that walkers and nodes are in one of the compartments, S (susceptible) and I (infected). In addition, walkers can be in compartment D (dead) whereas nodes never die.

Let $Z_S(t), Z_I(t)$ ($N_S(t), N_I(t)$) be the number of walkers (nodes) in compartments S and I, and $Z_D(t)$ the non-decreasing number of walkers (in compartment D) died from the disease up to time $t$. We consider $Z = Z_I(t)+ Z_S(t) + Z_D(t)$ walkers ($Z$ independent of time)
and a constant number of nodes $N = N_I(t)+N_S(t)$.
We assume at instant $t=0$ the first spontaneous occurrence of the disease of a few infected walkers $Z_I(0) \ll Z$ or nodes $N_I(0) \ll N$ (and no dead walkers $Z_D(0)=0$).
We introduce the compartmental fractions 
$S_w(t) = \frac{Z_S(t)}{Z}$, $J_w(t)=\frac{Z_I(t)}{Z}$, $d_w(t)= \frac{Z_d(t)}{Z}$ for the walkers (normalized with respect to $Z$) with $S_w(t)+J_w(t)+d_w(t)=1$, and $S_n(t)= \frac{N_S(t)}{N}$, $J_n(t)=\frac{N_I(t)}{N}$ with $S_n(t)+J_n(t)=1$.
To limit the complexity of our model we do not consider the demographic evolution, i.e. there are no natural birth and dead processes.
We denote with ${\cal A}_w(t), {\cal A}_n(t)$ the infection 
rates (rates of transitions S $\to$ I) of walkers and nodes, respectively.
We assume the following simple bi-linear forms
\beq
\label{infections_rate}
\begin{array}{clr}
\ds {\cal A}_w(t)&  = \ds {\cal A}_w[S_w(t), J_n(t)]   =  \beta_w S_w(t)J_n(t) & \\ \\
\ds {\cal A}_n(t) & = \ds {\cal A}_n[S_n(t), J_w(t)]   =  \beta_n S_n(t)J_w(t) &
\end{array}
\eeq
with constant rate parameters $\beta_w, \beta_n >0$ (independent of time). 
${\cal A}_w(t)$ indicates the infection rate of walkers where its dependence of $S_w,J_n$ is telling us that susceptible walkers can be infected only by (visiting) infected nodes. $ {\cal A}_n(t)$ stands for the infection rate of nodes depending on $S_n(t), J_w(t)$ indicating that susceptible nodes can only be 
infected by (visits of) infected walkers. There are no direct transmissions among walkers and among nodes.
Infections of walkers (nodes) take place with specific transmission probabilities in a contact of a node and a walker
which
are captured by (yet not identical with) the transmission rate constants $\beta_{w,n}$.

The infection time spans $t_I^{w,n} >0$ without mortality (waiting times in compartment I)
of walkers and nodes are assumed to be independent random variables drawn from specific distributions specified hereafter.
As only admitted dead process we assume that infected walkers may die within the time span of their infection.
To capture this kind of mortality caused by the disease, we introduce a further independent random variable $t_M>0$ which indicates the life span of an infected walker. Both the infection and life time spans $t_I^w, t_M$ are counted from the time instant of the infection.
A walker survives the disease if $t_M > t_I^w$ and dies from it for $t_M<t_I^w$. 
With these assumptions, we give first a stochastic formulation of the evolution equations
\beq
\label{evoleqs}
\begin{array}{clr}
\ds \frac{d}{dt}S_w(t) & = \ds   - {\cal A}_w(t) + 
\left\langle {\cal A}_w(t-t_I^w)
\Theta(t_M-t_I^w)\right\rangle  + J_w(0)\langle \delta(t-t_I^w)\Theta(t_M-t_I^w) \rangle & 
 \\ \\ 
\ds \frac{d}{dt} J_w(t) & = \ds  {\cal A}_w(t)    - \left\langle \,  {\cal A}_w(t-t_I^w) \Theta(t_M-t_I^w)\, \right\rangle  
 -J_w(0)\left\langle \,\delta(t-t_I^w)\Theta(t_M-t_I^w) \, \right\rangle    -\frac{d}{dt}d_w(t)  &   \\ \\
\ds \frac{d}{dt} d_w(t) = & \ds 
\left\langle \, {\cal A}_w(t-t_M) \Theta(t_I^w-t_M)\, \right\rangle +  J_w(0) \langle \delta(t-t_M)\Theta(t_I^w-t_M) \rangle
   &   \\ \\
\ds \frac{d}{dt}S_n(t) & = \ds - {\cal A}_n(t) +  \left\langle {\cal A}_n(t-t_I^n) \right\rangle + J_n(0)\langle \delta(t-t_I^n)\rangle  & \\ \\
\ds \frac{d}{dt}J_n(t) & =  \ds  -\frac{d}{dt}S_n(t)  &
\end{array}
\eeq
where $ \frac{d}{dt}d_w(t)$ indicates the (non-negative) mortality rate of walkers.
We indicate with $\left\langle ..\right\rangle$ average over the contained (set of independent) random variables $t_I^{w},t_I^{n},t_M$ outlined hereafter and in Appendix \ref {general}.
$\Theta(..)$ stands for the Heaviside function (\ref{Heaviside-stepfu}), and $\delta(..)$ for the Dirac's $\delta$-distribution.
An epidemic always starts from  “natural” initial conditions $S_w(0)=1$, $S_n(0)=1$  (globally healthy state)
where at $t=0$ the first infections occur spontaneously and can be “generated” by adding the source terms $J_{w,n}(0)\delta(t)$ to the infection rates of walkers and nodes, respectively.
Equivalently, we introduce initial conditions $S_{w,n}(0)=1-J_{w,n}(0)$ ($d_w(0)=0$) with $J_{w,n}(0) > 0$ consisting typically of a few infected walkers and/or nodes in a large susceptible population without dead walkers $d_w(0)=0$.

The interpretation of system (\ref{evoleqs}) is as follows. The instantaneous infection rate
${\cal A}_w(t)$ governs the transitions S $\to$ I of walkers (due to visits of infected nodes).
The term $\left\langle \,  {\cal A}_w(t-t_I^w) \Theta(t_M-t_I^w)\, \right\rangle$ 
describes the rate of walkers recovering at time $t$ and infected at $t-t_I^w$, i.e. their infection time span has elapsed and they survived as $t_M>t_I^w$ (indicated by $\Theta(t_M-t_I^w)=1$).
Then
$\left\langle \, {\cal A}_w(t-t_M) \Theta(t_I^w-t_M)\, \right\rangle$ captures the rate of walkers infected at $t-t_M$
dying at at time $t$ during the infection time span (indicated by $\Theta(t_I^w-t_M)=1$ for $t_I^w >t_M$). 
\paragraph{Remark I}
The infection time span of a walker (sojourn time in compartment I) is $min(t_I^w,t_M)$, i.e. $t_I^w$ if $t_M>t_I^w$ (where the walker survives the disease), and is $t_M$ if the walker dies within the infectious time span ($t_M < t_I^w$). $t_I^w$ is the walker's infection time span without mortality (retrieved for $t_M \to \infty$). The probability of persistence of a walker's infection at time $t$, given the infection starts at time $0$ is $\langle \, \Theta(t_I^w-t) \Theta(t_M-t)\, \rangle$ (see (\ref{obs_rand})). Note that $\Theta(t_I^w-t) \Theta(t_M-t) =1$ only if $t < min(t_I^w,t_M)$, i.e. when the walker is in compartment I. 
As a crucial element of our model, we will analyze the statistics of the walker's infection time span $min(t_I^w,t_M)$.
\\[1ex]
The initially infected walkers and nodes are as well subjected to the transition pathways, i.e. walkers either recover (alive) with rate $J_w(0)\langle \delta(t-t_I^w)\Theta(t_M-t_I^w) \rangle$ or they die with rate $J_w(0)\langle \delta(t-t_M)\Theta(t_I^w-t_M) \rangle$, and nodes always recover with rate $J_n(0)\langle \delta(t-t_I^n) \rangle$. For $t\to \infty$ these terms are evanescent thus the initial conditions do not affect large time limits (endemic state for zero mortality).
The importance of these terms can be seen by setting $\beta_{w,n}=0$ (no infections).
Without these terms the initially infected walkers and nodes would stay infected forever, inconsistent with our assumptions.

The rate equations for the nodes can be interpreted in the same way as interplay of instantaneous infections and delayed recovery without mortality. We emphasize that the evolution equations of the nodes and walkers are non-linearly coupled by the implicit dependencies of the infection rates (\ref{infections_rate}).
In order to derive an explicit representation of system (\ref{evoleqs}), we need to have a closer look on the averaging procedures and the involved distributions related to the independent random variables $T=\{t_I^{w},t_I^{n},t_M\} > 0$ drawn from specific probability density functions (PDFs) which we define by
\beq
\label{PDFs_random_delays}
Prob[T\in [\tau,\tau+{\rm d}\tau] = K(\tau){\rm d}\tau, 
\eeq
with their respective PDFs (kernels) $K(\tau)=\{K_{I}^{w,n}(\tau), K_M(\tau)\}$ which
are normalized $Prob[T >0] = \int_0^{\infty}K(\tau){\rm d}\tau = 1$.
Then recall the averaging rule for (suitable) functions $f(T)$ of the random variable $T$ which we use throughout the paper
\beq
\label{averaging_general}
\left\langle f(T) \right\rangle = \int_0^{\infty}K(\tau)f(\tau){\rm d}\tau ,
\eeq
see also Appendix \ref{general}. An important case is $\langle \delta(t-T) \rangle =K(t)$.
Then by applying (\ref{averaging_general}) we introduce the persistence probabilities of walker's (node's) infection (without mortality)
\beq
\label{persist_infection}
\Phi_I^{w,n}(t) = Prob(t_I^{w,n}>t) =  \langle \Theta(t_I^{w,n}-t)\rangle = \int_t^{\infty}K_I^{w,n}(\tau){\rm d}\tau 
\eeq
and the probability of walker's survival up to time $t$ (given $t_I^w =\infty$)
\beq
\label{persist_life} 
\Phi_M(t) =Prob(t_M>t) = \langle \Theta(t_M-t)\rangle = \int_t^{\infty}K_M(\tau){\rm d}\tau .
\eeq
The persistence probabilities fulfill the initial condition $\Phi_M(0)=\Phi_I^{w,n}(0)=1$ corresponding to the normalization of the waiting time
PDFs $K(\tau)=\{K_{I}^{w,n}(\tau), K_M(\tau)\}$ and are vanishing at infinity $\Phi_{M}(\infty) =\Phi_I^{w,n}(\infty) =0$. 
To evaluate the averages in (\ref{evoleqs}) we will use the following quantities:
\beq
\label{obs_rand}
\begin{array}{clr}
\ds \langle \delta(t-T) \rangle  & = \ds K(t) , \hspace{1cm}  T= \{t_I^w,  t_I^n, t_M\} & \\ \\
\ds \langle \Theta(t_M-t)\Theta(t_I^w-t) \rangle & = \ds \langle \Theta(t_M-t) \rangle \langle \Theta(t_I^w-t) \rangle =  
\Phi_I^w(t)\Phi_M(t) & \\ \\
\ds b_d(t) = \langle \delta(t-t_M) \Theta(t_I^w-t_M) \rangle & = \ds  \langle \delta(t-t_M)\rangle \langle \Theta(t_I^w-t) \rangle = K_M(t) \Phi_I^w(t) 
& \\ \\
\ds b_r(t) = \langle \delta(t-t_I^w)\Theta(t_M-t_I^w) \rangle & = \ds 
\ds \langle \delta(t-t_I^w)\rangle \langle \Theta(t_M-t) \rangle = K_I^w(t)\Phi_M(t) 
& \\ \\
\ds b_d(t)+b_r(t) =  {\cal K}_{I,M}^w(t)   & = \ds   -\frac{d}{dt}[ \langle \Theta(t_M-t) \rangle \langle \Theta(t_I^w-t) \rangle ]  = -\frac{d}{dt}\left[\Phi_I^w(t)\Phi_M(t)\right] & \\ \\
\ds \int_0^{\infty} {\cal K}_{I,M}^w(t){\rm d}t &= \ds 1 & \\ \\
\ds {\cal R}(t) = \langle \Theta(t-t_I^w) \Theta(t_M-t_I^w) \rangle & = \ds \int_0^t b_r(\tau){\rm d}\tau = \int_0^t K_I^w(\tau)\Phi_M(\tau){\rm d}\tau & \\ \\
\ds {\cal D}(t) =  \langle \Theta(t-t_M) \Theta(t_I^w-t_M) \rangle &  = \ds \int_0^t b_d(\tau){\rm d}\tau   = \int_0^t K_M(\tau) \Phi_I^w(\tau){\rm d}\tau & \\ \\
\ds {\cal R}(t) + {\cal D}(t) = \int_0^{t} {\cal K}_{I,M}^w(\tau){\rm d}\tau  & = \ds \int_0^t[b_d(\tau)+b_r(\tau)]{\rm d}\tau  = 1 - \Phi_I^w(t)\Phi_M(t) & \\ \\
\ds {\cal D}(\infty) + {\cal R}(\infty) = 1 & \\ \\
 \ds \langle A(t-t_I)\Theta(t_M-t_I) \rangle  & = \ds \langle A(t-t_I) \Phi_M(t_I)\rangle 
 = \int_0^t A(t-\tau)\Phi_M(\tau) K_I(\tau){\rm d}\tau & \\ \\
 \ds \langle A(t-t_M\Theta(t_I^w-t_M) \rangle  & = \ds \langle A(t-t_M) \Phi_I^w(t_M)\rangle =  
 \int_0^t A(t-\tau)\Phi_I^w(\tau) K_M(\tau){\rm d}\tau 
\end{array}
\eeq
In these averages we make use of the independence of the waiting times $t_M,t_{I}^{w,n}$, and of causality of $A(\tau)$ and the kernels $K(\tau)$ (i.e. $A(\tau), K(\tau)=0$ for $\tau <0$).
Of utmost importance are the "defective" PDFs (DPDFs) $b_{d,r}(t)$ of death and recovery. "Defective" means that $b_{d,r}(t)$
are no proper PDFs since they are not normalized to one, but rather to ${\cal D}(\infty), {\cal R}(\infty) <1$, respectively. Consult \cite{donofrio2024} for a recent account of defective distributions and related stochastic processes.
They have the following interpretation. $b_d(t){\rm d}t = K_M(t) \Phi_I^w(t){\rm d}t$ is the probability of
transition I $\to$ D within $[t,t+{\rm d}t]$ of an infected walker (infected at $t'=0$). 
$b_r(t){\rm d}t = K_I^w(t)\Phi_M(t){\rm d}t$ is the probability of transition I $\to$ S within 
$[t,t+{\rm d}t]$ of a walker infected at $t'=0$.
Therefore, 
\beq
\label{KMIW}
{\cal K}_{I,M}^w(t) = b_r(t)+b_d(t) =-\frac{d}{dt} \left\langle \Theta(t_M-t)\Theta(t_I^w-t) \right\rangle   = -\frac{d}{dt}[\Phi_I^w(t)\Phi_M(t)] 
\eeq
is non-negative (as are $K_I^w = -\frac{d}{dt}\Phi_I^w \geq 0$, $K_M =-\frac{d}{dt}\Phi_M \geq 0$) 
and is a proper well-normalized PDF of all exits of walkers from compartment I (i.e. I $\to$ S $+$ I $\to$ D). Without mortality ($\Phi_M(t) = 1$) this PDF retrieves ${\cal K}_{I,M}^w(t) = K_I^w(t)$.

The quantities ${\cal R}(t), {\cal D}(t)$ introduced in (\ref{obs_rand}) have the following interpretation.
${\cal R}(t)$ is the probability that a walker infected at instant $0$ is at time $t$ in compartment S  
(i.e. recovered prior or up to time $t$). 
${\cal D}(t)$ is the probability that a walker infected at instant $0$ is at time $t$ in compartment D
(i.e. died prior and up to time $t$).
Important are the infinite time limits: ${\cal R}(\infty)$ has the interpretation of the overall probability that an infected walker survives the infection, and ${\cal D}(\infty)$ is the overall probability for an infected walker to die from the disease. We refer ${\cal D}(\infty)$ also to as "overall mortality". It must not be confused with the infinite time limit of dead walkers fraction $d_w(\infty)$ which is different from ${\cal D}(\infty)$ as we will see in details subsequently. A small value ${\cal D}(\infty)$ may cause a high value of $d_w(\infty)$ for instance for short infectious periods where walkers may be repeatedly infected.

In most cases not all infected walkers die from their disease (in an infinite observation time), hence 
${\cal D}(\infty) < 1$ (as $b_d$ is defective). ${\cal D}(\infty) \to 1$ 
represents the limit of a fatal disease, and ${\cal D}(\infty) \to 0$ a disease without mortality.
${\cal R}(\infty) < 1 $ (as $b_r$ is defective) is the complementary probability with ${\cal D}(\infty)+{\cal R}(\infty)=1$.

With these remarks, system (\ref{evoleqs}) reads
\beq
\label{evoleqsB}
\begin{array}{clr}
\ds \frac{d}{dt}S_w(t) & = \ds - {\cal A}_w(t) +   \int_0^t {\cal A}_w(t-\tau)K_I^w(\tau)\Phi_M(\tau) {\rm d}\tau + 
J_w(0)K_I^w(t)\Phi_M(t)   & 
 \\ \\
\ds \frac{d}{dt} J_w(t) & =   \ds \frac{d}{dt} \int_0^t  {\cal A}_w(\tau) \Phi_M(t-\tau)\Phi_I^w(t-\tau){\rm d}\tau -J_w(0)\left[K_I^w(t)\Phi_M(t)+K_M(t) \Phi_I^w(t)\right] 
\ds     &  \\ \\
\ds \frac{d}{dt}S_n(t) & = \ds - {\cal A}_n(t) +  
\int_0^t {\cal A}_n(t-\tau)K_I^n(\tau) {\rm d}\tau + J_n(0)K_I^n(t) & \\ \\
\ds \frac{d}{dt}J_n(t) & =  \ds - \frac{d}{dt}S_n(t). & \\ & &
\end{array}
\eeq
The PDF (\ref{KMIW}) that a walker leaves compartment I (either by recovery or by death)
allows to rewrite the second equation of (\ref{evoleqsB}) as
\beq
\label{J_w-second-rep}
\frac{d}{dt} J_w(t) = {\cal A}_w(t) - \int_0^t  {\cal A}_w(t-\tau) {\cal K}_{I,M}^w(\tau){\rm d}\tau 
- J_w(0) {\cal K}_{I,M}^w(t) .
\eeq
Worthy of closer consideration is the mortality rate of the infected walkers (representing the total mortality -- entry rate into the D compartment)
\beq
\label{mortality-rate}
\begin{array}{clr}
\ds \frac{d}{dt} d_w(t)  =  -\frac{d}{dt} (S_w(t)+J_w(t))  & =  \ds \left\langle \, {\cal A}_w(t-t_M) \Theta(t_I^w-t_M)\, \right\rangle
+  J_w(0) \langle \delta(t-t_M) \Theta(t_I^w-t_M) \rangle 
& \\ \\ 
   & =\ds  \int_0^t{\cal A}_w(t-\tau) K_M(\tau)\Phi_I^w(\tau){\rm d}\tau + J_w(0)K_M(t)\Phi_I^w(t)   &
   \end{array}
\eeq
where clearly $\frac{d}{dt} d_w(t)\geq 0$.
Integrating this relation yields the fraction $d_w(t)$ of dead walkers
\beq
\label{dead_walkers}
\begin{array}{clr}
\ds d_w(t) & = \ds  1-S_w(t)-J_w(t) & \\ \\ 
&  = \ds  \int_0^t  {\cal A}_w(t-\tau) \langle \Theta(\tau -t_M)\Theta(t_I^w-t_M) \rangle {\rm d}\tau + J_w(0) \langle \Theta(t-t_M) \Theta(t_I^w-t_M) \rangle     & \\ \\
& =\ds  \int_0^t {\cal A}_w(t-\tau) {\cal D}(\tau) {\rm d}\tau +  J_w(0){\cal D}(t). &
\end{array}
\eeq
An interesting quantity is the cumulative recovery rate of walkers 
(integrated entry rates of walkers into the S compartment, see first equation in (\ref{evoleqs})) 
\beq
\label{survived_walkers}
\begin{array}{clr}
\ds r_w(t) & = \ds \int_0^t {\cal A}_w(t-\tau) \left\langle \Theta(\tau-t_I^w)  \Theta(t_M-t_I^w)\right\rangle  {\rm d}\tau + J_w(0) \left\langle \Theta(t-t_I^w)  \Theta(t_M-t_I^w)\right\rangle  & \\ \\ & = \ds \int_0^t {\cal A}_w(t-\tau) {\cal R}(\tau){\rm d}\tau +  J_w(0){\cal R}(t). &
\end{array}
\eeq
The quantity $r_w(t)$ records all recovery events of walkers up to time $t$, where individual walkers may
suffer repeated infections and recoveries.
We observe that (see (\ref{obs_rand}))
\beq
\label{r_and_d}
r_w(t)+d_w(t)= \int_0^t {\cal A}_w(t-\tau)[1 - \Phi_I^w(\tau)\Phi_M(\tau)]{\rm d}\tau + 
J_w(0)\left(1-\Phi_I^w(t)\Phi_M(t)\right] .
\eeq
Relation (\ref{dead_walkers}) records all dead events of walkers up to time $t$. Since each walker may die only once, it follows indeed that $d_w(t) \in [0,1]$.
Contrarily, the quantity $r_w(t)$ is not restricted to this interval as walkers may be repeatedly infected and recovered but due to mortality eventually only a finite number of times ($r_w(\infty) < \infty$, see (\ref{inf_rw})). Mortality renders the chain of
infection and recovery events transient (due to the defective feature of $b_r=K_I^w \Phi_M$).
To shed more light on the behavior of $r_w(t)$ consider for a moment zero mortality (${\cal R}(\infty)=1$)
and $t\to \infty$: We then have ${\cal A}_w(\infty) = \beta_wS_w^eJ_n^e >0 $ (shown in Section \ref{zero_mortality}) thus $r_w(\infty)= \infty$ coming along with an infinite chain of recurrent
infection and recovery events (as $b_r(t)$ turns into the proper non-defective PDF $b_r=K_I^w$).

Using (\ref{obs_rand}) we can rewrite (\ref{evoleqs}) in equivalent integral form 
\beq
\label{evoleqsB_intgrate}
\begin{array}{clr}
\ds S_w(t) & = \ds 1-J_w(0)\left[\Phi_M(t)\Phi_I^w(t) + {\cal D}(t)\right] - 
\int_0^t {\cal A}_w(\tau)[\Phi_M(t-\tau)\Phi_I^w(t-\tau) + {\cal D}(t-\tau) ]{\rm d}\tau   &
 \\ \\
\ds J_w(t) & =\ds  J_w(0)\Phi_M(t)\Phi_I^w(t) + \int_0^t  {\cal A}_w(\tau) \Phi_M(t-\tau)\Phi_I^w(t-\tau){\rm d}\tau 
\ds     &  \\ \\
\ds S_n(t) & = \ds 1 - J_n(t)  & \\ \\
\ds J_n(t) & =  \ds J_n(0)\Phi_I^n(t) \ds 
+ \int_0^t {\cal A}_n(\tau)\Phi_I^n(t-\tau) {\rm d}\tau & 
\end{array}
\eeq
and with (redundant) Eq. (\ref{dead_walkers}) for the fraction of dead walkers.
(\ref{evoleqsB_intgrate}) is a self-consistent system since the infection rates are implicit functions of susceptible 
and infected population fractions ${\cal A}_w(t)= A_w[S_w(t),J_n(t)]$, ${\cal A}_n(t)= A_w[S_n(t),J_w(t)]$ (see (\ref{infections_rate})).
Explore now the infinite time limit of (\ref{evoleqsB_intgrate}) where it is convenient to consider 
the Laplace transformed equations.
We introduce the Laplace transform (LT, denoted with a hat) of a function $g(t)$ as
$$
{\hat g}(\lambda) = \int_0^{\infty} g(t) e^{-\lambda t}{\rm d}t
$$
where $\lambda$ denotes the (suitably chosen) Laplace variable. In order to retrieve infinite time limits
we use the asymptotic feature
\beq
\label{asym}
g(\infty) = \lim_{\lambda \to 0} \lambda \, {\hat g}(\lambda)  \hspace{1cm} \left( = 
\lim_{\lambda \to 0}\int_0^{\infty}g(\frac{\tau}{\lambda}) e^{-\tau}{\rm d}\tau \to g(\infty) \int_0^{\infty} e^{-\tau}{\rm d}\tau \right).
\eeq
Observing that the LT of $\Phi_I^w(t)\Phi_M(t)$ is $\lambda^{-1}[1- {\hat {\cal K}}_{I,M}(\lambda)]$
and ${\hat D}(\lambda) =\lambda^{-1}{\hat b}_d(\lambda)$ where ${\hat b}_d(0)={\cal D}(\infty)$ (see (\ref{obs_rand})), we arrive at
\beq
\label{Sw_infty}
\begin{array}{clr}
\ds J_w(\infty) & = \ds  \lim_{\lambda \to 0} \lambda {\hat J}_w(\lambda) = [1-{\hat {\cal K}}_{I,M}(0)][J_w(0)+ {\hat A}_w(0)] = 0 & \\ \\
\ds J_n(\infty) &  =\ds  \lim_{\lambda \to 0} \lambda {\hat J}_n(\lambda) = [1-{\hat K}_{n}(0)][J_n(0)+ {\hat A}_n(0)] =0  \\ \\
\ds d_w(\infty) & = \ds  \lim_{\lambda \to 0} \lambda {\hat d}_w(\lambda) = {\cal D}(\infty) [J_w(0)+ {\hat A}_w(0)]    , & {\hat A}_w(0) = \beta_w \int_0^{\infty} S_w(\tau) J_n(\tau) {\rm d}\tau  \\ \\
\ds S_w(\infty) & =\ds  1- d_w(\infty)  & \\ \\
\ds S_n(\infty) & =\ds  1  &
\end{array}
\eeq
where ${\hat A}_{w,n}(0) =  \int_0^{\infty}{\cal A}_{w,n}(t){\rm d}t < \infty$.
In the same way one obtains
\beq
\label{inf_rw}
r_w(\infty) =  {\cal R}(\infty) (J_w(0) +{\hat A}_w(0)) .
\eeq
Since ${\cal D}(\infty)$ is non-zero, the asymptotic values $S_w(\infty), d_w(\infty)$ 
depend on the initial condition $J_w(0)$ and the infection (rate) history. This is not any more true for zero mortality (${\cal D}(\infty)=0$) and considered in Section \ref{zero_mortality}.
We define the overall probability $P_D$ that a walker dies ($P_R$ survives) the disease
\beq
\label{overlall_mortality}
P_D = \frac{d_w(\infty)}{r_w(\infty)+d_w(\infty)} = {\cal D}(\infty)  , \hspace {1cm} P_R=1-P_D = \frac{r_w(\infty)}{r_w(\infty)+d_w(\infty)} = {\cal R}(\infty)
\eeq
consistent with our previous interpretation of ${\cal D}(\infty), {\cal R}(\infty)$, and the ratio 
\beq
\label{r-d-ratio}
\frac{d_w(\infty)}{r_w(\infty)} = \frac{{\cal D}(\infty)}{{\cal R}(\infty)} .
\eeq
The quantities (\ref{overlall_mortality}) and (\ref{r-d-ratio})
depend only on the stochastic features of $t_I^w$ and $t_M$. They are 
independent of the infection rates and initial conditions and therefore of the topological properties of the network.
In addition, they also do not depend on the stochastic features of the node's infection time span $t_I^n$.

\paragraph{Markovian (memoryless) case}
\label{Markovian}
Generally the system (\ref{evoleqsB}) contains the history of the process (memory) which makes the process non-Markovian. 
The only exception is when all waiting times are exponentially distributed, namely
$\Phi_I^w(t)=e^{-\xi_I^wt}$, $\Phi_M(t)=e^{-\xi_M t}$, $\Phi_I^n(t)= e^{-\xi_I^n t}$  
($\langle t_I^{w,n} \rangle= \xi_I^{w,n})^{-1}$, $\langle t_M \rangle = (\xi_M)^{-1}$).
Then (\ref{evoleqsB}) takes with (\ref{evoleqsB_intgrate}) the memoryless form 
\beq
\label{memoryless}
\begin{array}{clr}
\ds \frac{d}{dt}S_w(t) & = \ds  -\beta_wS_w(t)J_n(t) +\xi_I^wJ_w(t) & \\ \\
 \ds \frac{d}{dt}J_w(t) & = \ds  \beta_wS_w(t)J_n(t) -(\xi_I^w+\xi_M)J_w(t) & \\ \\
\ds \frac{d}{dt}d_w(t) & = \ds \xi_M J_w(t) & \\ \\
\ds \frac{d}{dt}S_n(t) & = \ds  -\beta_nS_n(t)J_w(t) + \xi_I^nJ_n(t)  &  \\ \\
\ds \frac{d}{dt}J_n(t) & = \ds  \beta_nS_n(t)J_w(t)-\xi_I^nJ_n(t).&  
\end{array}
\eeq
Putting the left-hand sides to zero yields the stationary state 
\beq
\label{stationary}
\begin{array}{clr}
\ds J_{w}(\infty)  = J_n(\infty) = {\cal A}_{w}(\infty) = {\cal A}_{n}(\infty) = 0  & & \\ \\
\ds  S_w(\infty)  = \ds 1-d_w(\infty)  ,  & \ds  d_w(\infty) =  \xi_M \int_0^{\infty} J_w(\tau){\rm d}\tau &\\ \\
\ds S_n(\infty) =1.  & & 
\end{array}
\eeq 
Let us check whether this result is consistent with (\ref{Sw_infty}). To this end, we integrate the second equation in (\ref{memoryless})
knowing that $J_w(\infty)=0$ leading to
\beq
\label{checking_mortality}
0 = J_w(0) + \int_0^{\infty} {\cal A}_w(t){\rm d}t -(\xi_I^w+\xi_M)\int_0^{\infty} J_w(t){\rm d}t
\eeq
thus $\int_0^{\infty} J_w(t){\rm d}t = \frac{1}{\xi_M+\xi_I^w}(J_w(0)+{\hat {\cal A}}_w(0))$. Plugging this relation into (\ref{stationary}) and accounting for
${\cal D}(\infty) =\frac{\xi_M}{\xi_I^w+\xi_M}$ recovers indeed the representation of expression (\ref{Sw_infty}).

For zero mortality $\xi_M=0$ one can straight-forwardly obtain the constant endemic equilibrium values
$J_w^e, J_n^e$ by setting the left-hand side of (\ref{memoryless}) to zero leading to subsequent Eq. (\ref{simple_representation}) derived in Section \ref{zero_mortality}
for general waiting time distributions with finite means.

\begin{figure}[H]
\centerline{
\includegraphics[width=0.5\textwidth]{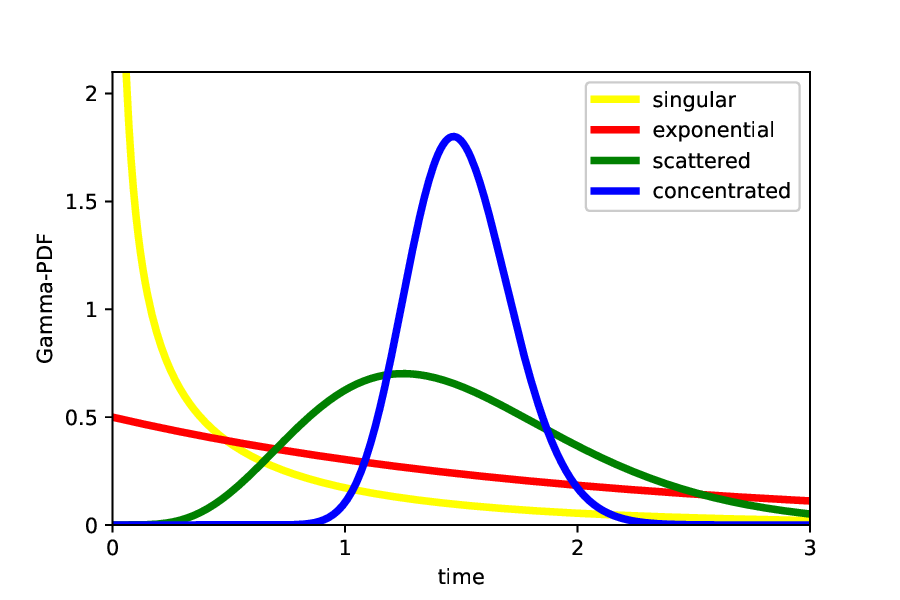}
\includegraphics[width=0.5\textwidth]{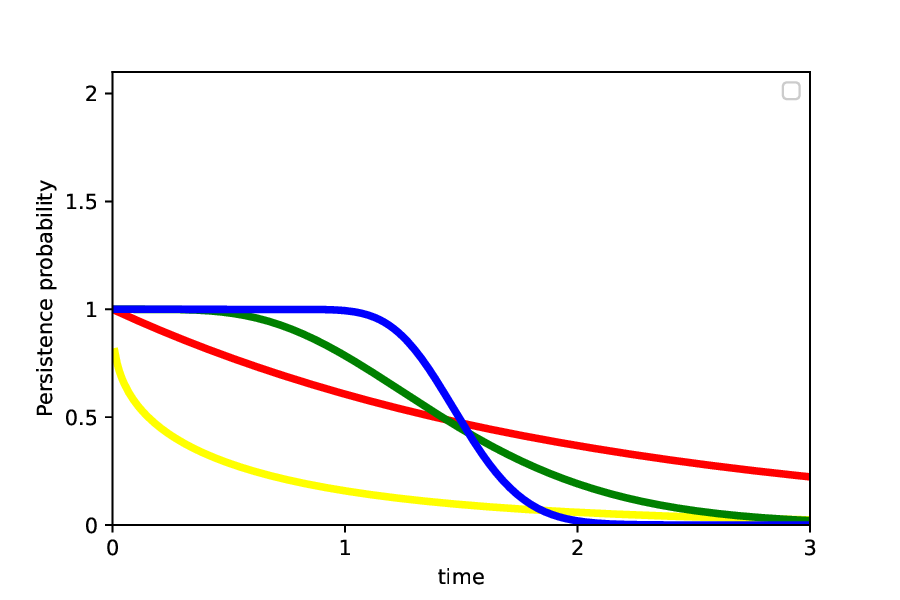}
}
\caption{Left frame: Gamma distribution for four different cases: Weakly singular (at $t=0$) [$\langle t \rangle =0.5$, $\xi=0.7$], exponential [$\langle t \rangle = 2$, $\xi=0.5$], broad [$\langle t \rangle =1.5$, $\xi=4$], and narrow [$\langle t \rangle =1.5$, $\xi=30$].\\
Right frame: Their Persistence (survival) probability distributions of Eq. (\ref{incomplete_Gamma}) where the same color code is used.}
\label{Gamma-PDFs}
\end{figure}

\paragraph{A few more words on waiting time distributions}
\label{Gammadistribution}
In our simulations we assume that the time spans $t_I^w, t_I^n,t_M$ are independent random variables drawn from specific Gamma distributions. 
The advantage to use Gamma distributions is that they may realize a large variety of shapes, see Fig. \ref{Gamma-PDFs} for a few examples.
To generate Gamma distributed random numbers we employ the PYTHON random number generator (library numpy.random). Recall the Gamma distribution 
\beq
\label{Gamma_dist}
K_{\alpha,\xi}(t) = \frac{\xi^{\alpha}t^{\alpha-1}}{\Gamma(\alpha)} e^{-\xi t}, \hspace{1cm}\xi ,\, \alpha > 0
\eeq
where $\alpha$ is the so called `shape parameter' and $\xi$ the rate parameter (often is used the term 'scale parameter' $\theta=\xi^{-1}$) and $\Gamma(\alpha)$ stands for the Gamma function.
We also will subsequently use the LT of the Gamma PDF
\beq
\label{LT_Gamma}
{\hat K}_{\alpha,\xi}(\lambda) = \int_0^{\infty} K_{\alpha,\xi}(t) e^{-\lambda t}{\rm d}t   = \frac{\xi^{\alpha}}{(\lambda +\xi)^{\alpha}} 
\eeq
The Gamma PDF has finite mean $\langle t \rangle_{\alpha,\xi} = \int_0^{\infty} t K_{\alpha,\xi}(t){\rm d}t = \frac{\alpha}{\xi}$ and
for $\alpha <1$ the Gamma-PDF is weakly singular at $t=0$ and $\alpha=1$ recovers exponential PDFs. For $\alpha \leq 1$ the Gamma PDF is completely monotonic (CM) (see Appendix, (\ref{CM_def}) for a definition). For the range $\alpha >1$ the Gamma-PDF has a maximum at $t_{max}=\frac{\alpha-1}{\xi}$ and becomes narrower the larger $\xi$ (while keeping its mean $\alpha/\xi$ fixed), especially we can generate sharp waiting times using the feature
\beq
\label{delta_distribution}
\lim_{\xi \to \infty} K_{\alpha=\xi T_0 ,\xi}(t) =\delta(t-T_0) .
\eeq
We also will use subsequently the persistence probability of the Gamma distribution (see right frame of Fig. \ref{Gamma-PDFs})
\beq
\label{incomplete_Gamma}
\Phi_{\alpha,\xi}(t) = \int_t^{\infty} \frac{\xi^{\alpha}t^{\alpha-1}}{\Gamma(\alpha)} e^{-\xi t}{\rm d}t = \frac{\Gamma(\alpha,\xi t)}{\Gamma(\alpha)}
\eeq
where $\Gamma(\alpha,x)$ indicates the upper incomplete Gamma function with $\Gamma(\alpha,0)=\Gamma(\alpha)$. (\ref{incomplete_Gamma}) fulfills necessarily the initial condition $\Phi_{\alpha,\xi}(0)=1$ and is vanishing at infinity $\Phi_{\alpha,\xi}(\infty)=0$.

\section{Endemic equilibrium for zero mortality}
\label{zero_mortality}
Here we consider the large time asymptotics of the compartment populations without mortality ($S_w(t)+J_w(t)=1$) where the self-consistent system (\ref{evoleqsB_intgrate}) reads
\beq
\label{evoleqsB-integrated-zero_mort}
\begin{array}{clr}
\ds  J_w(t) & =  
\ds     J_w(0) \Phi_I^w(t) +  \int_0^t {\cal A}_w(t-\tau)\Phi_I^w(\tau) {\rm d}\tau  &  \\ \\
\ds J_n(t) & =  \ds  J_n(0)\Phi_I^n(t) + \int_0^t {\cal A}_n(t-\tau)\Phi_{I}^n(\tau){\rm d}\tau  &
\end{array}
\eeq
The endemic state emerging in the large time asymptotics does not depend on the initial conditions $J_{w,n}(0)$ as
$\Phi_I^{w,n}(t) \to 0$. For what follows it is convenient to consider the LTs of (\ref{evoleqsB-integrated-zero_mort}) which read
\beq
\label{evoleqsB-integrated-laplace_no_mort}
\begin{array}{clr}
\ds  {\hat J}_w(\lambda) & =  \ds  \left[J_w(0) + {\hat {\cal A}}_w(\lambda)\right] \frac{1-{\hat K}_I^w(\lambda)}{\lambda}   &  \\ \\
\ds {\hat J}_n(\lambda) & =  \ds  \left[J_n(0) + {\hat {\cal A}}_n(\lambda)\right] \frac{1-{\hat K}_I^n(\lambda)}{\lambda} & 
\end{array}  
\eeq
where $\Phi_I^{w,n}(\lambda)=\frac{1-{\hat K}_I^{w,n}(\lambda)}{\lambda}$ are the LTs of the persistence distributions, 
and ${\hat S}_{w,n}(\lambda) +{\hat J}_{w,n}(\lambda)= \frac{1}{\lambda}$ reflecting constant populations of walkers and nodes.
In order to determine the endemic equilibrium, we
assume that the mean infection time spans for the nodes and walkers exist 
\beq
\label{mean_inf_time}
\langle t_I^{w,n}\rangle = \lim_{\lambda \to 0} \frac{1-{\hat K}_I^{w,n}(\lambda)}{\lambda} = -\frac{d}{d\lambda}{\hat K}_I^{w,n}(\lambda)\bigg|_{\lambda=0} =\int_0^{\infty}\Phi_I^{w,n}(t){\rm d}t  = 
\int_0^{\infty} \tau K_I^{w,n}(\tau){\rm d}\tau < \infty 
\eeq
thus the admissible range of the Laplace variable is $\lambda \geq 0$ (if chosen real).
Now using (\ref{asym}) we obtain the endemic equilibrium from 
$J_{w,n}(\infty) = \lim_{\lambda\to 0} \lambda {\hat J}_{w,n}(\lambda)$ where
the initial conditions are wiped out at infinity as ${\hat K}_{w,n}(\lambda)\bigg|_{\lambda=0}=1$.
Assuming that the infection rates are constant in the endemic equilibrium we have 
${\cal A}_{w,n}(\lambda) \sim {\cal A}_{w,n}(\infty)/\lambda$, ($\lambda \to 0$) and arrive at
\beq
\label{ende_eqs}
\begin{array}{clr}
\ds  J_w(\infty) & = \ds {\cal A}_w(\infty)\left\langle t_I^w \right\rangle , \hspace{1cm} ({\cal A}_w(\infty)=\beta_wS_w(\infty) J_n(\infty))   &  \\ \\
\ds J_n(\infty) & =  \ds  {\cal A}_n(\infty)\left\langle t_I^n \right\rangle , \hspace{1cm} ({\cal A}_n(\infty)=
\beta_nS_n(\infty)J_w(\infty))   & \\ & &
\end{array}
\eeq
thus
\beq
\label{simple_representation}
\begin{array}{cl}
\ds \frac{J_w(\infty)}{1-J_w(\infty)} -\beta_w\langle t_I^w\rangle J_n(\infty) = 0 
& \\ \\ 
\ds \frac{J_n(\infty)}{1-J_n(\infty)} -\beta_n\langle t_I^n\rangle J_w(\infty) = 0 .
\end{array}
\eeq
One can see that the globally healthy state $J_{w,n}(0)=0$ is also a solution of this equation. 
Beside that, only solutions  $J_n(\infty), J_w(\infty) \in (0,1)$ correspond to an endemic equilibrium.
One gets
\beq
\label{endemic_equil}
\begin{array}{cl}
\ds J_w(\infty) = J_w^e = \frac{\beta_w\beta_n\langle t_I^w\rangle \langle t_I^n\rangle -1}{\beta_n\langle t_I^n\rangle[1+\beta_w\langle t_I^w\rangle]} = \frac{R_0-1}{R_0} \frac{\beta_w\langle t_I^w\rangle}{1+\beta_w\langle t_I^w\rangle} \\ \\
\ds J_n(\infty) = J_n^e = \frac{\beta_w\beta_n\langle t_I^w\rangle \langle t_I^n\rangle -1}{\beta_w\langle t_I^w\rangle[1+\beta_n\langle t_I^n\rangle]} = \frac{R_0-1}{R_0} \frac{\beta_n\langle t_I^n\rangle}{1+\beta_n\langle t_I^n\rangle}    &
\end{array}
\eeq
for the endemic equilibrium which is
independent of the initial conditions $J_{w,n}(0)$. It depends only on the phenomenological rate constants $\beta_{w,n}$ and the mean infection time spans $\langle t_I^{w,n} \rangle$.
We point out that the endemic equilibrium (\ref{endemic_equil}) has exchange symmetry $w \leftrightarrow n$ between walkers and nodes
reflecting this feature in the system (\ref{evoleqs}) of evolution equations without mortality.
The endemic values $J_{w,n}^e$ are within $(0,1)$, i.e. existing only if $R_0=\beta_w\beta_n\langle t_I^w\rangle \langle t_I^n\rangle > 1$. We interpret $R_0$ as the basic reproduction number (average number of new infections at $t=0$ -- nodes or walkers -- due to one infected node or walker at $t=0$). That this is really the appropriate interpretation can be seen by the following somewhat rough consideration of the infection rates at $t=0$.
Assume we have initially one single infected node $J_n(0)=1/N$ and no infected walkers $S_w(0)=1$.
The expected number of walkers infected by this first infected node during its infectious period $t_I^n$ is
$$\langle Z_I(t_I^n) \rangle \sim Z {\cal A}_w(0) \langle t_I^n \rangle  =  Z \langle t_I^n \rangle \beta_w/N \sim Z \langle J_w(t_I^w)\rangle .$$
The number of nodes getting infected by these $\langle Z_I(t_I^n) \rangle$ infected walkers during their infectious time $t_I^w$ is
$$
N_I(\langle t_I^n \rangle +\langle t_I^w \rangle )\sim  N \langle {\cal A}_n(t_I^n) \rangle \langle t_I^w \rangle \sim N \beta_n \langle J_w(t_I^w)\rangle = \beta_n\beta_w \langle t_I^n \rangle \langle t_I^w \rangle = R_0 .
$$
Hence $R_0$ is the average number of infected nodes caused by the first infected node (with zero initially infected walkers). Due to the exchange symmetry (nodes $\leftrightarrow$ walkers) of the infection rates, this argumentation remains true when we start with one infected walker and no infected nodes. 

We infer that the globally healthy state is unstable for $R_0>1$ where the endemic equilibrium (\ref{endemic_equil}) is a unique stable fixed point and attractor for all initial conditions $J_{w,n}(0)$. We will confirm this in the next section by a linear stability analysis of the globally healthy state. The stability of the endemic state is demonstrated in the next section with Appendix \ref{endemic_stability}.

Remarkable is the limit $\beta_w\langle t_I^w \rangle \to \infty$ (while $\beta_n\langle t_I^n\rangle$ are kept constant) where all walkers become infected $J_{w}^e \to 1$ but not all nodes $J_n^e \to \frac{\beta_n\langle t_I^n \rangle}{1+\beta_n\langle t_I^n} < 1$ and vice versa.
\begin{figure}[H]
\centerline{\includegraphics[width=0.65\textwidth]{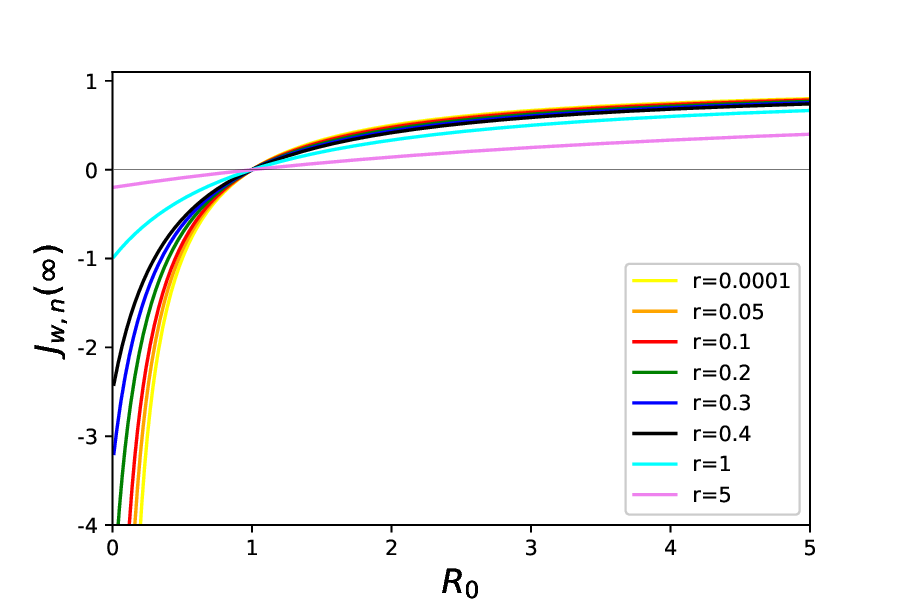}}
\caption{\label{endemic_state_W} Endemic states of infected walkers/nodes $J_{w,n}(\infty)= (R_0-1)/(R_0+r)$ versus $R_0$ 
for various values of parameter $r$ which has to be read $r= \beta_n\langle t_I^n\rangle $ ($r= \beta_w\langle t_I^w\rangle $) for the walker's (node's) endemic states.}
\end{figure}
We plot the endemic state in Fig. \ref{endemic_state_W} versus $R_0$ where positive values for $J_{w,n}(\infty)$ occur only for $R_0>1$ which
correspond to endemic states. Next we perform a linear stability analysis of the endemic and healthy state where we will indeed identify $R_0$ as crucial control parameter. 
\section{Stability analysis of endemic and healthy state without mortality}
\label{stability_healthy}
\begin{figure}[H]
\centerline{\includegraphics[width=0.65\textwidth]{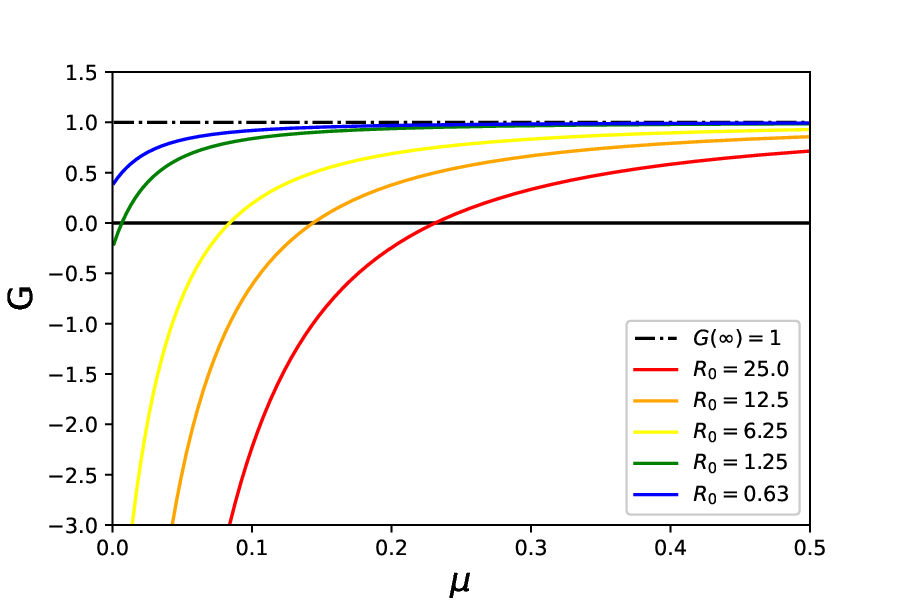}}
\caption{$G(\mu)$ of (\ref{detnull_heathy}) for a some Gamma distributed $t_I^{w,n}$. 
Positive zeros of $G(\mu)$ exist only for $R_0>1$ (instability of globally healthy state).}
\label{healthy}
\end{figure}

Here we are interested in the condition of spreading for zero mortality, or equivalently in the condition for which the globally healthy state (endemic state) is unstable (stable). 
To check stability of the endemic fixed point $S_e^w=1-J_e^w, J_e^w$, $S_e^n=1-J_e^n,J_e^n$ we set
\beq
\label{linear}
\begin{array}{clr}
S_w(t)&= S_w^e +u_w e^{\mu t} ,\hspace{1cm} J_w(t) = J_w^e-u_w e^{\mu t} & \\ \\
S_n(t)&= S_n^e +u_n e^{\mu t} ,\hspace{1cm} J_n(t) = J_n^e-u_n e^{\mu t}
\end {array}
\eeq
where $u_w,u_n$ are `small' constant amplitudes. This ansatz accounts for the constant populations of nodes and walkers.
Then we have for the infection rates up to linear orders in the amplitudes
\beq
\label{Awns}
\begin{array}{clr}
{\cal A}_w(t) & =\beta_w S_w(t)J_n(t) = \beta_w S_w^eJ_n^e + \beta_w (u_w J_n^e-u_nS_w^e)e^{\mu t} & \\ \\
{\cal A}_n(t) & =\beta_n S_n(t)J_w(t) = \beta_n S_n^eJ_w^e + \beta_n (u_n J_w^e-u_wS_n^e)e^{\mu t}
\end{array}
\eeq
Plugging these relations in our evolution equations (\ref{evoleqs}) without mortality, omitting two redundant equations leads to the system
\beq
\label{mat_equations}
\left[\begin{array}{ll} \mu + \beta_wJ_n^e[1-{\hat K}_I^w(\mu)] ;&  -\beta_wS_w^e[1-{\hat K}_I^w(\mu)]   \\ \\
 -\beta_nS_n^e[1-{\hat K}_I^n(\mu)] ;&  \mu + \beta_nJ_w^e[1-{\hat K}_I^n(\mu)] \end{array}\right] \cdot \left(\begin{array}{l} u_w \\ \\ u_n \end{array}\right) = \left(\begin{array}{l} 0 \\ \\ 0 \end{array}\right)
\eeq
where we have used $\langle e^{-\mu t_I^{w,n}} \rangle = {\hat K}_{I}^{w,n}(\mu)$ and the cases of $\delta$-distributed $t_I^{w,n}$ are contained for ${\hat K}_{I}^{w,n}(\mu) = e^{-\mu t_I^{w,n}}$.   
We point out that in ansatz (\ref{Awns}) we relax causality i.e. we admit ${\cal A}_{w,n}(t-\tau) \neq 0$ for $t-\tau <0$ thus
\beq
\label{non-causal-average}
\langle e^{\mu(t-t_I^{w,n})} \rangle = e^{\mu t}\langle e^{-\mu t_I^{w,n}} \rangle = e^{\mu t} {\hat K}_I^{w,n}(\mu) .
\eeq
The solvability of this matrix equation requires the determinant to vanish leading to a transcendental characteristic equation for $\mu$
\beq
\label{determinant_null}
\mu^2 + \mu\left( \beta_wJ_n^e[1-{\hat K}_I^w(\mu)]+ \beta_nJ_w^e[1-{\hat K}_I^n(\mu)]\right) +
\beta_w\beta_n[1-{\hat K}_I^w(\mu)][1-{\hat K}_I^n(\mu)] (J_n^e J_w^e-S_n^e S_w^e) = 0 .
\eeq
Generally, a fixed point is unstable if solutions with
positive real part of $\mu$ exist.
Consider this first for the globally healthy state $J_n=0,J_w=0$ for which Eq. (\ref{determinant_null}) reads
\beq
\label{detnull_heathy}
G(\mu) = 1 - \beta_w\beta_n\frac{[1-{\hat K}_I^w(\mu)]}{\mu}\frac{[1-{\hat K}_I^n(\mu)]}{\mu} = 
1- \beta_w\beta_n {\hat \Phi}_I^{w}(\mu){\hat \Phi}_I^{n}(\mu) = 0
\eeq
where we notice that $\frac{[1-{\hat K}_I^{w,n}(\mu)]}{\mu} ={\hat \Phi}_I^{w,n}(\mu)$ are the LTs of the persistence probabilities of the infection time spans.
Consider this equation for $\mu \to 0$ and take into account (\ref{mean_inf_time}) we arrive at
\beq
\label{Gfunction}
 G(0) = 1- \beta_w\beta_n \langle t_I^{w} \rangle \langle  t_I^{n} \rangle .
\eeq
We observe that $G(0) < 0$ for $R_0=\beta_n \beta_w \langle t_I^{w} \rangle \langle  t_I^{n} \rangle >1$. On the other hand, 
we have for $\mu \to \infty$ that ${\hat \Phi}_I^{w,n}(\mu) \to 0$ and hence
\beq
\label{ginfty}
G(\infty) = 1 .
\eeq
One can hence infer from complete monotony of ${\hat \Phi}_I^{w,n}(\mu)$ and therefore of ${\hat \Phi}_I^{w}(\mu){\hat \Phi}_I^{n}(\mu)$ (see Appendix \ref{endemic_stability}, Eq. (\ref{CM_def}) for a precise definition), that $\frac{d}{d\mu}G(\mu) > 0$ ($\mu \geq 0$) thus $G(\mu)=0$ has one single positive zero only if $G(0) <0$,
i.e. for $R_0 >1$ which therefore is the condition of instability of the healthy state (spreading of the disease). Conversely, for $R_0 <1$ the healthy state turns into a stable fixed point where there is no spreading of the disease.
In particular, the healthy state is always unstable ($R_0=\infty$) if at least one of mean infection time spans
$\langle t_I^{w,n} \rangle = \infty$. This is true for fat-tailed kernels scaling as $K_I^{w,n}(t) \propto t^{-\alpha-1}$ ($\alpha \in (0,1)$) for $t \to \infty$. We consider such a distribution briefly in subsequent section.
We plot function $G(\mu)$ versus $\mu$ for different $R_0$ for Gamma-distributed $t_I^{w,n}$ in Fig. \ref {healthy}.
Now we consider the stability of the endemic state with $G_e(\mu)=0$ where from (\ref{determinant_null}) this function reads
\beq
\label{endem_function}
G_e(\mu)= 1-\beta_w\beta_n{\hat \Phi}_I^w(\mu){\hat \Phi}_I^n(\mu)
+ \beta_wJ_n^e{\hat \Phi}_I^w(\mu) + \beta_nJ_w^e{\hat \Phi}_I^n(\mu) +
\beta_w\beta_n(J_w^e+J_n^e){\hat \Phi}_I^w(\mu) {\hat \Phi}_I^n(\mu)
\eeq
with
\beq
\label{endem_function_G}
\begin{array}{clr}
\ds G_e(0) &= \ds 1-R_0
+ \beta_w  \langle t_I^{w} \rangle J_n^e + \beta_n  \langle  t_I^{n} \rangle J_w^e +
(J_w^e+J_n^e)R_0 & \\ \\
 & = \ds R_0 - 1   .            &
 \end{array}
\eeq
On the other hand, $G_e(\infty) = 1$ (as ${\hat \Phi}_I^{w,n}(\infty)=0$) and from monotony of $G_e(\mu)$ follows that there is no positive solution of $G_e(\mu)=0$ for $R_0>1$. We plot $G_e(\mu)$ in Fig. \ref{endemic_stability_Fig} for different values of $R_0$ and Gamma distributed $t_I^{w,n}$. In Appendix \ref{endemic_stability} we complete the analytical proof that $G_e(\mu) > 0$ for $R_0>1$. 
\begin{figure}[H]
\centerline{\includegraphics[width=0.65\textwidth]{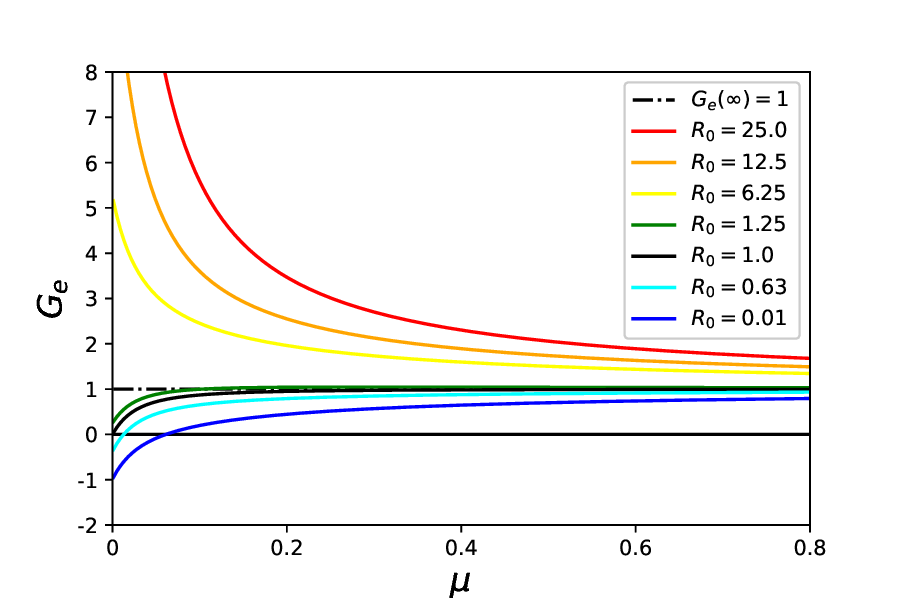}}
\caption{$G_e(\mu)$ of (\ref{endem_function}) for different values of $R_0$ where $G_e(\mu) >0$ for $R_0>1$ (stability of the endemic state).}
\label{endemic_stability_Fig}
\end{figure}
\section{Stability analysis of the healthy state with mortality}
\label{stability_healthy_mortality}

\begin{figure}[H]
\centerline{\includegraphics[width=0.65\textwidth]{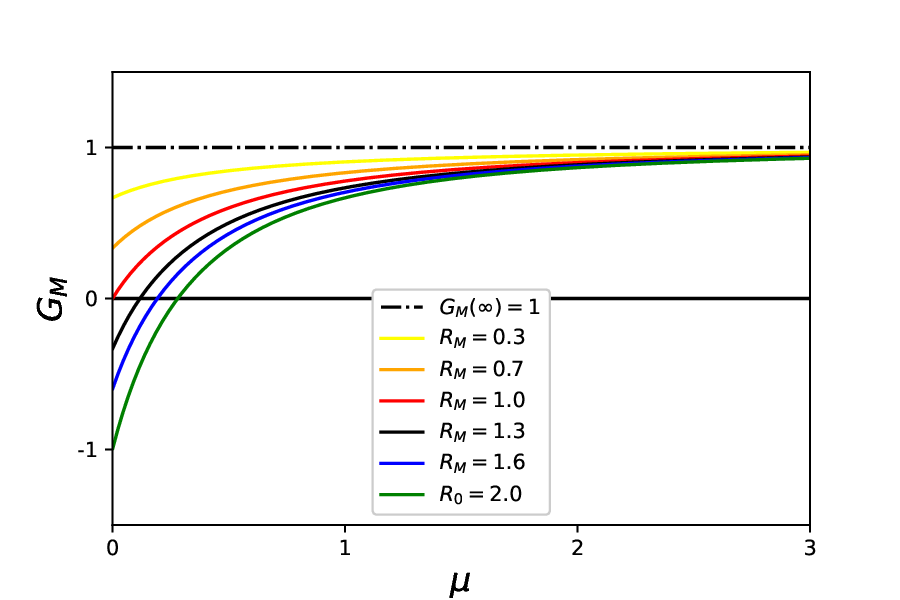}}
\caption{We depict function $G_M$ of Eq. (\ref{G-M-function}) for a few values of $R_M$
for exponentially distributed $t_I^{w,n}, t_M$. The basic reproduction number $R_M$ is monotonously decreasing 
with increasing mortality parameter $\xi_M$ (see Fig. \ref{ROM_with_morta}). 
The parameters are $\beta_{w,n}=1$, $\alpha_I^w=1$, $\xi_I^w=1$, $\alpha_n=1$, $\xi_I^n=0.5$ with $R_0=2$ where here $\langle t_{I,M} \rangle =R_0/(1+\xi_M)$.}
\label{GM_plot}
\end{figure}

\begin{figure}[H]
\centerline{
\includegraphics[width=0.65\textwidth]{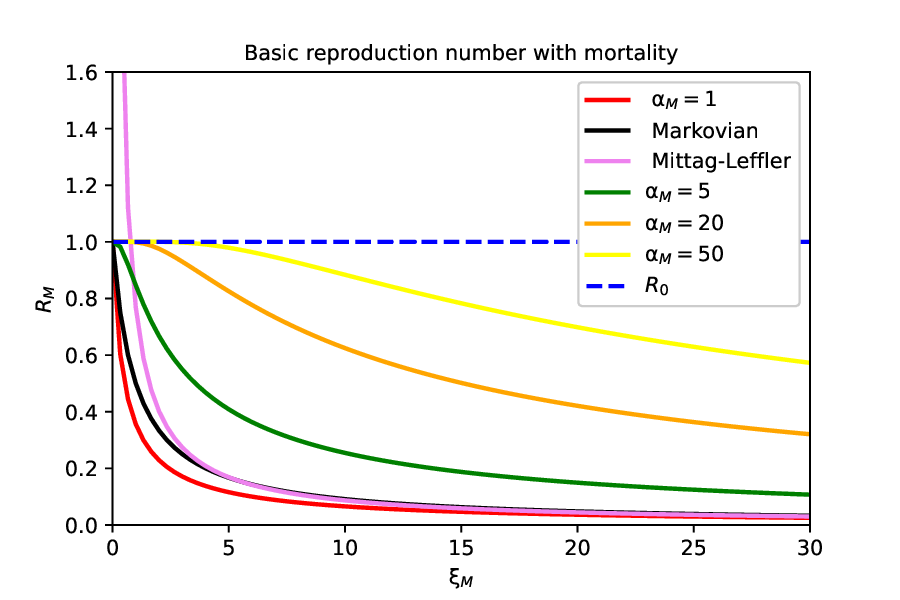}
}
\caption{Basic reproduction number $R_M$ of Eq. (\ref{basic_rep_with_mortality}) versus mortality rate parameter $\xi_M$ for Gamma distributed $t_I^{w,n},t_M$ for various $\alpha_M$ 
where we have set $\beta_n=\beta_w = \langle t_I^w \rangle = \langle t_I^n \rangle= 1$, ($\alpha_I^w=\xi_I^w=0.3$) and $\alpha_M=1$, $\alpha_w =\xi_I^w=1$ for the Markovian case which is recovered by Eq. (\ref{expon_gen}).}
\label{ROM_with_morta}
\end{figure}

An important question is, how mortality does modify the instability of the healthy state and the basic reproduction number.
To shed light on this question we perform a linear stability analysis of the healthy state $S_{w,n}=1$ where we set
\beq
\label{lin-stab-mort}
S_w(t) =1 + a \, e^{\mu t}, \hspace{0.5cm} J_w(t)= b \, e^{\mu t} , \hspace{0.5cm} d_w(t) = -(a+b)\, e^{\mu t} , 
\hspace{0.5cm} S_n(t) = 1 - c e^{\mu t}, \hspace{0.5cm} J_n(t) =  c e^{\mu t}
\eeq
with ${\cal A}_w(t) = \beta_w c\, e^{\mu t}$ and ${\cal A}_n(t) = \beta_n b\, e^{\mu t}$. 
Plugging this ansatz for $\mu \geq 0$ into three independent Eqs. of (\ref{evoleqs}), say the first, third and fourth one, and performing the averages (relaxing causality as previously) we arrive at
\beq
\label{mat_equations_mort}
\left[\begin{array}{lll} \mu \, ;& \hspace{0.5cm} 0 ;& \beta_w[1-{\hat b}_r(\mu)] ]   \\ \\
\mu ;& \hspace{0.5cm} \mu \, ;& \beta_w {\hat b}_d(\mu) \\ \\
 0 \, &  -\beta_n[1-{\bar K}_I^n(\mu)] \,  ;  & \mu
 \end{array}\right] \cdot \left(\begin{array}{l} a \\ \\ b \\ \\ c \end{array}\right) = \left(\begin{array}{l} 0 \\ \\ 0 \\ \\0 \end{array}\right) .
\eeq
Putting the determinant of the matrix to zero yields the condition
\beq
\label{solvability-modified}
\mu^2 -\beta_n\beta_w[1-{\bar K}_I^n(\mu)][1-{\hat b}_r(\mu)-{\hat b}_d(\mu)] = 0
\eeq
where the LTs ${\hat b}_r(\mu), {\hat b}_d(\mu)$ of the DPDFs $b_{r,d}(t)$ defined in (\ref{obs_rand}) come into play.
We are interested under which conditions there 
is a positive solution (instability of healthy state) of (\ref{solvability-modified}).
Since  ${\hat b}_r(0) = {\cal R}(\infty)$ and ${\hat b}_d(0) ={\cal D}(\infty)$ with ${\cal R}(\infty)+{\cal D}(\infty)=1$
we see that $\mu=0$ is solution of (\ref{solvability-modified}). 
Recall from (\ref{obs_rand}) that $b_d(t)+b_r(t) = {\cal K}_{I,M}^w(t) $ is the (properly normalized) PDF (\ref{KMIW}).
Condition (\ref{solvability-modified}) then reads
\beq
\label{G-M-function}
G_M(\mu) = 1- \beta_n\beta_w\frac{[1-{\bar K}_I^n(\mu)]}{\mu}\frac{[1-{\hat {\mathcal K}}^w_{I,M}(\mu)]}{\mu} = 0 
\eeq
where $\frac{[1-{\hat {\cal K}}^w_{I,M}(\mu)]}{\mu}$ is the LT of the persistence probability $\Phi_M(t)\Phi_I^w(t)$ of the walker's infection, i.e. the probability that $t< min(t_I^w,t_M)$ (see Remark I).
For zero mortality we have ${\cal K}^w_{I,M}=K_I^w$, ($b_r = K_I^w$ and $b_d=0$) retrieving condition (\ref{detnull_heathy}). 
The mean sojourn time in compartment $I$ with mortality yields
\beq
\label{ROM_res}
\langle  min(t_I^w,t_M) \rangle  = \langle t^w_{IM} \rangle  = \frac{[1-{\hat {\mathcal K}}^w_{I,M}(\mu)]}{\mu}\bigg|_{\mu=0} =\int_0^{\infty} t  {\cal K}_{I,M}^w(t){\rm d}t =
\int_0^{\infty}  \Phi_M(t)\Phi_I^w(t) {\rm d}t \, \leq \,\int_0^{\infty}  \Phi_I^w(t) {\rm d}t = \langle t_I^w \rangle
\eeq
where we arrive at
\beq
\label{G_M_mu_zero}
G_M(0) = 1- \beta_n\beta_w \langle t_I^n \rangle \langle t^w_{IM} \rangle  .
\eeq
Relation (\ref{ROM_res}) shows that 
$\langle t^w_{IM} \rangle  \leq \langle t_I^w \rangle $ (equality only for zero mortality).
On the other hand we have $G_M(\infty)= 1$, so there is a positive solution of $G_M(\mu)=0$ only if 
\beq
\label{basic_rep_with_mortality}
R_M =\beta_n\beta_w \langle t_I^n \rangle \langle t^w_{IM} \rangle \, > \, 1
\eeq
where $R_M$ is the basic reproduction number modified by mortality with $R_M \leq R_0$ (equality only for zero mortality). 
To visualize the effect of mortality on the instability of the healthy state we plot $G_M(\mu)$ for a few values
of $R_M$ in Fig. \ref{GM_plot}. Increasing mortality turns an unstable healthy state into a stable one.
In the random walk simulations we deal with Gamma distributed $t_I^{w,n},t_M$ where the persistence probabilities are then normalized incomplete Gamma functions (\ref{incomplete_Gamma}). To explore the effect of mortality for such cases, we determine $R_M$ by numerical integration of (\ref{ROM_res})
as a function of the mortality rate parameter $\xi_M$ and plot the result in Fig. \ref{ROM_with_morta} where one can see that $R_M$ is monotonous decreasing with mortality rate $\xi_M$.
We also include a case of a fat-tailed (Mittag-Leffler) distributed $t_I^w$ which we discuss hereafter.
The parameters in Fig. \ref{ROM_with_morta} are such that the zero mortality case occurs with $R_0=1$ as the upper bound.
The essential feature is that $R_M$ decays monotonically with increasing mortality rate parameter $\xi_M$ approaching zero for $\xi_M \to \infty$. Diseases with high mortality stabilize the healthy state even for $\langle t_I^w \rangle \to \infty$.

Consider briefly the case where $t_M$ is exponentially distributed (i.e. $\alpha_M=1$ in the Gamma distribution of $t_M$) with $\Phi_M(t)= e^{-\xi_M t}$. Then we have $\langle t_{IM}^w \rangle = {\hat \Phi}_I^w(\xi_M) $ thus
\beq
\label{expon_tM}
R_M = \beta_w\beta_n \langle t_I^n \rangle {\hat \Phi}_I^w(\xi_M) .
\eeq
The zero mortality case is recovered for $\xi_M=0$ with ${\hat \Phi}_I^w(0) = \langle t_I^w \rangle$.
For Gamma distributed $t_I^{w,n}$ this yields

\beq
\label{expon_tM_Gamma}
R_M = \beta_w\beta_n \frac{\alpha_I^n}{\xi_I^n \xi_M}\left(1- \frac{(\xi_I^w)^{\alpha_I^w}}{(\xi_M+\xi_I^w)^{\alpha_I^w}}\right) .
\eeq
where $\alpha_I^{w,n}$, $\xi_I^{w,n}$ are the parameters of the respective Gamma distributions of the infection times of nodes and walkers.
The Markovian case where all waiting times are exponentially distributed is covered for
$\alpha_I^{w,n} =1$ and yields
\beq
\label{expon_gen}
R_M = \frac{\beta_w\beta_n}{\xi_I^n(\xi_I^w+\xi_M)} 
\eeq
containing the zero mortality case for $\xi_M=0$.
\\[2ex]
{\it Fat-tailed distributed $t_I^w$:} \\
Finally, an interesting case emerges if $t_I^w$ follows a fat-tailed distribution, i.e. $\Phi_I^w(t) \propto t^{-\alpha}$ for $t$ large ($\alpha \in (0,1)$) and $\langle t_I^w\rangle =\infty$, $R_0=\infty$. Let us have a look, how mortality is affecting this situation. Fat tailed $t_I^w$ distributions describe diseases where the infectious periods are very long and the healthy state without mortality is extremely unstable ($R_0=\infty$). Infected walkers can infect many nodes during their long infection time spans. An important case of this class is constituted by the Mittag-Leffler (ML) distribution $\Phi_I^w(t) = E_{\alpha}(-\xi_I^w t^{\alpha})$ where $E_{\alpha}(\tau)$ 
indicates the Mittag-Leffler function, see \cite{Metzler-Klafter2000,Mainardi-etal2004} and references therein for representations and connections with fractional calculus. The ML function recovers the exponential for $\alpha=1$ ($E_1(-\tau)=e^{-\tau}$). 
Assuming exponential mortality $\Phi_M(t) = e^{-\xi_M t}$ one obtains with (\ref{expon_tM}) 
\beq
\label{fractional_case}
R_M =\beta_w\beta_n \langle t_I^n \rangle \frac{(\xi_M)^{\alpha-1}}{\xi_I^w + (\xi_M)^{\alpha}} , \hspace{1cm} \alpha \in (0,1)
\eeq
containing the LT of the ML persistence probability distribution ${\hat \Phi}_I^w(\lambda) = 
\lambda^{\alpha-1}/(\xi_I^w+\lambda^{\alpha})$.
The essential feature here is that $R_M$ is weakly singular at $\xi_M=0$ with a monotonously decreasing $\xi_M^{\alpha-1}$ scaling law,
where the healthy state becomes stable for mortality parameters larger as $\xi_M \approx 1$.
We depict this case in Fig. \ref{ROM_with_morta} for $\alpha= \xi= 0.3$ (violet curve). 

\section{Random walk simulations}
\label{simulations}
The remainder of our paper is devoted to test the mean field model under "real world conditions" which we mimic by
$Z=Z_S(t)+Z_I(t)+Z_D(t)$ random walkers navigating independently on an undirected connected (ergodic) graph. 
In our simulations we focused on Barabási-Albert (BA), Erdös-Rényi (ER) and Watts-Strogatz (WS) graphs
\cite{RiascosSanders2021,NohRieger2004,fractional_book2019} (see Appendix \ref{complex_graphs} for a brief recap) and implemented the compartments and transmission pathway for walkers and nodes outlined in Section \ref{SIS-motality}. 
A susceptible walker gets infected with probability $p_w$ by visiting an infected node, and a susceptible node gets infected
with probability $p_n$ at a visit of an infectious walker. We assume that the infection probabilities $p_{n,w}$ are constant for all nodes and walkers, respectively.
They are related yet not identical with the macroscopic rate constants $\beta_{w,n}$.
A critical issue is whether the simple bi-linear forms for the mean field infection rates (\ref{infections_rate}) 
still capture well the complexity of the spreading in such "real world" networks. 
One goal of the subsequent case study is to explore this question. 

We characterize the network topology by $i=1,\ldots N$ nodes with the $N\times N$ adjacency matrix $(A_{ij})$ where $A_{ij}=1$ if the pair of nodes $i,j$ is connected by an edge, and $A_{ij}=0$ if the pair is disconnected. 
Further, we assume $A_{ii}=0$ to avoid self-connections of nodes. We confine us to undirected networks where the edges have no direction and the adjacency matrix is symmetric. The degree $k_i$ of a node $i$ counts the number
of neighbor nodes (edges) of this node.
Each walker $z=1,..,Z$ performs simultaneous independent random steps at discrete time instants $t=\Delta t, 2 \Delta t, \ldots $ from one to another connected node. The steps from a node $i$ to one of the neighbor nodes are chosen with probability $1/k_i$, following for all walkers the same transition matrix 
\beq
\label{markovian walk}
\Pi(i \to j) = \frac{A_{ij}}{k_i} , \hspace{1cm} z=1,\ldots, Z , \hspace{1cm} i,j=1,\ldots, N
\eeq
which is normalized $\sum_{j=1}^N\Pi(i \to j) =1$.
This is a common way to connect the network topology with simple Markovian random walks \cite{NohRieger2004,Newman2010}. In the simulations the departure nodes at $t=0$ of the walkers are randomly chosen. The path of each walker is independent and not affected by contacts with other walkers
or by transition events from one to another compartment.

\begin{figure}[H]
\centerline{\includegraphics[width=1.3\textwidth]{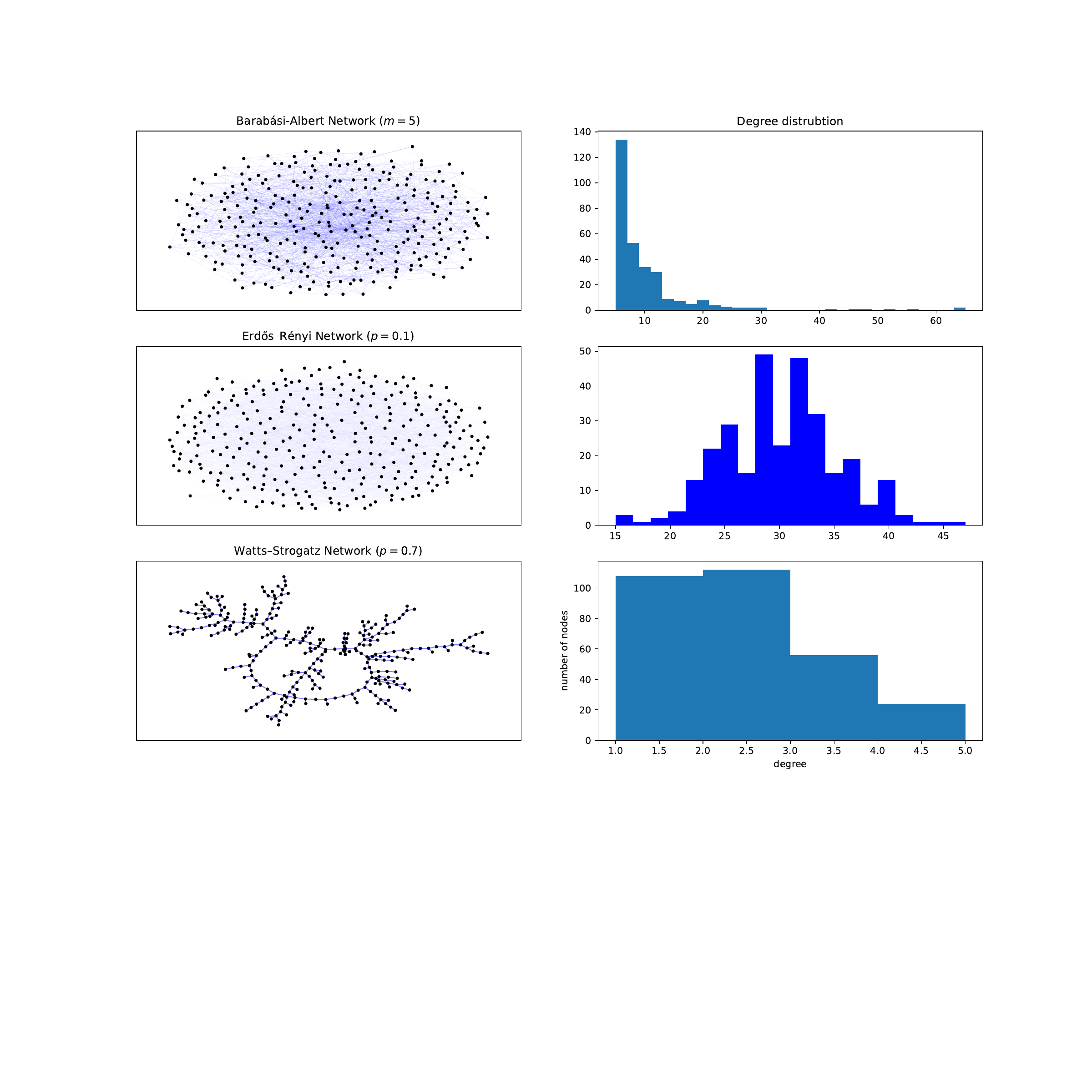}}
\vspace{-6cm}
\caption{Barab\'asi-Albert, Erd\"os-Renyi and Watts-Strogatz types with $300$ nodes and connectivity parameters used in some of the simulations. The WS graph for connectivity parameter $m=2$ lacks the small world property resembling a complex real world structure. The ER network
has a broad degree distribution and the small world property. The BA graph is for $N\to \infty$ asymptotically scale-free with a power law degree distribution and the small world feature where a large number of nodes have small degrees and a few (hub) nodes with very large degrees. Almost all nodes are only a few links away from hub nodes.}
\label{networks}
\end{figure}

\paragraph{Case study and discussion}
In order to compare the epidemic dynamics of the mean field model and random walk simulations
we integrate the stochastic evolution Eqs. (\ref{evoleqs}) numerically as follows.
We average the increments of the compartmental fractions in a generalized Monte-Carlo sense converging towards 
the convolutions of the right hand side of (\ref{evoleqsB}) where we use the Monte-Carlo convergence feature
\beq
\label{MC_feature}
\lim_{n\to \infty} \frac{1}{n} \sum_{k=1}^n A(t-T_k) = 
\int_0^tA(t-\tau)K(\tau){\rm d}\tau 
\eeq
for random variables $T$ drawn from PDFs $K(\tau)$. We perform this average for any time increment ${\rm d}t$ with respect to all involved independent random time spans
$t_I^{w,n},t_M$ (see Appendix \ref{general}) and integrate the averaged compartmental increments in a fourth order Runge-Kutta scheme (RK4). 
We use in the random walk simulations and the Monte-Carlo (mean field) integration  
exactly the same (Gamma distributed) random values (PYTHON seeds) for the $t_I^{w,n},t_M$. The values of the infection rate parameters $\beta_{w,n}$ used in the mean field integration are determined from Eq. (\ref{simple_representation}) by plugging in the large time asymptotic values of the random walk simulation with identical parameters (without mortality). The compartmental fractions in the random walk simulations are determined by simply counting the compartmental populations at each time increment $\Delta t$ of walker's steps.
The so determined rate parameters $\beta_{w,n}$ plugged into the mean field integration depend in a complex manner on the infection probabilities $p_{w,n}$ and topology of the network. In this way this information 
is also contained in the basic reproduction numbers with and without mortality.
We explore the spreading in random graphs of different complexity such as represented in Fig. \ref{networks}.
The BA graph is small world with power law distributed degree (Appendix \ref{complex_graphs})
which means that there are many nodes having a few connections, and a few (hub) nodes with a huge number of connections.
The average distance between nodes becomes small, as it is sufficient that almost every node is only a few links away from a hub node.
The ER graph is small world due to a broad degree distribution. The WS graph with the choice of connectivity parameter $m=2$ in Fig. \ref{networks} has long average distances and is large world. Intuitively, one infers that a small world structure is favorable for spreading processes, a feature which was already demonstrated in the literature 
\cite{Satoras-Vespignani-etal2015,Pastor-SatorrasVespignani2001}. In our simulations spreading in network architectures  with increased connectivity comes along with increased values of $R_0$ and $R_M$, respectively.

\begin{figure}[H]
\centerline{
\includegraphics[width=0.8\textwidth]{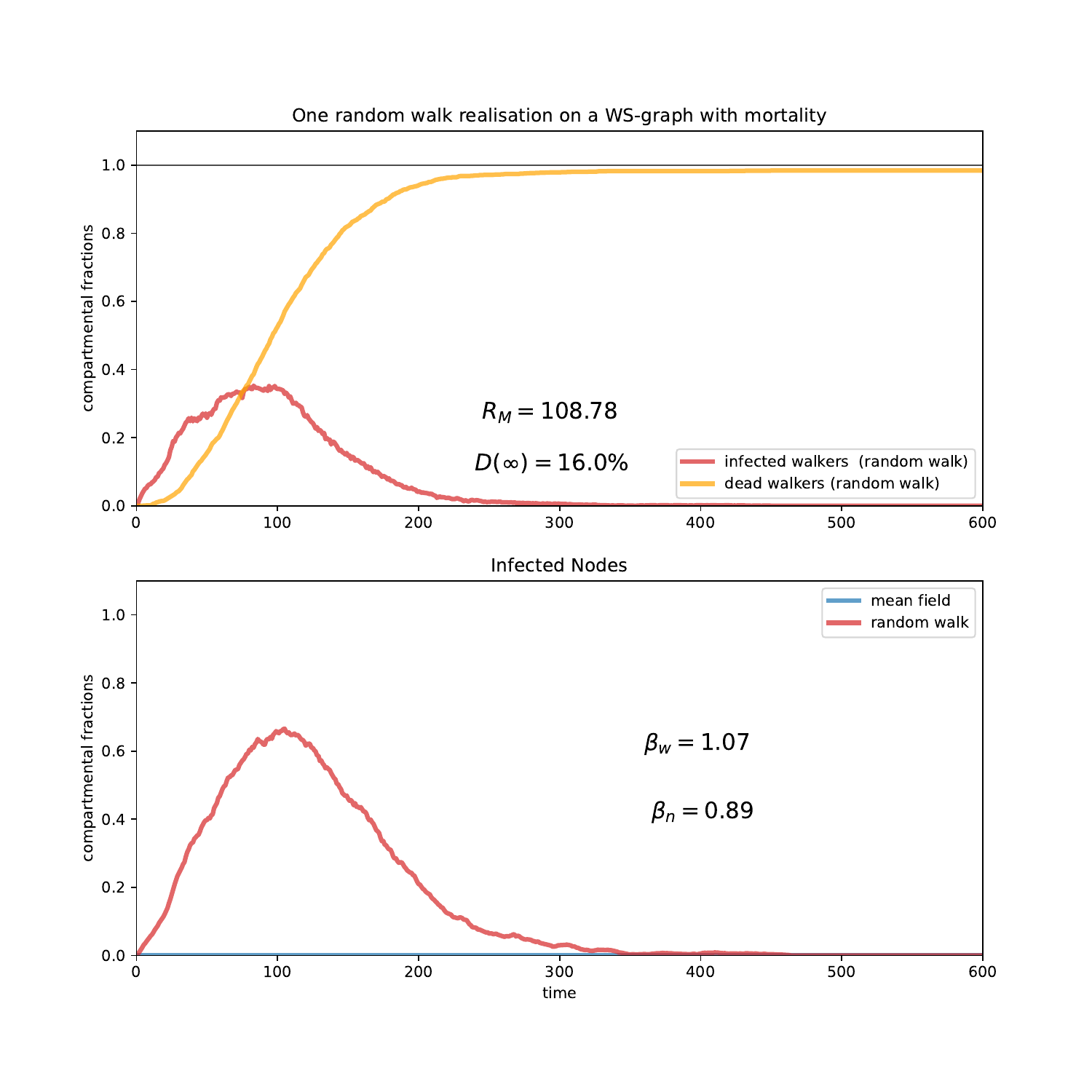}
}
\centerline{
\includegraphics[width=0.35\textwidth]{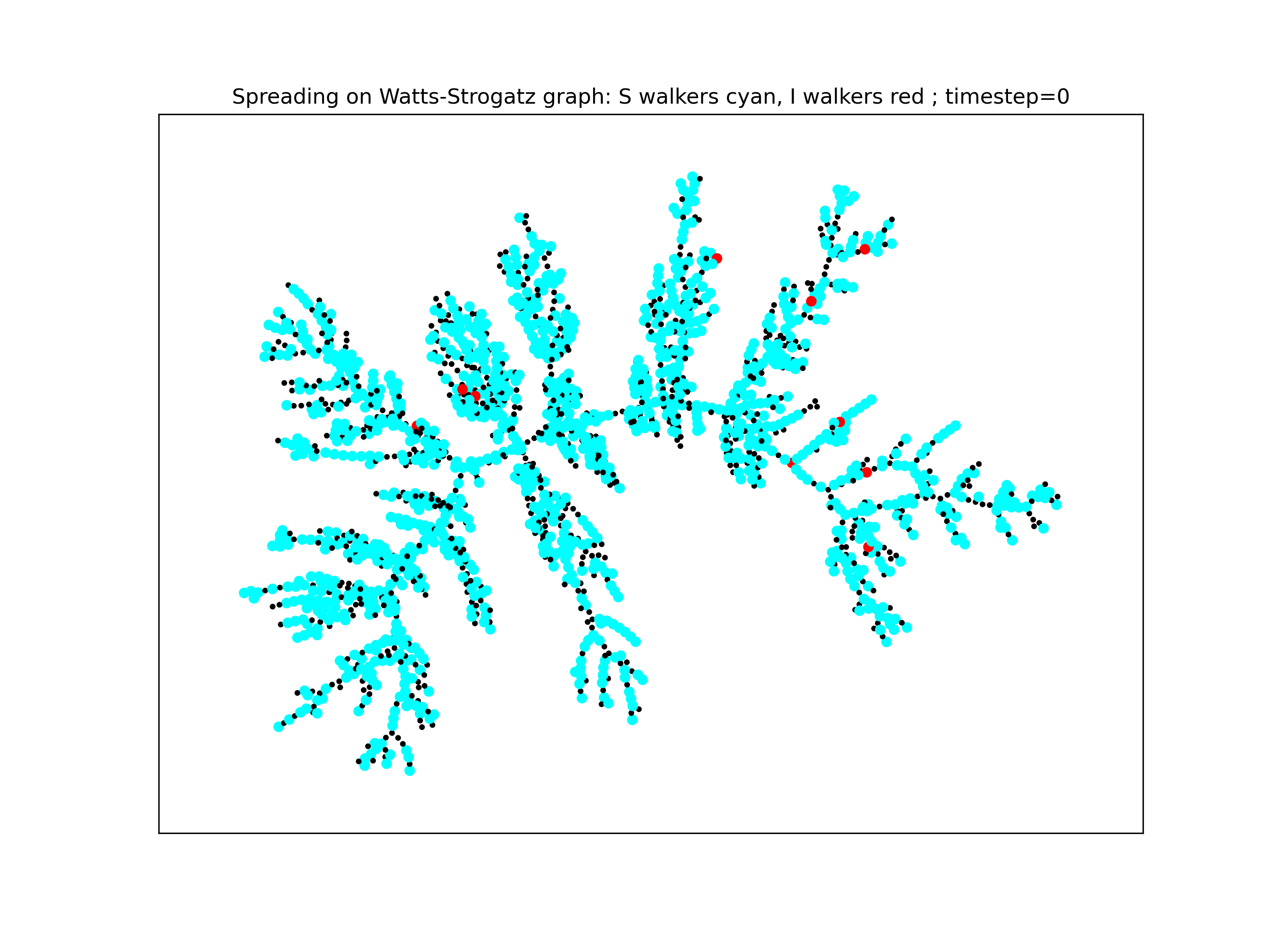}
\includegraphics[width=0.35\textwidth]{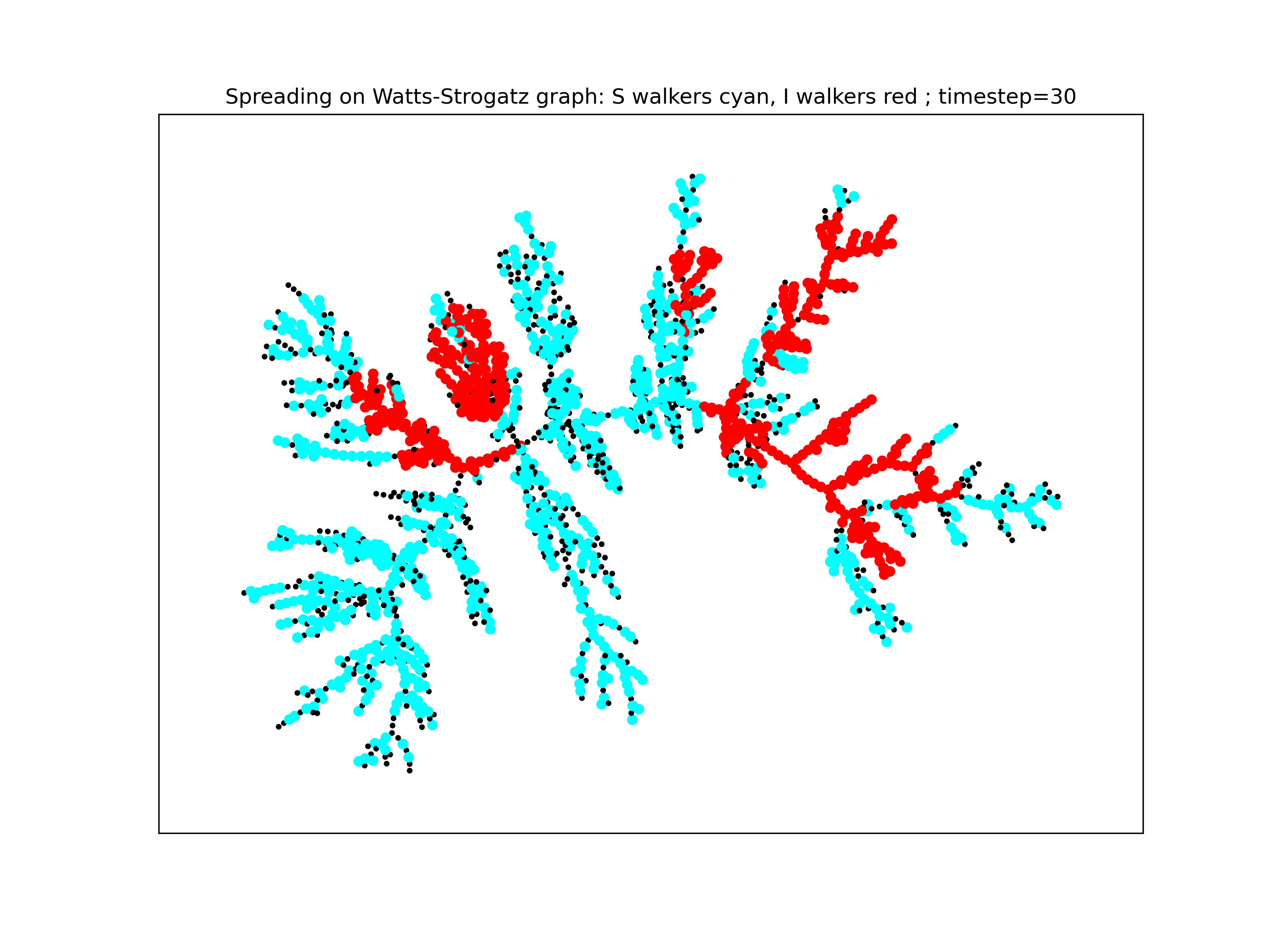}
\includegraphics[width=0.35\textwidth]{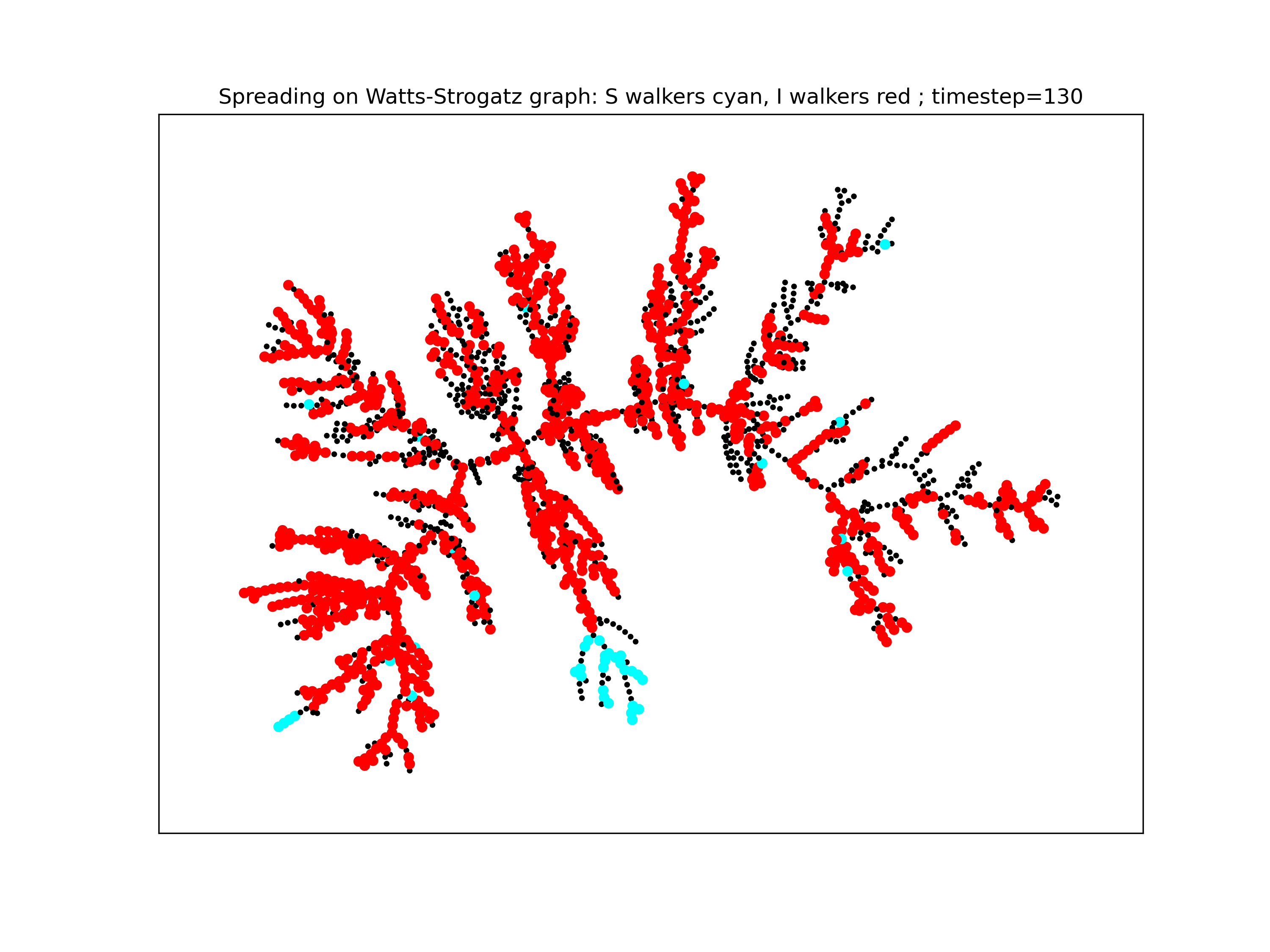}
}
\centerline{
\includegraphics[width=0.35\textwidth]{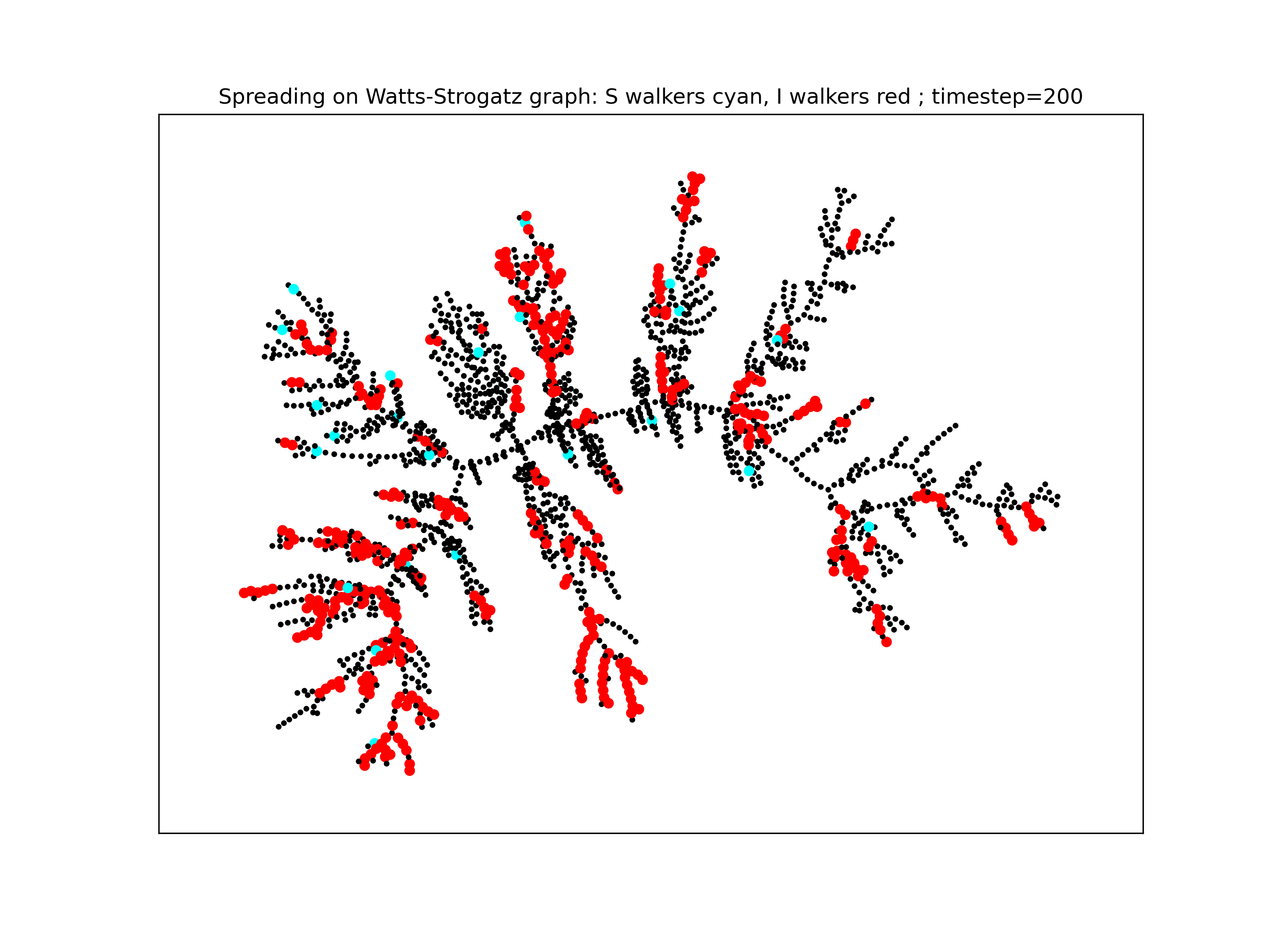}
\includegraphics[width=0.35\textwidth]{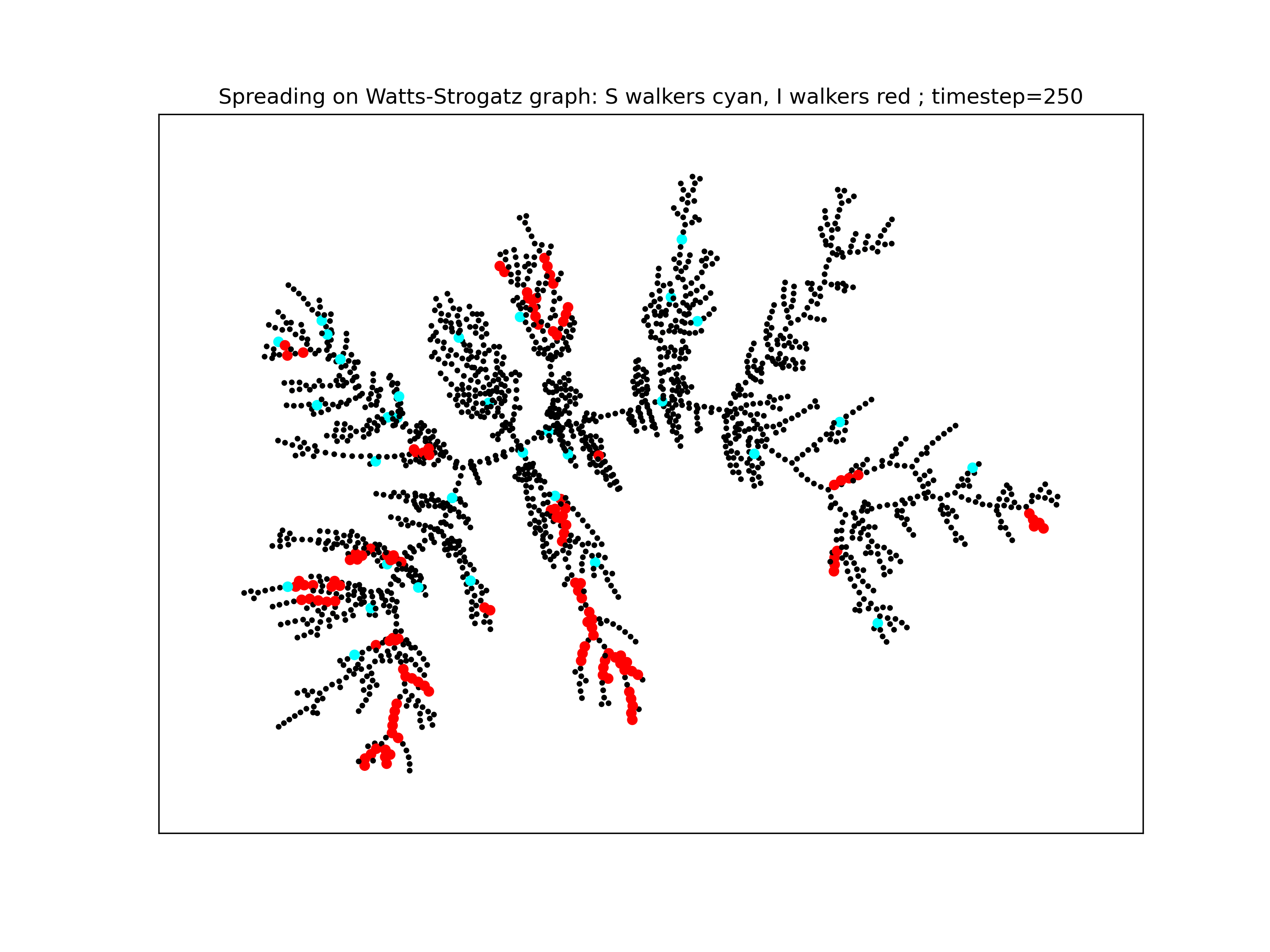}
\includegraphics[width=0.35\textwidth]{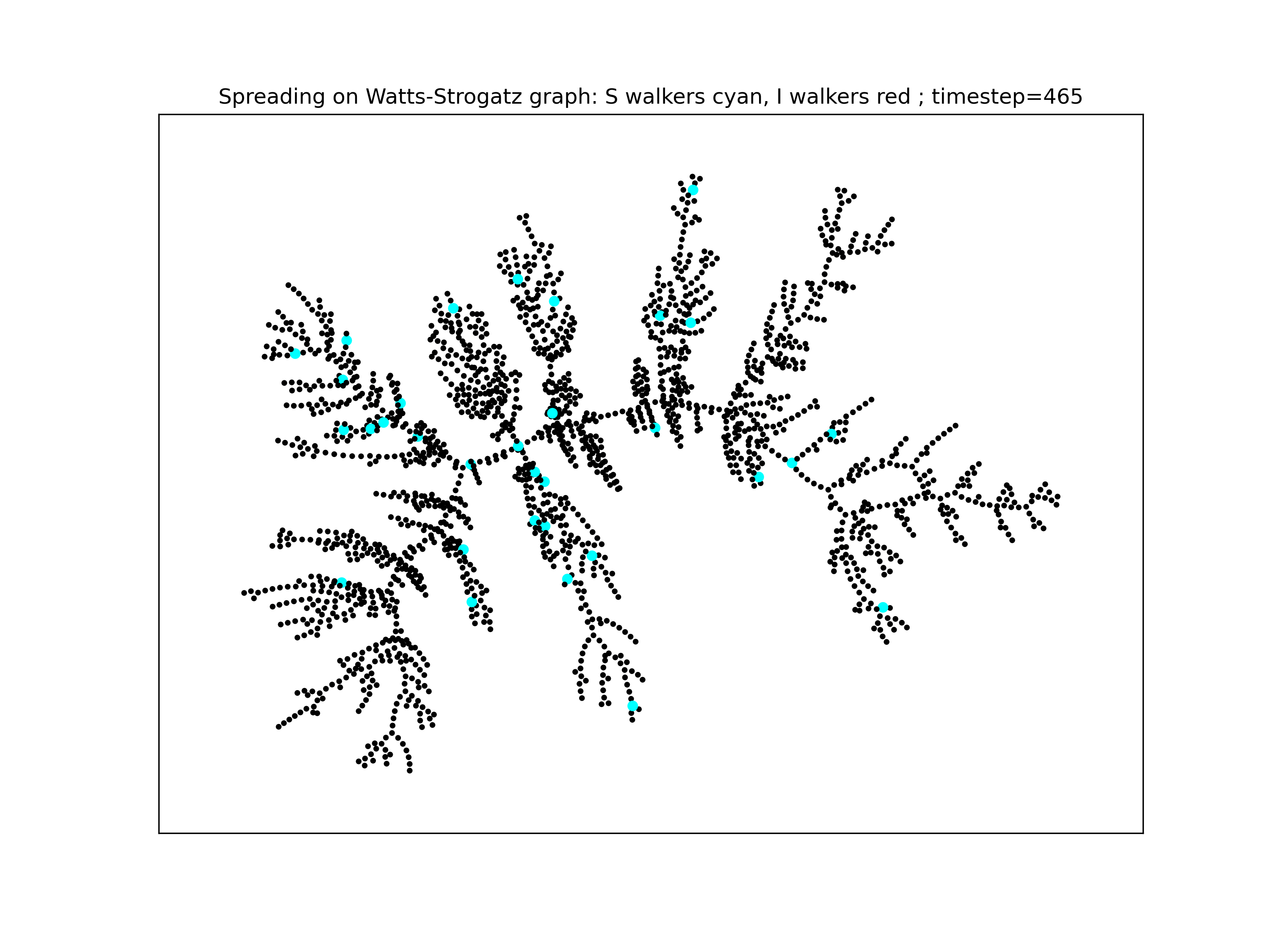}
}
\caption{Snapshots of spreading in a WS graph ($Z=2000$ walkers, $N=2000$ nodes, connectivity parameter $m=2$) and mortality parameter $\xi_M=0.4$ with ${\cal D}(\infty)\approx 16\%$. The remaining parameters are the same as in Fig. \ref{WS3}. One observes
$d_w(\infty)\approx 0.99$ and $S_w(\infty) \approx 1\%$ with only about 20 surviving walkers after extinction of the disease. S walkers are in cyan color, I walkers red, D walkers invisible and nodes without walkers are represented in black. Consult here an
\href{https://drive.google.com/file/d/1-fhroAsoAVDKGR5H9yWtqjD7A1ZU5pQt/view}{\textcolor{blue}{animated video of this process}}.}
\label{WS4}
\end{figure}

We identify the starting time instant ($t=0$) of the evolution in the mean field model with the time instant of the first infection of a walker in the random walk simulations. 
In all cases we start with a small number of randomly chosen initially infected nodes $N_I(0)= 10 \ll N$ ($N_I(0) \approx 10$) and no infected or dead walkers.
To reduce the numbers of parameters and to concentrate on topological effects we have put in all simulations the transmission probabilities $p_w=p_n=1$. 
We refer to \cite{supplementaries} for the PYTHON codes\footnote{Free to download and non-commercial use.} and animated simulation videos related to the present study. 
In order to visualize a typical spreading process, we depict in Fig. \ref{WS4} a few snapshots
in a Watt-Strogatz graph with rather high overall mortality probability of ${\cal D}(\infty) \approx 16\%$.
In this case a single infection wave emerges where a large part of walkers gets
repeatedly infected increasing their probability to die. 
This leads to a very high fraction of eventually dead walkers $d_w(\infty) \approx 99\%$ and small fraction 
$S_w(\infty) \approx 1\%$ of surviving walkers corresponding to the stationary state (\ref{Sw_infty}) which is taken as soon as the disease gets extinct $J_w=J_n=0$.
Fig. \ref{WS4} shows that first the infection gains large parts of the network 
consistent to the large value of $R_M$ observed in this case.
After the first wave the disease gets extinct by the high mortality of the walkers. 
A disease with a similar high mortality characteristics is for instance Pestilence.
The process of Fig. \ref{WS4} is visualized in an \href{https://drive.google.com/file/d/1-fhroAsoAVDKGR5H9yWtqjD7A1ZU5pQt/view}{\textcolor{blue}{animated video}}. 

\begin{figure}[H]
\centerline{
\includegraphics[width=0.55\textwidth]{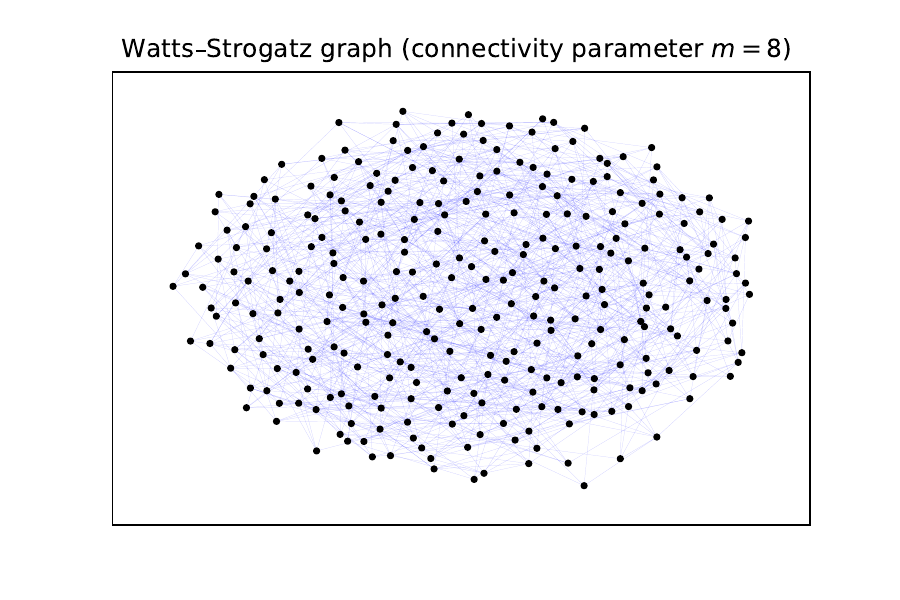}
\includegraphics[width=0.55\textwidth]{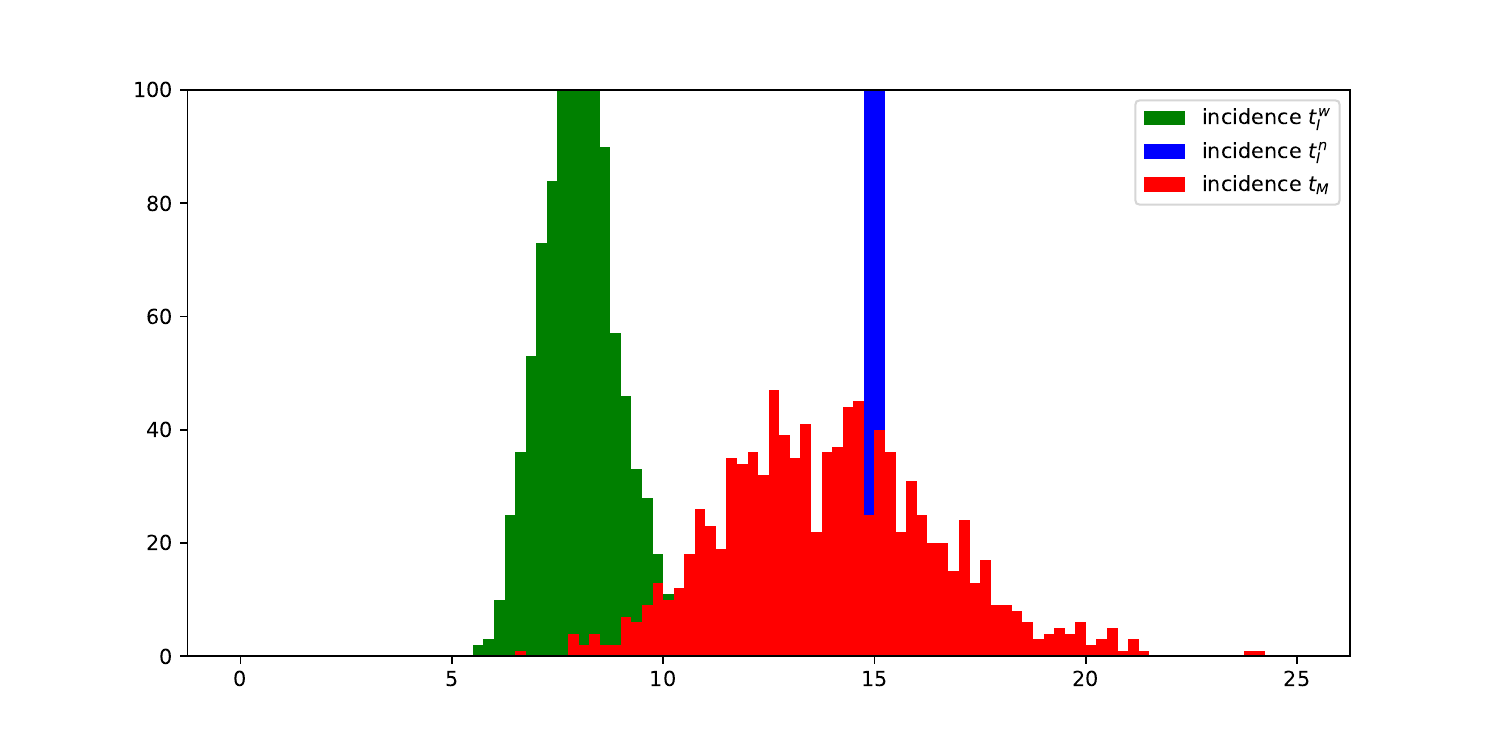}
}
\centerline{
\includegraphics[width=0.55\textwidth]{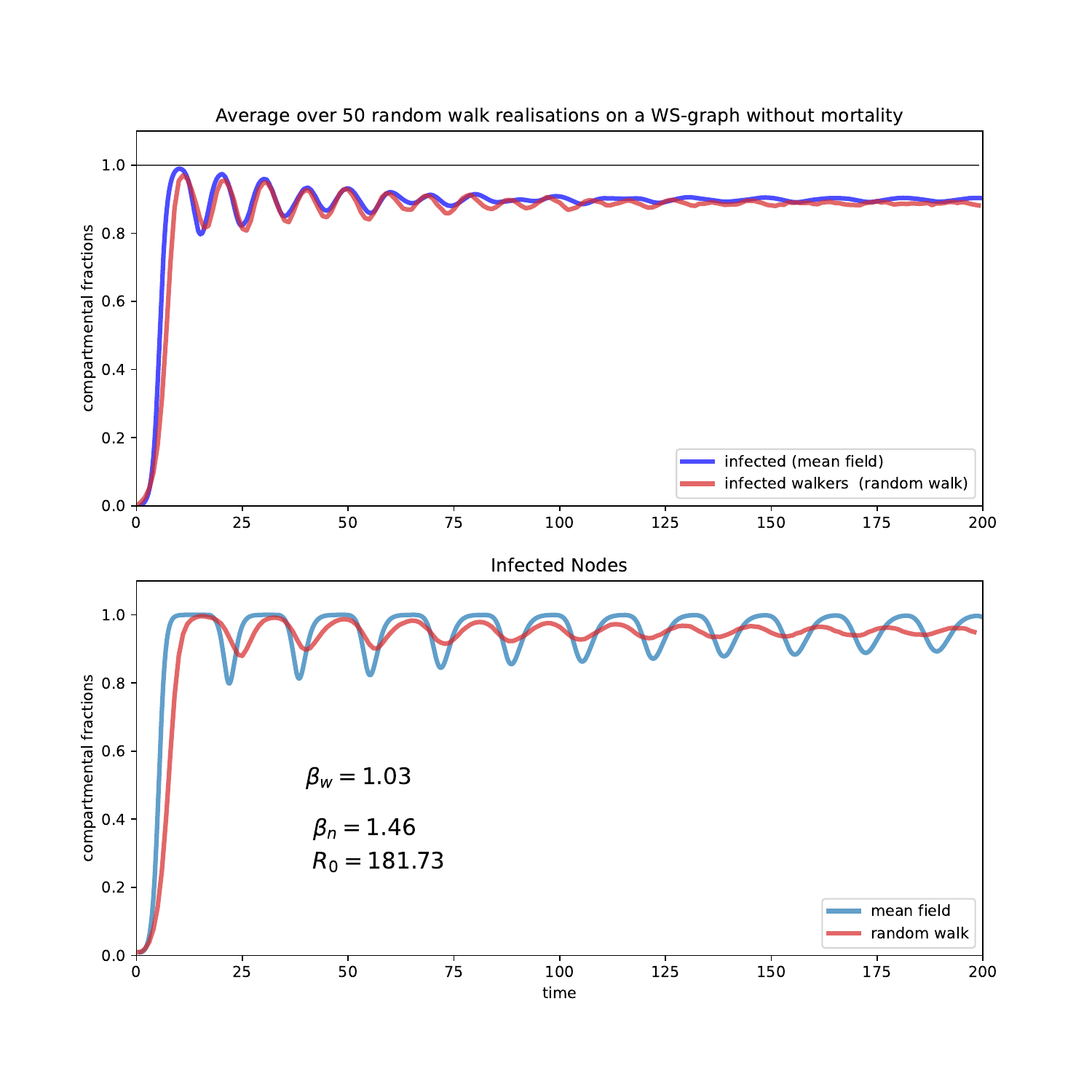}
\includegraphics[width=0.55\textwidth]{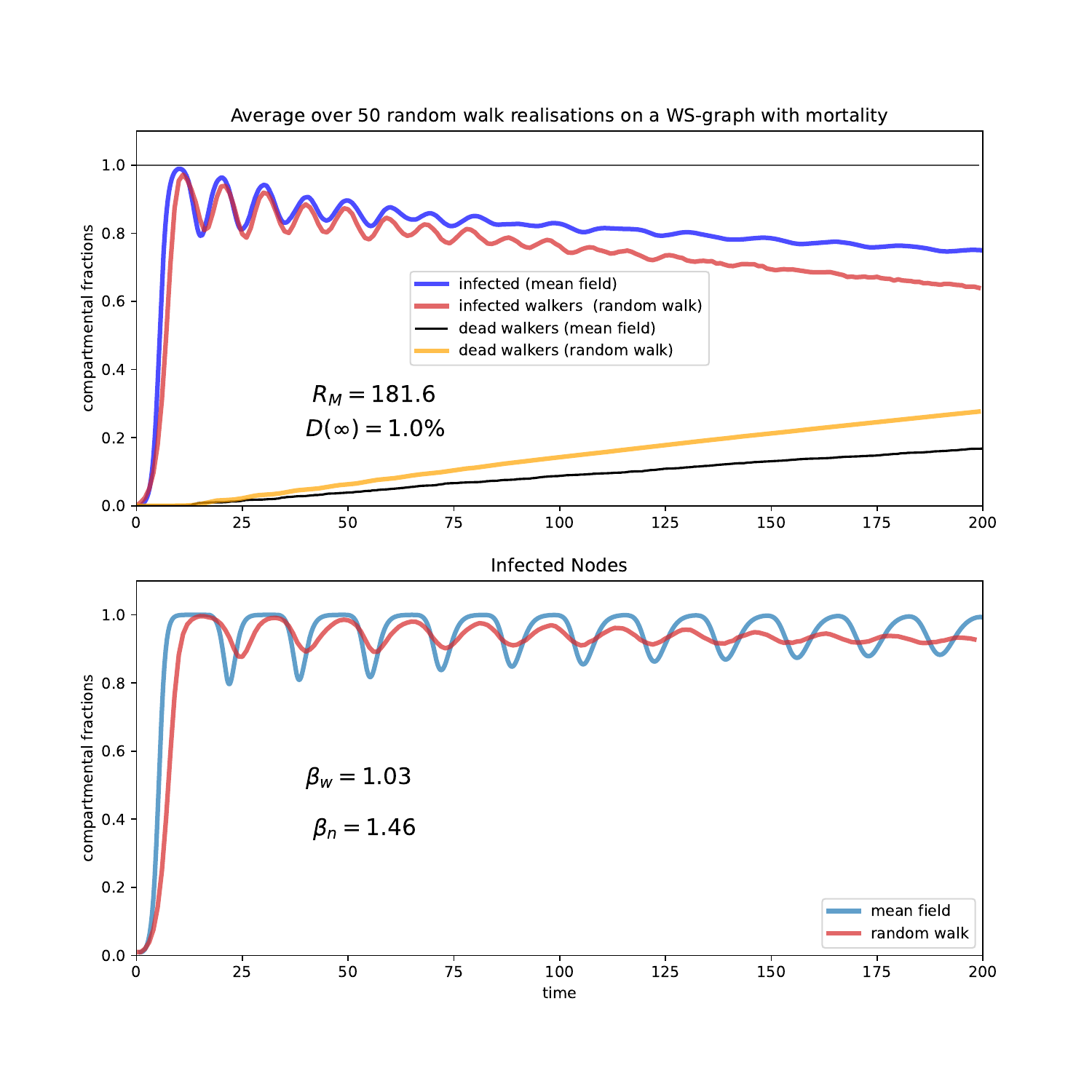}
}
\caption{The plots show the evolution on WS graph with $Z=1000$ walkers for connectivity parameter $m=8$ and rewiring probability $p=0.7$ ($nx.connected\_watts\_strogatz\_graph(N =1000 , m=8 , p=0.7, seed=seed)$) without mortality (left frame) and with mortality (right frame). $t_I^{w,n},t_M$ are Gamma distributed with the parameters $\langle t_M \rangle =14$,
$\xi_M=2$, $\langle t_I^{w}\rangle =8$, $\xi_I^w=10$, and $\langle t_I^n\rangle=15$, $\xi_{n} =10^5$, see histogram. ${\cal D}(\infty) \approx 0.01$ and is determined by numerical integration of (\ref{obs_rand}). }
\label{WS8}
\end{figure}

\begin{figure}[H]
\centerline{
\includegraphics[width=0.55\textwidth]{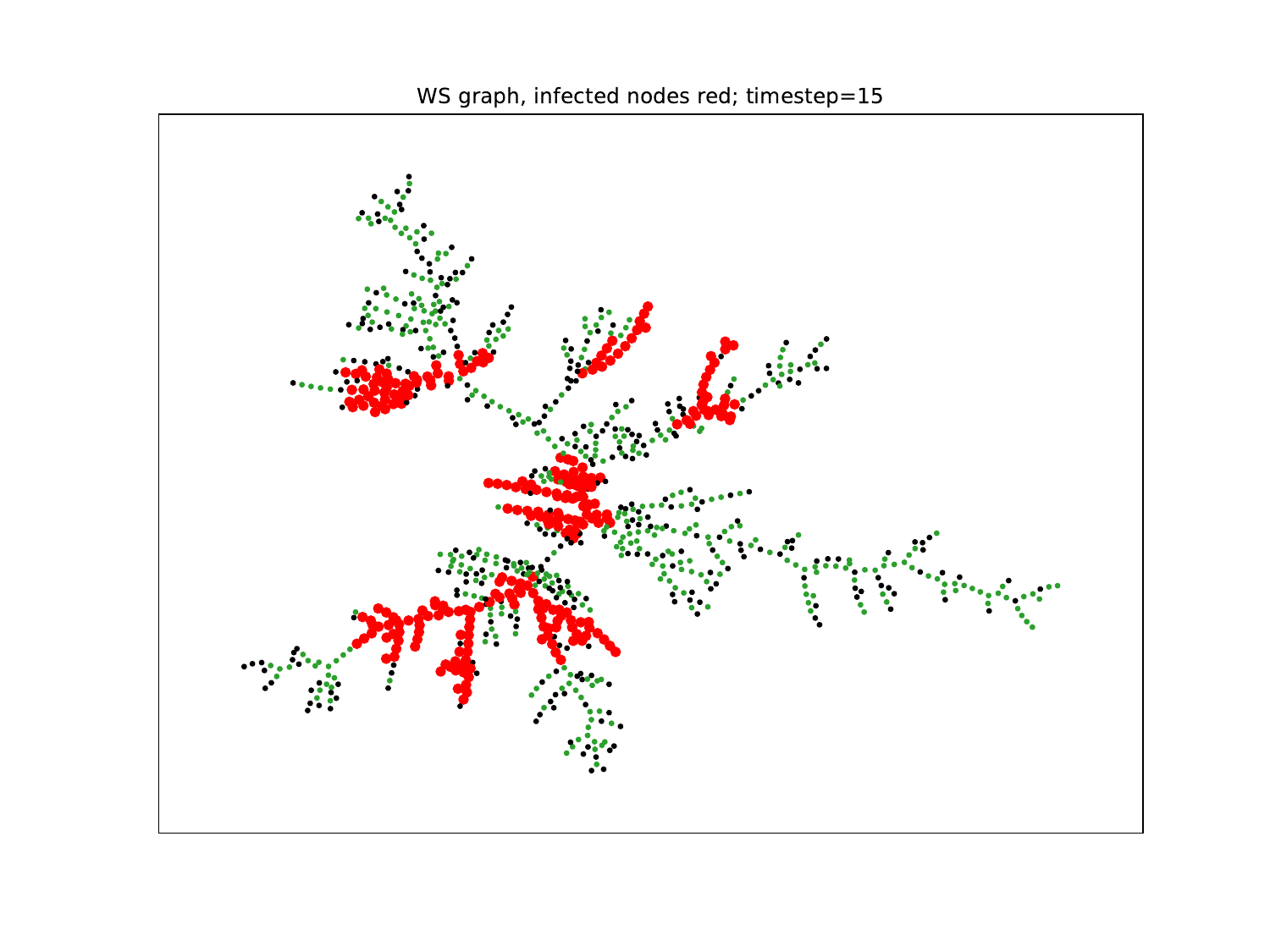}}
\centerline{
\includegraphics[width=0.55\textwidth]{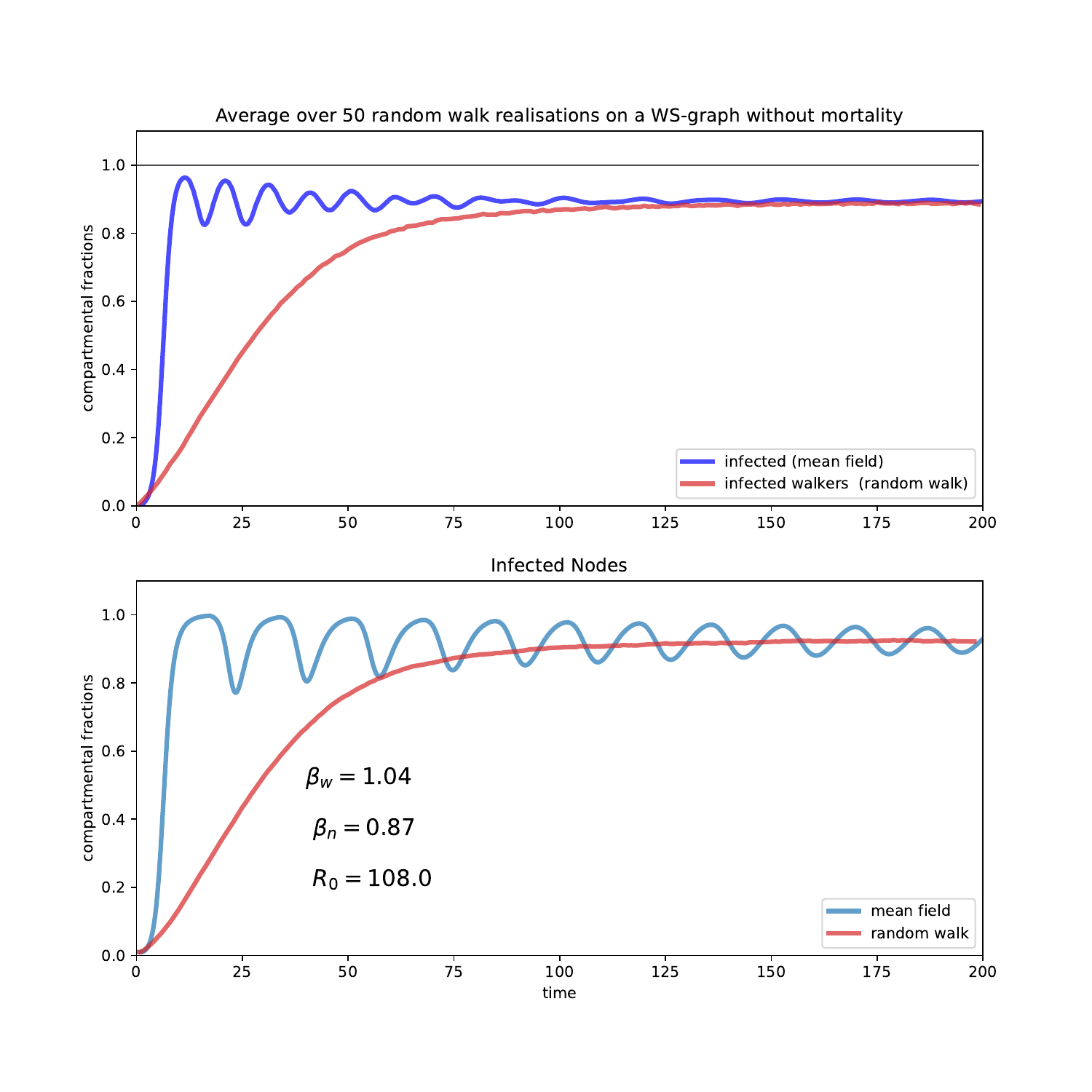}
\includegraphics[width=0.55\textwidth]{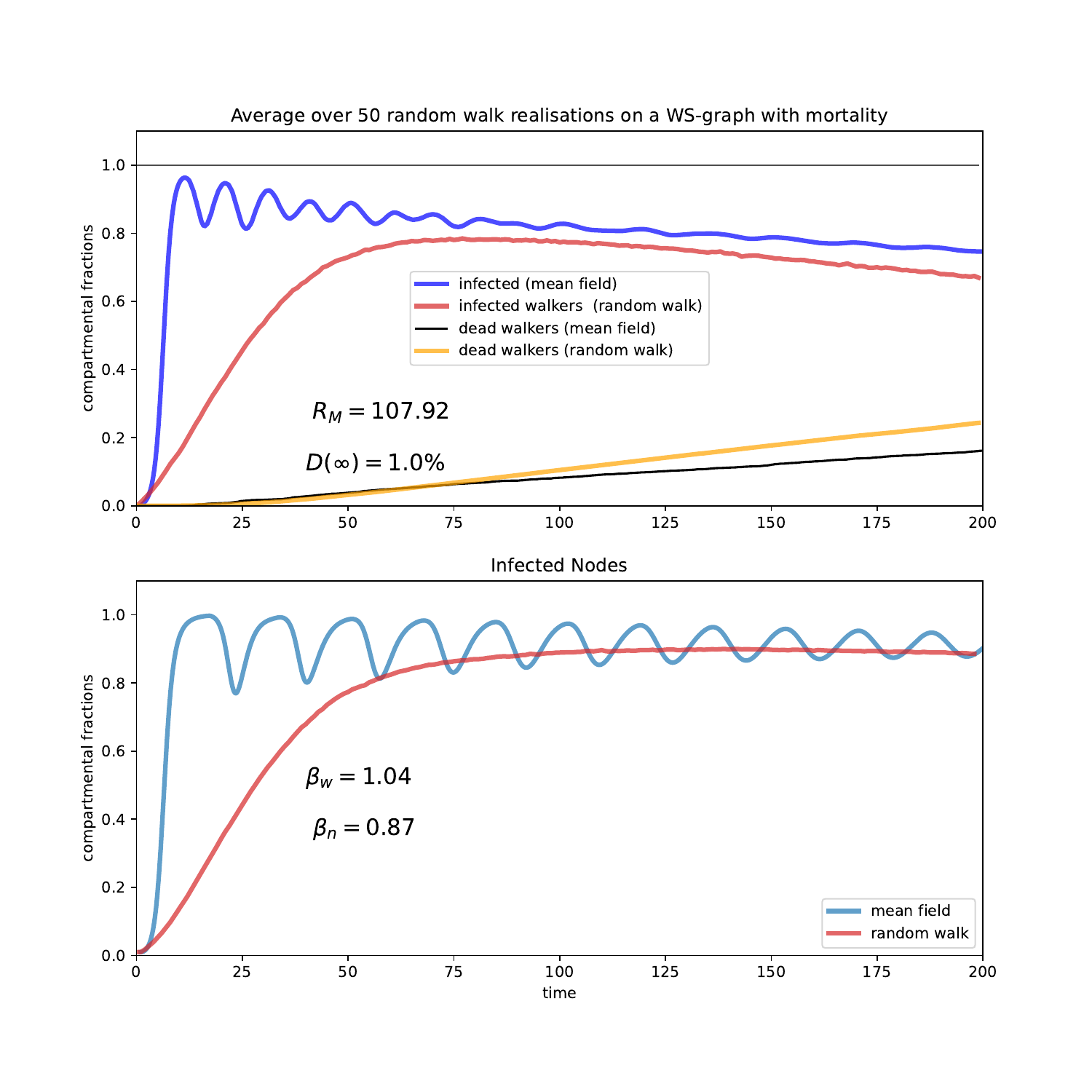}}
\caption{Evolution on WS graph with $Z=1000$ walkers and $N=1000$ nodes for the same parameters as in Fig. \ref{WS8} but with reduced connectivity parameter $m=2$.
The upper frame shows a snapshot ($t=15$) of the spreading in one random walk realization (susceptible walkers green, 
susceptible nodes black, infected nodes red).}
\label{WS3}
\end{figure}

Figs. \ref{WS8} and \ref{WS3} show the evolution in WS graphs with identical parameters and Gamma distributions of $t_I^{w,n},t_M$ as in Fig. \ref{WS4}
but with different mortality rate parameter $\xi_M$ and a much smaller overall mortality $ {\cal D}(\infty) \approx 1\%$. 
The different network connectivity leads to different values of $\beta_{w,n}$ and mean field solutions in Figs. \ref{WS8} and \ref{WS3}.
In addition, the networks
of Figs. \ref{WS8} and \ref{WS3} have different connectivity features. The graph of Fig. \ref{WS8} is small world (highly connected) whereas the WS graph in \ref{WS3} is weakly connected and large world.
One observes in Fig. \ref{WS8} that the infection numbers exhibit strong and immediate increases followed by attenuated oscillations around the endemic equilibrium (for zero mortality) 
with high values $J_w^e \approx 0.9$ and $J_n^e \approx 0.95$. 
The basic reproduction numbers with mortality are in both graphs only slightly smaller as $R_0$. 
This is due to a rather small overall mortality of ${\cal D}(\infty)\approx 0.01$. This effect can also be seen in the small overlap of the Gamma distributions of $t_I^w$ and $t_M$ in the histogram. 
Recall that a small value of ${\cal D}(\infty)$ does not necessarily mean small $d_w(\infty)$ as this quantity depends also on the infection rates and network topology (see (\ref{Sw_infty}))
and is sensitive to repeated infections. repeated infections may indeed play an important role here
as $\langle t_I^w \rangle =8$ is rather small.

In Fig. \ref{WS3} the infections of the random walk simulations are increasing slower (red curves) compared to Figs. \ref{WS8}. The structure with higher connectivity 
Fig. \ref{WS8} shows excellent quantitative agreement of random walk and mean field solutions for the walkers and nodes capturing well the attenuated oscillations, especially for zero mortality. In the network with smaller connectivity of Fig. \ref{WS3} the increase of the infections is delayed compared to the mean field. On the other hand, for non-zero mortality the mean field and random walk dynamics for the walkers diverge 
slightly with time. We infer that mortality may deviate the infection rates from (\ref{infections_rate}).

The comparison of the spreading in Figs.\ref{WS8}, \ref{WS3} shows clearly the role of the connectivity: The mean field model captures better the spreading in networks with higher connectivity (short average distances between nodes) and with low mortality. The following cases give further evidence for these observations.
\begin{figure}[H]
\centerline{
\includegraphics[width=0.55\textwidth]{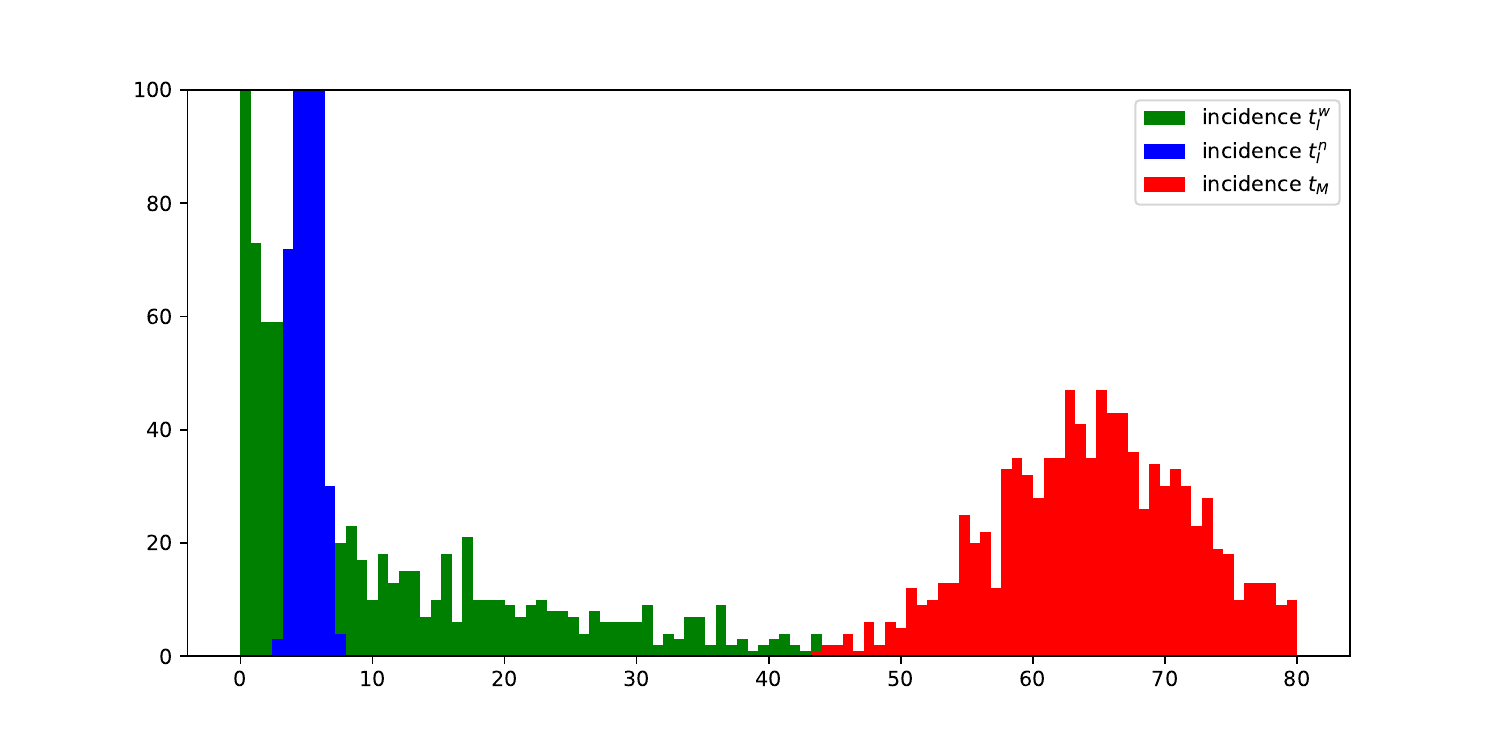}}
\centerline{
\includegraphics[width=0.55\textwidth]{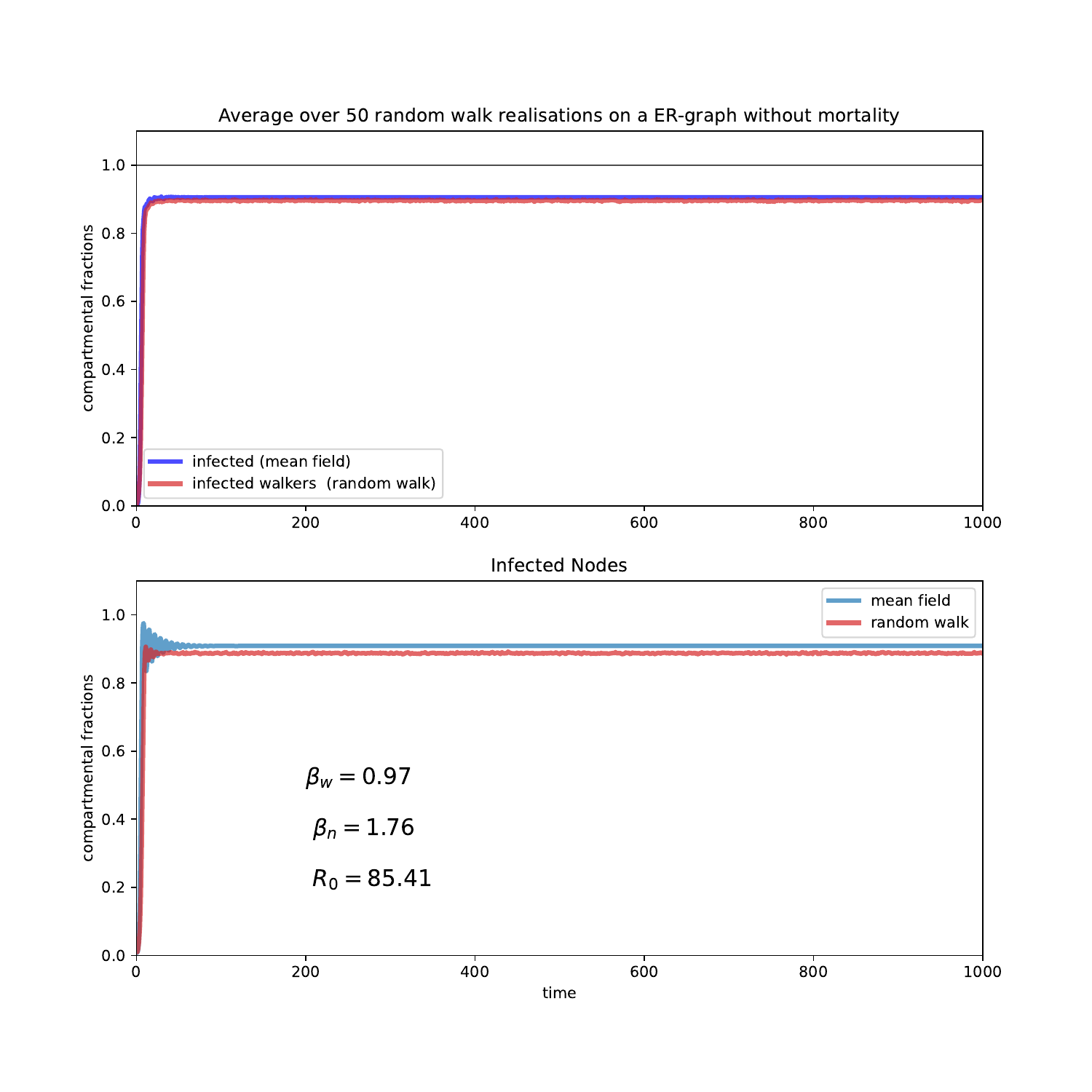}
\includegraphics[width=0.55\textwidth]{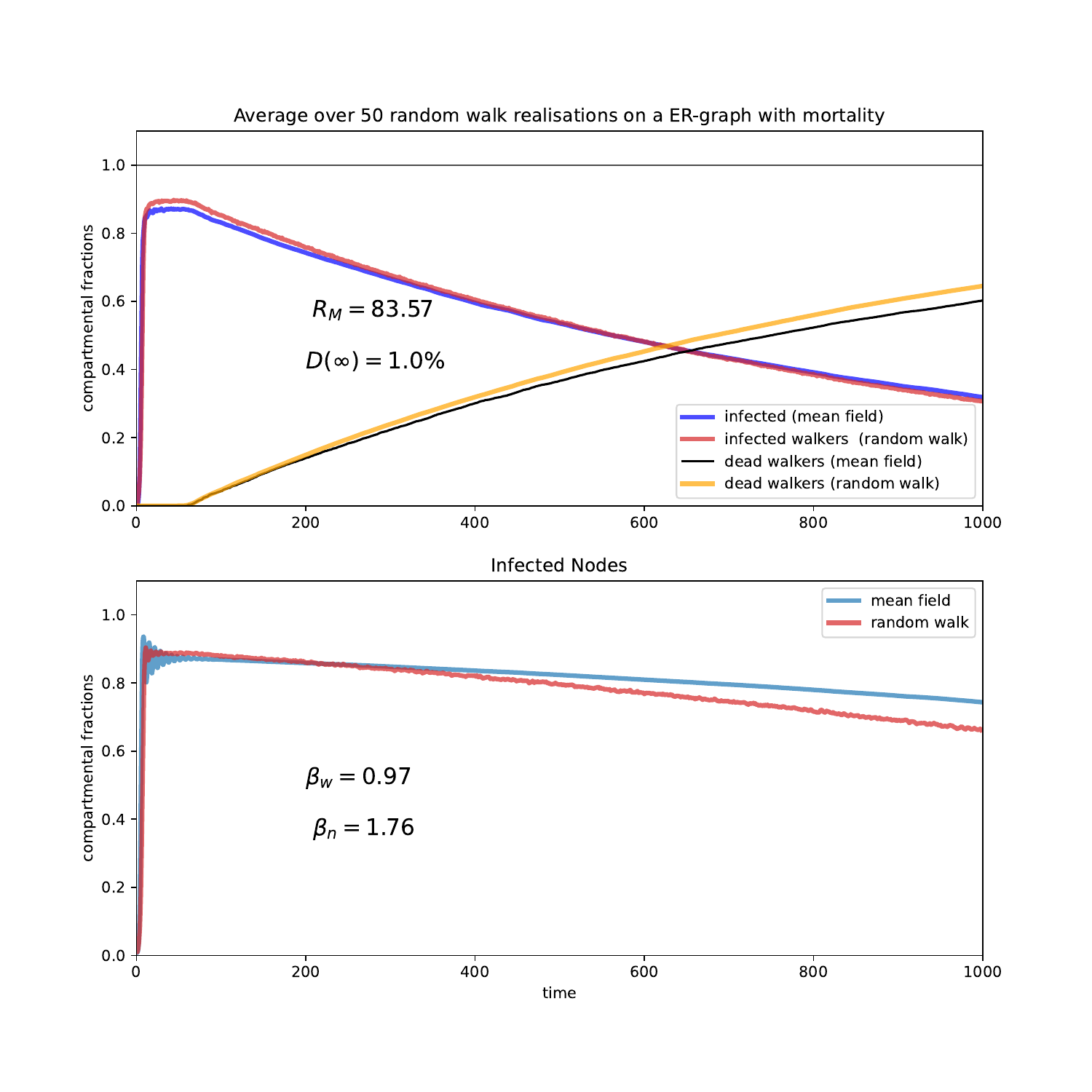}}
\caption{Evolution on ER graph ($nx.erdos\_renyi\_graph(N=1000,p=0.1,seed=seed)$) with $Z=1000$ walkers and small rewiring probability $p=0.1$ (above the percolation limit $p_c=0.01$ to ensure a connected structure). The parameters are 
$\langle t_I^n \rangle = 5$, $\xi_I^n =10$,  $\langle t_I^w \rangle = 10$, $\xi_I^w=0.05$, $\langle t_M \rangle = 65$, $\xi_M=1$. The left upper frame shows a snapshot of the evolution (same color code as in Fig. \ref{WS3}).}
\label{WS_ERm8}
\end{figure}
\noindent Next we explore the spreading on an ER graph in Fig. \ref{WS_ERm8}. The agreement of random walk simulations and mean field model is impressive where this holds for both with and without mortality. One can see by the degree distribution in Fig. \ref{networks} that for these connectivity parameters the graph is well connected and small world giving strong evidence that the mean field approach is here well capturing the spreading dynamics.
\begin{figure}[H]
\centerline{
\includegraphics[width=0.38\textwidth]{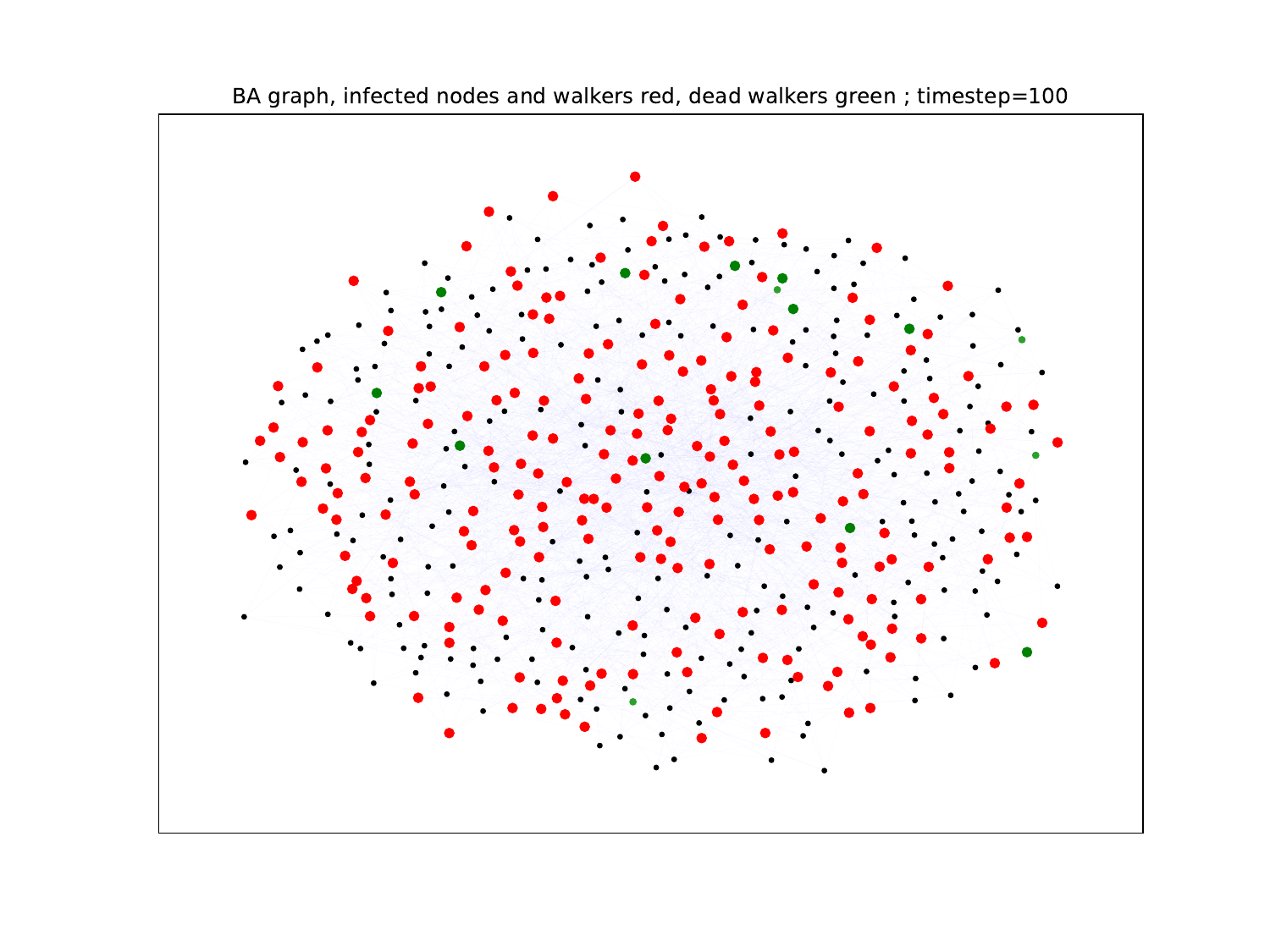}
\includegraphics[width=0.55\textwidth]{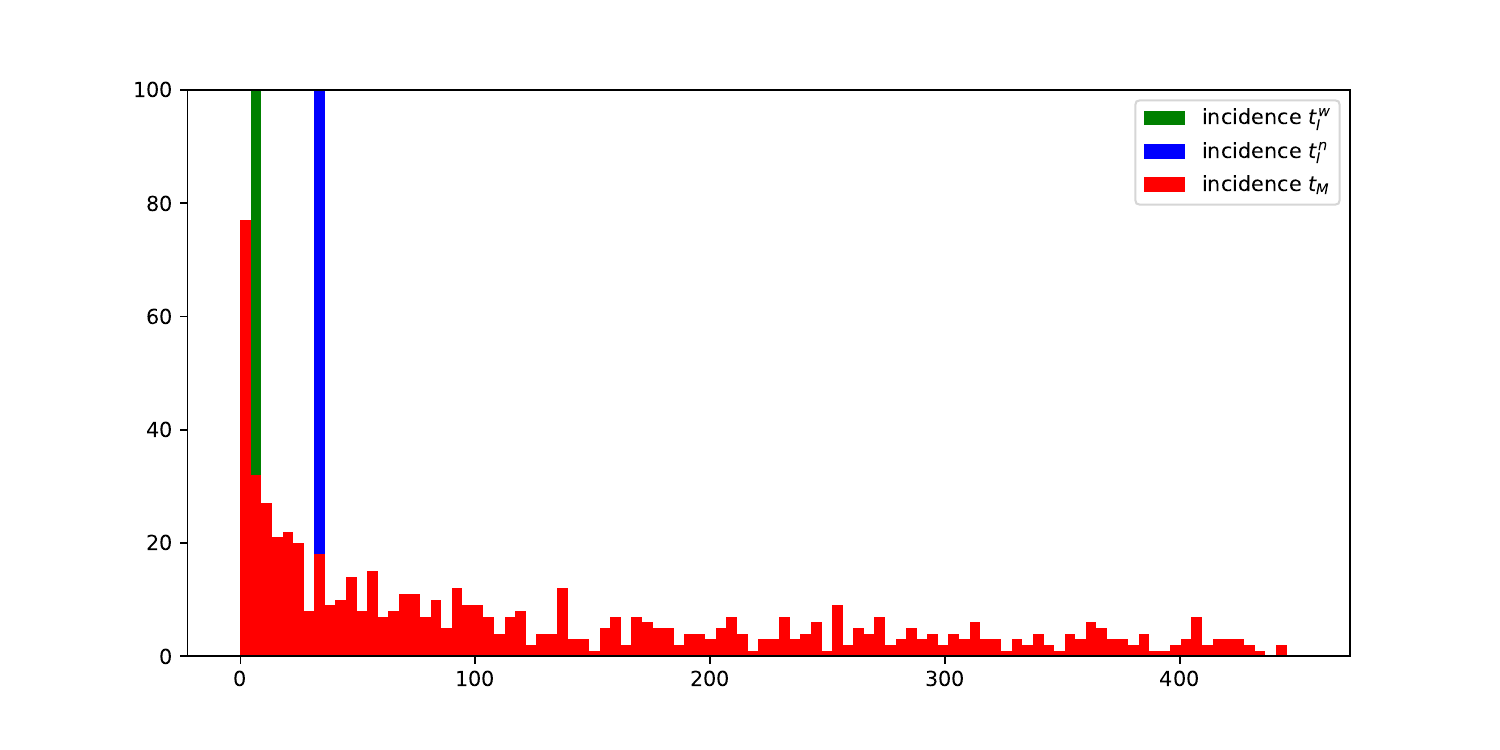}}
\centerline{
\includegraphics[width=0.55\textwidth]{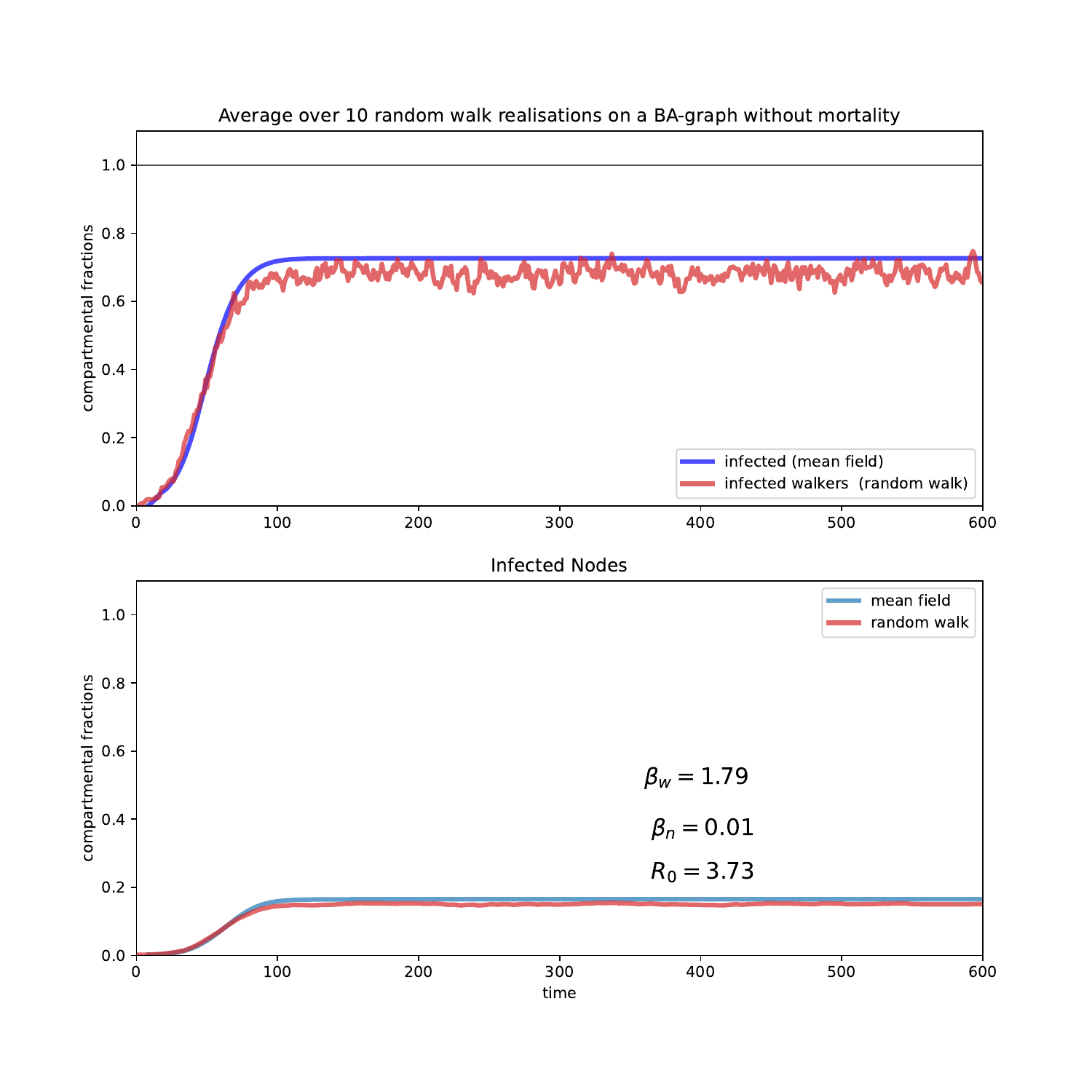}
\includegraphics[width=0.55\textwidth]{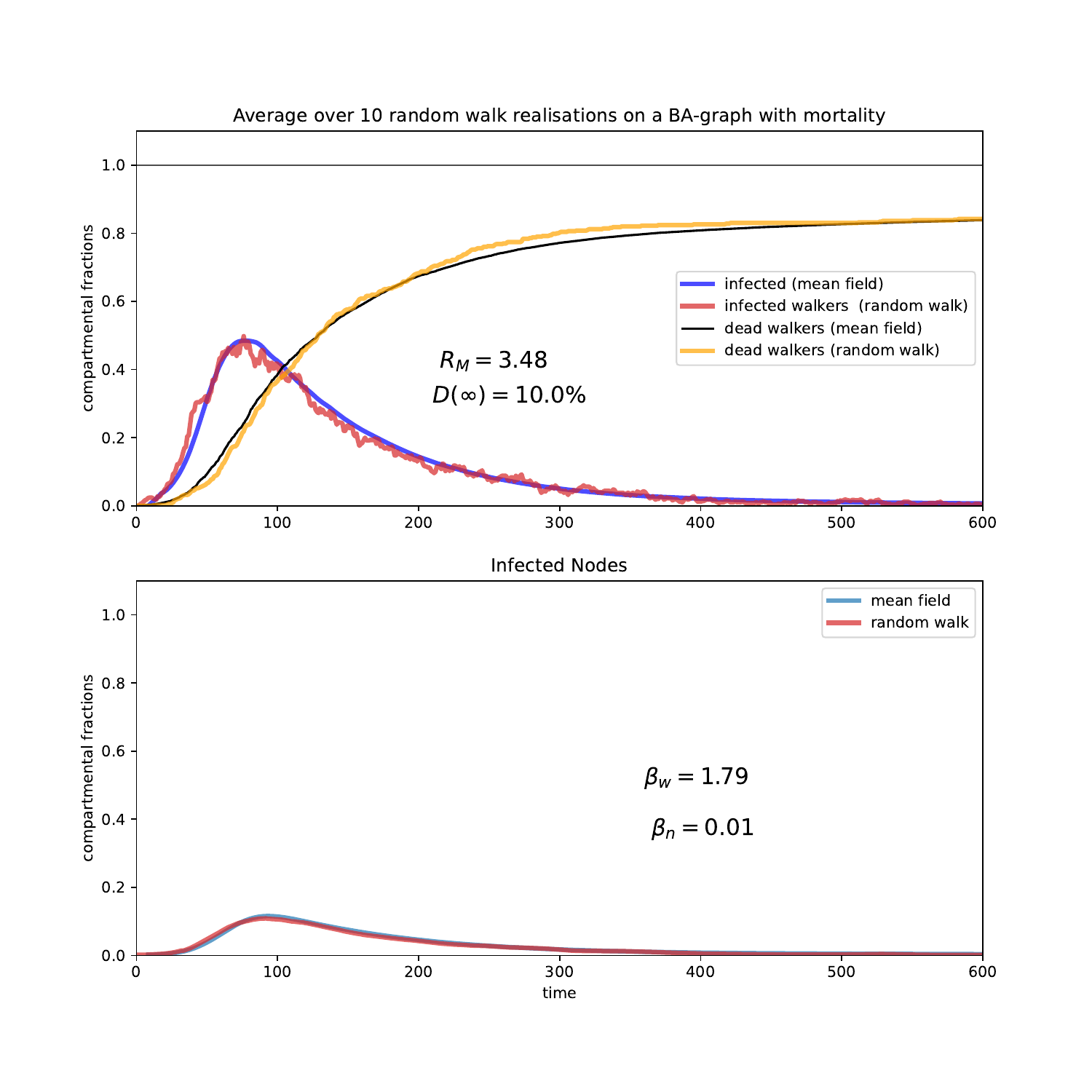}}
\caption{Evolution on BA graph with $Z=50$ walkers and $N=5000$ nodes ($nx.barabasi\_albert\_graph(N=5000, m=5, seed=seed)$) with parameters $\langle t_I^n \rangle = 32$, $\xi_I^n=10^4$, $\langle t_I^w\rangle = 8$, $\xi_I^w=10^4$ (sharp $t_I^{w,n}$), $t_M=500$, $\xi_M=10^{-3}$. The basic reproduction number $R_M$ is here only slightly smaller than $R_0$ without mortality. The left upper frame shows a snapshot of the evolution (same color code as in Fig. \ref{WS_ERm8}).}
\label{BA}
\end{figure}
\noindent Finally we explore in Fig. \ref{BA} the dynamics on a BA network.
In the right frame we have high overall mortality of ${\cal D}(\infty) \approx 10\%$ probability for a walker to die from an infection. 
In this example the disease is starting to spread as $R_M\approx 3.48 >1$ where only a single infection wave emerges which is extinct by the high mortality. 
Recall that that $R_M>1$ is only telling us that the healthy state is unstable, i.e. that the disease is starting to spread. It does not contain the information whether the spreading is persistent or whether the disease is eventually extinct. 
To explore the role of topological features such as average distances between nodes we perform the same simulation experiment with identical parameters and less ($N=2100$) nodes, i.e. higher density of walkers (Fig. \ref{BA2}).
\begin{figure}[H]
\centerline{
\includegraphics[width=0.55\textwidth]{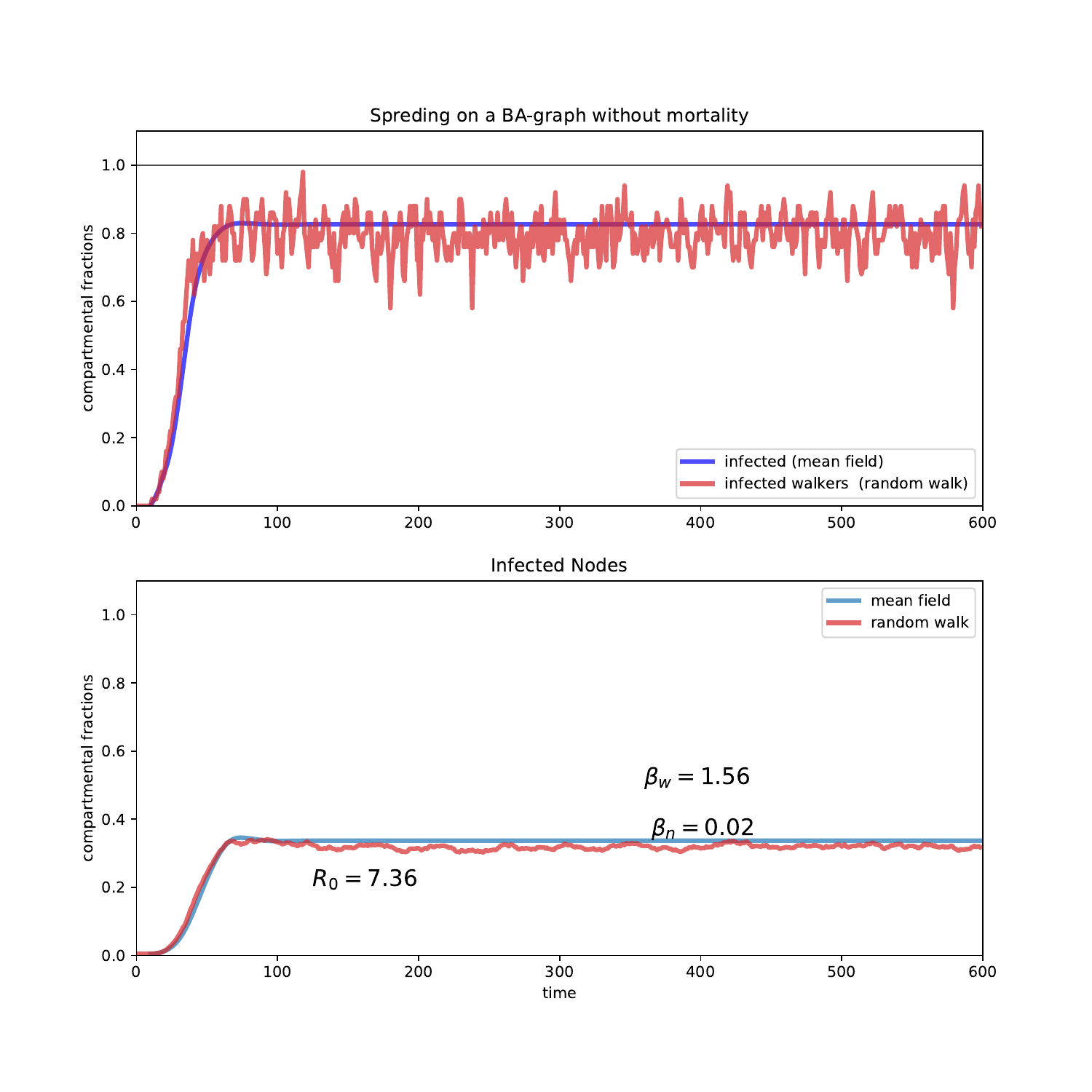}
\includegraphics[width=0.55\textwidth]{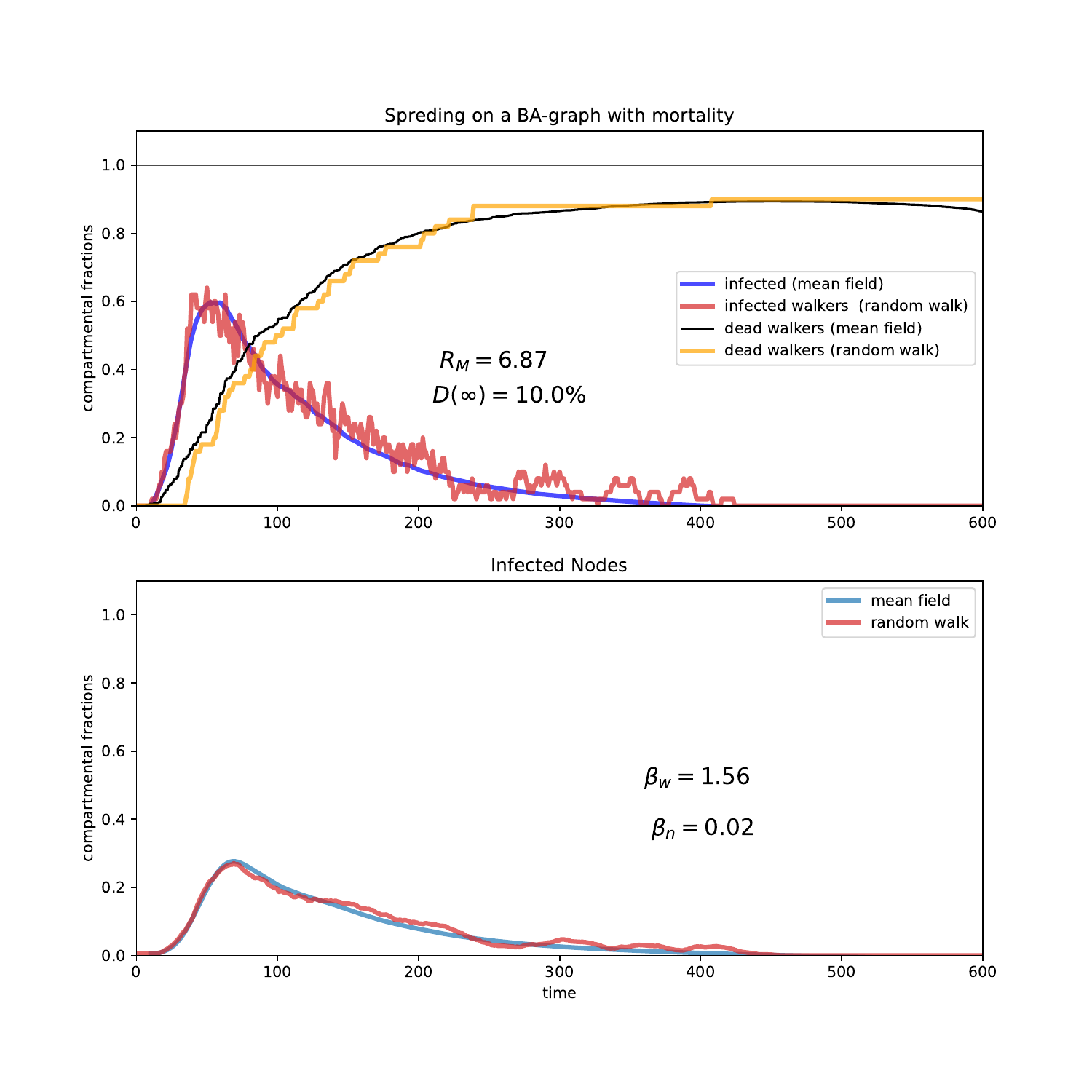}}
\caption{Evolution with the same parameters and number of walkers ($Z = 50$) as in Fig. \ref{BA} but less nodes ($N=2100$) for one random walk realization. We interpret the increase of $R_{M}$ and $R_0$ due to more frequent passages of susceptible walkers on infected nodes (higher infection rates).}
\label{BA2}
\end{figure}

The accordance of mean field model and random walk simulation is also in Fig. \ref{BA2} indeed excellent. We explain this by the fact that the BA network is a strongly connected structure
with pronounced small world property. The higher density of walkers lead to increased $R_M$ and $R_0$ compared to Fig. \ref{BA}.
There is also only a single infection wave occurring with a higher maximum value compared to Fig. \ref{BA}.
In both cases  (Figs. \ref{BA}, \ref{BA2}, right frames) the infection waves are extinct by the high mortality of walkers where stationary states (\ref{Sw_infty}) with $d_w(\infty) \approx 80\% $ of dead walkers are taken. 
When we switch off 
mortality (left frames), stable endemic states emerge more rapidly in Fig. \ref{BA2} (case with higher density of walkers).

Further simulation experiments (not shown here) reveal that
the mean field model and random walk simulations exhibit excellent accordance when we further increase the attachment parameters $m$ or the density of walkers with otherwise identical parameters. 
For higher mortality the agreement becomes less well and diverges with increasing observation time. 
This observation suggests that mortality modifies the infection rates in the network for larger observation times. We leave this issue for future research.

Our overall finding from this case study is that the mean field approach (with infection rates (\ref{infections_rate})) is particularly well suited to mimic spreading in strongly connected environments with pronounced small world feature, but is less well for higher mortality.

\section{Conclusions}
We studied epidemic spreading in complex graphs where we focused on transmission pathway via vectors mimicking the spreading of a certain class of diseases
such as Dengue, Malaria or Pestilence and others.
We developed a stochastic compartment model for the walkers and nodes with mortality for the walkers. For zero mortality we obtained 
the endemic equilibrium in explicit form (Eqs. (\ref{endemic_equil})). Stability analysis of the endemic and healthy states reveal the crucial control parameter for spreading, the basic reproduction number.
We obtained the basic reproduction numbers $R_M$ and $R_0$ with and without mortality, respectively, 
where we proved that $R_M \leq R_0$, (relations (\ref{basic_rep_with_mortality}) with (\ref{ROM_res})).
For $R_M, R_0 >1$ the healthy state is an unstable fixed point where the endemic equilibrium exists for zero mortality as a unique stable fixed point independent of the initial conditions.
The basic reproduction numbers depend on the means of the compartmental sojourn times in compartment I of the nodes and walkers
and on the topology of the network captured by the mean field rate constants $\beta_{w,n}$. 

Our model has applications beyond epidemic dynamics, for instance in chemical reaction models \cite{Simon2020}.
An interesting question to explore in a follow-up project is whether our class of compartment models with indirect transmission pathways may exhibit (for zero mortality) persistent oscillations (Hopf instabilities) (See our brief discussion at the end of Appendix \ref{endemic_stability}.).  
An open problem also remains how large world network topology may be included into such a mean field model, modifying the infection rates. Further promising directions include an account for immune and incubation compartments, effects of mitigation measures, vaccination among many others. 

\section{Acknowledgements}
T.G. gratefully acknowledges to have been hosted at the
Institut Jean le Rond d’Alembert in the framework of an internship (stage M1 Physique $n^o \, 28281$) for the development of the PYTHON simulation codes and participation in the present study.

\begin{appendix}
\section{Appendix}

\subsection{Some basic notions}
\label{general}
Here we recall briefly some basic notions used in the paper. 
The infection rates are a causal functions 
\beq
\label{probab}
{\cal A}(t) = A(t)\Theta(t)
\eeq
where $\Theta(t)$ indicates the Heaviside step function defined by 
\beq
\label{Heaviside-stepfu}
\Theta(\zeta) =\left\{\begin{array}{l} 1 , 
\hspace{1cm} \zeta \geq 0  \\ \\  0 , \hspace{1cm} \zeta < 0 
\end{array}\right.
\eeq
with $\Theta(\zeta)+\Theta(-\zeta)=1$ and in or definition $\Theta(0)=1$. 
Its derivative yields the Dirac $\delta$-distribution $\frac{d}{d\zeta}\Theta(\zeta)= \delta(\zeta)$.
We use throughout the paper mutually independent strictly positive random variables $T_1,\ldots, T_n \in \mathbb{R}_{+}$ 
The random variables $T_j$ are assumed to follow their specific PDFs
\beq
\label{PDFs}
Prob[T_j\in [u,u+{\rm d}u]] =\left\langle \, \delta(u-T_j)\,  \right\rangle 
= K_j(u){\rm d}u
\eeq
where the PDFs $K_j$ are causal functions as a consequence of the positiveness of the $T_j$.
Then applies the averaging rule
\beq
\label{averaging}
\left\langle \, f(t;T_1,T_2,\ldots, T_n) \, \right\rangle = 
\int_0^{\infty}\ldots\int_0^{\infty}{\rm d}t_1\ldots{\rm d}t_n
 f(t;t_1,t_2,\ldots, t_n)
K_1(t_1)\ldots K_n(t_n)
\eeq
for suitable functions $f$. For $f(t;T_1,T_2,\ldots, T_n) =g_1(T_1) \ldots g_n(T_n) $ using independence of the $T_j$ this
yields 
$$ \left\langle \, g_1(T_1) \ldots g_n(T_n) \, \right\rangle = \left\langle \, g_1(T_1), \right\rangle \ldots 
\left\langle \, g_n(T_n) \, \right\rangle .$$
Important cases emerge by applying (\ref{averaging}) to exponentials
\beq
\label{averages}
\left\langle \, e^{-\lambda(T_1+\ldots+T_n)} \, \right\rangle  = \int_0^{\infty}e^{-\lambda t}
\left\langle \delta(t-T_1-\ldots-T_n) \right\rangle {\rm d}t 
={\hat K}_1(\lambda)\ldots {\hat K}_n(\lambda), \hspace{0.75cm} \Re\{\lambda\} \geq 0
\eeq
In this relation the LTs of the PDFs come into play
\beq
\label{Laplace_trafos}
{\hat K}_j(\lambda) =  \int_0^{\infty} e^{-\lambda t} K_j(t){\rm d}t
\eeq
where ${\hat K}_j(\lambda)\bigg|_{\lambda=0}=1$ reflects the normalization 
of PDFs (\ref{PDFs}). A further observation is
\beq
\label{convolution_causal}
\left\langle \delta(t-T_1-\ldots-T_n) \right\rangle = (K_1 \star \ldots \star K_n)(t)
\eeq
where $\star$ stands for convolution $(K_1 \star K_2)(t) = \int_0^tK_1(\tau)K_2(t-\tau){\rm d}\tau$ of the causal PDFs.
\subsection{Proof of stability of the endemic equilibrium}
\label{endemic_stability}
Here we develop the rest of the proof stability of the endemic equilibrium, i.e. we show that
the function (\ref{endem_function}) is strictly positive for $\mu>0$ with $R_0-1 >0$. First, we observe that
${\hat \Phi}_I^{w,n}(\mu) \leq \langle t_I^{w,n} \rangle $ with ${\hat \Phi}_I^{w,n}(0)= \langle t_I^{w,n} \rangle$ and ${\hat \Phi}_I^{w,n}(\infty) =0$. For our convenience, we introduce the 
functions
\beq
\label{lambdas}
\lambda_{w,n}(\mu) = \frac{{\hat \Phi}_I^{w,n}(\mu)}{\langle t_I^{w,n} \rangle} \in (0,1]
\eeq
which the LTs of the normalized $\Phi_I^{w,n}(t)/\langle t_I^{w,n}\rangle $ and
which are by virtue of Bernstein's theorem \cite{WidderSchilling2010} completely monotonic (CM) with respect to $\mu$, i.e. 
\beq
\label{CM_def}
(-1)^n\frac{d^n}{d\mu^n} \lambda_{w,n}(\mu) \geq 0 , \hspace{1cm} \mu \in (0,\infty)
\eeq
inheriting this feature from the exponential $e^{-\mu \tau}$ ($t >0$).
Therefore, $$\frac{d}{d\mu}\lambda_{w,n}(\mu) = - \frac{1}{\langle t_I^{w,n} \rangle} \int_0^{\infty} e^{-\mu t} t \Phi_I^{w,n}(t){\rm d}t < 0$$ (as $\Phi_{I}^{n,w}(t) \in (0,1]$) exists thus $\lambda_{w,n}(\mu)$ is monotonously decreasing with $\mu$ with $\lambda_{w,n}(0) = 1 \geq \lambda_{w,n}(\mu) >0$.
Further we observe in Eqs. (\ref{endemic_equil}) that $J_n^e\beta_w \langle t_I^{w} \rangle =\frac{R_0-1}{1+\beta_n \langle t_I^{n} \rangle}$, $J_w^e\beta_n \langle t_I^{n} \rangle =\frac{R_0-1}{1+\beta_w \langle t_I^{w} \rangle}$ thus
$$
R_0(J_n^e+J_w^e) = (R_0-1)\left( \frac{\beta_n \langle t_I^{n} \rangle}{1+\beta_n \langle t_I^{n} \rangle}+ \frac{\beta_w \langle t_I^{w} \rangle}{1+ \beta_w \langle t_I^{w} \rangle} \right)
$$
Then $G_e(\mu)$ reads
\beq
\label{Ge_conv}
\begin{array}{clr}
\ds G_e(\mu) & = \ds  1+ \lambda_w(\mu)\lambda_n(\mu)\left\{-R_0 + (R_0-1)\left( \frac{\beta_n \langle t_I^{n} \rangle}{1+\beta_n \langle t_I^{n} \rangle}+ \frac{\beta_w \langle t_I^{w} \rangle}{1+ \beta_w \langle t_I^{w} \rangle} \right)\right\} & \\ \\ & \ds \hspace{0.5cm} + (R_0-1)
\left(\frac{\lambda_w(\mu)}{1+\beta_n \langle t_I^{n} \rangle}+\frac{\lambda_n(\mu)}{1+\beta_w \langle t_I^{w} \rangle}\right) &
\end{array}
\eeq
Now we observe that $ \lambda_w(\mu)\lambda_n(\mu) \leq \lambda_{w,n}(\mu)$ thus a lower bound function $H(\mu) \leq G_e(\mu)$ 
is generated by replacing $\lambda_{w,n}(\mu) \to \lambda_w(\mu)\lambda_n(\mu)$
in the second line. Now it is sufficient to prove that $0< H(\mu)$. 
We hence get for this lower bound function
\beq
\label{H_function}
\begin{array}{clr}
H(\mu) & = 1+ \lambda_w(\mu)\lambda_n(\mu)\left\{-R_0 + (R_0-1)\left( \frac{\beta_n \langle t_I^{n} \rangle}{1+\beta_n \langle t_I^{n} \rangle}+ \frac{\beta_w \langle t_I^{w} \rangle}{1+ \beta_w \langle t_I^{w} \rangle} \right)+ (R_0-1)
\left(\frac{1}{1+\beta_n \langle t_I^{n} \rangle}+\frac{1}{1+\beta_w \langle t_I^{w} \rangle}\right)\right\} & \\ \\
 & = \ds  1+ \lambda_w(\mu)\lambda_n(\mu)(R_0-2)  & \\ \\
 & = \ds 1 - \lambda_w(\mu)\lambda_n(\mu) + (R_0-1)\lambda_w(\mu)\lambda_n(\mu) 
 \end{array}
\eeq
and with $1 - \lambda_w(\mu)\lambda_n(\mu) \geq 0$ and $(R_0-1)\lambda_w(\mu)\lambda_n(\mu) >0$
it follows that $0 < H(\mu) < G_e(\mu)$ concluding the proof of stability of the endemic equilibrium.
\paragraph{A few remarks on the possibility of oscillatory (Hopf) instabilities of the endemic equilibrium}
Let us briefly explore whether an oscillatory (Hopf) instability of the endemic equilibrium
is possible. To that end we write $G_e(\mu)$ as
\beq
\label{Ge-gen}
G_e(\mu)= 1 + \sigma {\hat g}(\mu)  \geq 1 - {\hat g}(\mu) 
\eeq
where $\sigma =\pm 1$ and ${\hat g}(\mu)$ is a non-negative CM function (see Fig. \ref{endemic_stability_Fig})
with maximum value ${\hat g}(0) = |R_0-2|$.
Then the following two cases may occur. \newline\newline
\noindent {\it Case (i)} $\sigma=-1$; $0< G_e(0) =R_0-1 < 1$ ($1< R_0 < 2$): \newline\newline
Then ${\hat g}(\mu)$ can be represented as LT of a non-negative function $g(t)$ and
consider now $\mu=\mu_1+i\mu_2$ with $\mu_1 \geq 0$ thus the real part of ${\hat g}(\mu_1+i\mu_2)$ can be written as
\beq
\label{real_g}
\Re g(\mu_1+i\mu_2)  = \int_0^{\infty} g(t) e^{-\mu_1 t} \cos(\mu_2 t) {\rm d}t , \hspace{1cm} \mu_1 \geq 0
\eeq
with ${\hat g}(0)= 2-R_0$
and clearly $-{\hat g}(\mu_1) \leq \Re g(\mu_1+i\mu_2) \leq {\hat g}(\mu_1)$. Therefore,
\beq
\label{strictly_pos}
\Re G_e(\mu_1+i\mu_2) = 1 - \Re g(\mu_1+i\mu_2) \geq 1- {\hat g}(\mu_1) \geq 1-{\hat g}(0) = R_0-1 > 0 .
\eeq
Hence in the range of case (i) $1< R_0 < 2$ there is no oscillatory (Hopf) instability of the 
endemic state possible\footnote{A similar consideration of function $G(\mu)$ of (\ref{detnull_heathy}) shows as well that the healthy state for $R_0<1$ does not exhibit an oscillatory instability.}.
\newline\newline
{\it Case (ii)} $\sigma= +1$; $R_0-1 >1$: 
\newline\newline
Here we have two pertinent ranges of $R_0$.
The first is the range (a) ${\hat g}(0) = R_0-2 < 1$ (i.e. $R_0<3$) and the second one (b) is ${\hat g}(0) =R_0-2 >1$.
Clearly in the range (a) (\ref{strictly_pos}) remains true and $\Re G_e(\mu_1+i\mu_2)$ strictly positive. Hence for $1 < R_0 < 3$ no Hopf instability is possible. \newline
This changes in the range (b) since  ${\hat g}(0)=R_0-2 >1$ thus  $\Re G_e(\mu_1+i\mu_2) = 1 + \Re g(\mu_1+i\mu_2)$ may become negative. Therefore, for $R_0>3$ a Hopf instability of the endemic state becomes {\it possible}. However, the possibility that $\Re G_e(\mu_1+i\mu_2)=0$
is only necessary but not sufficient for a Hopf instability. One also needs simultaneously that the imaginary 
part $\Im G_e(\mu_1+i\mu_2)=0$ is vanishing for the same $\mu=\mu_1+i\mu_2$. In the simulations performed for this paper, we did not observe persistent oscillations. 
We leave the  exploration of this issue for future research in a follow-up project.

\subsection{A very brief recap of random graphs}
\label{complex_graphs}
Here we recall briefly some essential features of the three classes of random graphs, which we use in the random walk simulations.
For an extended outline, consult e.g. \cite{BarabasiAlbertJeong1999}. The three classes of random graph models depicted herafter are motivated by the observation that complex random network structures are encountered ubiquitously and crucially determine human and animal mobility 
patterns including epidemic propagation. 
\paragraph{(i) Erd\"os and R\'enyi (ER) graph}

The ER graph is one of the most basic variants of a random graph, which was introduced in 1959 by Erd\"os and R\'enyi \cite{ER1959}. We use is the so-called $G(N,p)$ variant of random ER graph model (which is actually due to Gilbert \cite{Gilbert1959}) which is generated as follows \cite{NewmanWattsStrogats2002,Barabasi2016}.
Given are $N$ labeled nodes. Any pair of nodes is connected independently by an edge with uniform probability $p$. 
The probability $P_N(k)$ that a node has $0 \leq k \leq N-1 $ connections is given by a binomial distribution 
$$
P_N(k) = \left(\begin{array}{c} N-1 \\ k \end{array} \right) p^k(1-p)^{N-1-k} \to \frac{\langle k\rangle^k}{k!}e^{-\langle k\rangle}
$$
where $\langle k \rangle = (N-1)p \sim Np$ denotes the average degree. For $N \to \infty$ 
(while $Np$ is kept constant) the degree distribution $P_N(k)$ converges to a Poisson distribution representing the infinite graph limit of the ER $G(N,p)$-model. Therefore, $P_N(k)$ is rapidly decaying with degree $k$, so the number of nodes with a high number of connections is very small. In order to obtain in the $G(N,p)$-model a connected graph, it is necessary that
$p>p_c=\frac{log N}{N}$ is above the percolation limit \cite{ER1959,Newman2010}.

\paragraph{(ii) Watts-Strogatz (WS) network}
The WS graph model \cite{ErdoesRen1960,WattsStrogats1998,NewmanWattsStrogats2002} starts 
with a ring of $N$ nodes where each node is connected symmetrically with a number $m << N$ to left and right neighbor nodes by an edge such that each node has $2m$ connections. In the second step, each of the connections $i,j$ of node $i$ is replaced with probability $p$ by a randomly chosen connection $i,k$ uniformly among other nodes avoiding self-connections and link duplication, so that each connection $i,k$ is chosen only once. There are two 
noteworthy limits for a WS graph. For $p=0$ (no rewiring of links) we have a regular ring with constant degree $2m$ for all nodes. In the limit $p=1$ an ER graph is emerging with probability $2m/(N-1)$ for a link.
The WS graph has the small-world property (short average distances between pairs of nodes) and a high tendency to develop clusters of nodes \cite{WattsStrogats1998}.

\paragraph{(iii) Barab\'asi-Albert (BA) graph}
The BA graph is generated by a preferential attachment mechanism for newly added nodes \cite{Barabasi2016,BarabasiAlbert1999,BarabasiAlbertJeong1999,Jeong-etal200}. One starts with $m_0$ nodes and  adds new nodes. Any newly added node is connected with $m\leq m_0$ existing nodes ($m$ is referred to as the attachment parameter) where most likely with nodes of high degrees. In this way nodes with high degree receive further links. This leads to an asymptotically scale-free network with a power law degree-distribution
$$
P(k) \propto k^{-2-\mu} , \hspace{1cm} \mu \approx 1 .
$$
As the decrease in this power law is relatively slow, there might exist quite a few nodes with many links (hub nodes) and many nodes with few links.
BA graphs are believed to mimic a large class of real-world networks including the world wide web, 
citation-, social-, and metabolic networks.

Realizations of these three types of random graph types used in our multiple random walkers simulations are shown in Fig. \ref{networks}.

\end{appendix}


\begin{thebibliography}{100}

\bibitem{public_health} P. Rhodes, J.H. Bryant, Public Health. Encyclopedia Britannica, (2024)
https://www.britannica.com/topic/public-health 

\bibitem{KermackMcKendrick1927} W.O. Kermack, A.G. McKendrick, A contribution to the mathematical 
theory of epidemics, Proc. Roy. Soc. A 115, 700–721 (1927).

\bibitem{LiuHeathcote1987} W. M. Liu, H.W. Hethcote HW, S.A. Levin, Dynamical behavior of epidemiological
models with non-linear incidence rate. J. Math. Biol. 25, 359–380 (1987).

\bibitem{Li-etal1999} M.Y. Li, J.R. Graef, L. Wang, J. Karsai, Global dynamics of a SEIR model with
varying total population size. Math. Biosci. 1999 160:191–213.

\bibitem{Anderson1992} R. M. Anderson, and R. M. May, 1992, Infectious Diseases in Humans, (Oxford University
Press, Oxford).

\bibitem{Martcheva2015} M. Martcheva, An Introduction to Mathematical Epidemiology, Springer (2015).

\bibitem{Harris2023} J. E. Harris, Population-Based Model of the Fraction of Incidental
COVID-19 Hospitalizations during the Omicron BA.1 Wave in the United States, COVID 3(5), 728-743 (2023).
Doi: 10.3390/covid3050054

\bibitem{Whitehead2007} S. S. Whitehead, J. E. Blaney, A. P. Durbin,
B.R. Murphy, Prospects for a dengue virus vaccine. Nat Rev Microbiol 5, 518–528 (2007). Doi: 10.1038/nrmicro1690

\bibitem{Satoras-Vespignani-etal2015} R. Pastor-Satorras, C. Castellano, P. Van Mieghem, A. Vespignani,
Epidemic processes in complex networks, Rev. Mod. Phys. 87, 925-979 (2015).

\bibitem{Pastor-SatorrasVespignani2001} R. Pastor-Satorras, A. Vespignani, Epidemic dynamics and endemic states in complex networks, Phys. Rev. E 63, 066117 (2001).

\bibitem{OkabeShudo2021} Y. Okabe Y, A. Shudo, Microscopic Numerical Simulations of Epidemic Models on Networks. Mathematics 9, 932 (2021).

\bibitem{BasnakovSandev-etal2020} L Basnarkov, I. Tomovski, T. Sandev, L. Kocarev, on-Markovian SIR epidemic spreading model of COVID-19, Chaos, Solitons and Fractals 160, 112286 (2022).

\bibitem{donofrio2024} G. d'Onofrio, T.M. Michelitsch, F. Polito, A.P. Riascos, On discrete-time arrival processes and related random motions, (preprint) arXiv:2403.06821 

\bibitem{Metzler-Klafter2000} R. Metzler, J. Klafter, The random walk’s guide to anomalous diffusion : A fractional dynamics approach. Phys. Rep. 339:1--77 (2000).

\bibitem{Mainardi-etal2004} F. Mainardi, R. Gorenflo, E. Scalas, A fractional generalization of the Poisson processes. 
Vietnam J. Math. 32 SI:53--64 (2004).

\bibitem{Barabasi2016}  A.-L. Barab\'asi, Network Science. Cambridge University Press, Cambridge (2016).

\bibitem{BarabasiAlbert1999} A.-L. Barab\'asi and R. Albert, Emergence of Scaling in Random Networks, Science 286, 509 (1999).

\bibitem{BarabasiAlbertJeong1999}
Barab\'asi, R. Albert, and H. Jeong, Mean-field theory for scale-free random networks, Physica A 272, 173-187 (1999).

\bibitem{Jeong-etal200} 
H. Jeong, B. Tombor, R. Albert, Z. N. Oltvai, A L Barab\'asi, The large-scale organization of metabolic networks, Nature 
407(6804):651-4 (2000).

\bibitem{ER1959} P. Erd\"os,A. R\'enyi, On Random Graphs I, Publ. Math. 6, 290–297 (1959).

\bibitem{ErdoesRen1960} P. Erd\"os, P. R\'enyi, On the evolution of random graphs,
Publ. Math. Inst. Hung. Acad. Sci 5, 17 (1960).

\bibitem{Gilbert1959} E. N. Gilbert, Random Graphs. Annals Math. Sci. 30, 1141-1144 (1959). 

\bibitem{Ross1996} S.M. Ross, Stochastic Processes (John Wiley \& Sons,
New York), 1996.

\bibitem{Newman2010} M.E.J. Newman, Networks: An Introduction, Oxford University Press, Oxford, 2010.

\bibitem{NewmanWattsStrogats2002} M. E. J. Newman, D. J. Watts, S. H. Strogatz, Random graph models of social networks, PNAS 99, 2566-2572 (2002).

\bibitem{WattsStrogats1998} D. J. Watts, S. H. Strogatz, Collective dynamics of ‘small-world’ networks, Nature, London,  393, 440 (1998).

\bibitem{Eraso-Hernandez-etal2023} L. K. Eraso-Hernandez, A. P. Riascos, T.M. Michelitsch, J. Wang-Michelitsch, Evolution of transport under cumulative damage in metro systems, Int. J. Mod. Phys. C 2450037 (2023). DOI: 10.1142/S0129183124500372

\bibitem{Barrat-etal2008} A. Barrat, M. Barth\'elemy, A. Vespignani, Epidemic spreading in population networks, In: Dynamic Processes on Complex Networks, pp. 180 --215, Cambridge University Press (2008), DOI: 10.1017/CBO9780511791383.010 

\bibitem{RiascosSanders2021} A. P. Riascos, D. P. Sanders, Mean encounter times for multiple random walkers on networks,
Phys. Rev. E 103, 042312 (2021). Doi: 10.1103/PhysRevE.103.042312

\bibitem{BesMi-etal2021}
M. Bestehorn, A. P. Riascos, T. M. Michelitsch, B. A. Collet, A Markovian random walk model of epidemic spreading,
Continuum Mech. Thermodyn. 33:1207–1221 (2021), Doi: doi.org/10.1007/s00161-021-00970-z

\bibitem{fractional_book_MiRia2019}  T.M. Michelitsch, A.P. Riascos, B.A. Collet, A.F. Nowakowski, F.C.G.A. Nicolleau, Fractional Dynamics on Networks and Lattices, ISTE/Wiley, London,
2019.

\bibitem{RiaMat2012} A. P. Riascos, J. L. Mateos, Long-range navigation on complex networks using L\'evy random walks,
Phys. Rev. E 86, 056110 (2012).


\bibitem{BesMiRias2022}
M. Bestehorn, T. M. Michelitsch, B. A. Collet, A. P. Riascos,
and A. F. Nowakowski, Simple model of epidemic dynamics with memory effects, Phys. Rev. E 105, 024205
(2022).

\bibitem{Granger-et-al2023} T. Granger, T. M. Michelitsch, M. Bestehorn, A. P. Riascos, B. A. Collet,
Four-compartment epidemic model with retarded transition rates, Phys. Rev. E 107 044207 (2023).

\bibitem{BesMi2023} M. Bestehorn, T. M. Michelitsch, Oscillating Behavior of a Compartmental 
Model with Retarded Noisy Dynamic Infection Rate, 
Int. J. Bifurcation Chaos 33 2350056 (2023). 

\bibitem{vanKampen1981} van Kampen, N. G. (1981), Stochastic processes in chemistry
and physics (North Holland, Amsterdam).

\bibitem{VanMiegem2014}
P. Van Mieghem, Exact Markovian SIR and SIS epidemics on networks and an upper
bound for the epidemic threshold, arXiv:1402.1731 (2014).

\bibitem{ZhuShenWang2023}Y. Zhu, R. Shen, H. Dong, W. Wang, Spatial heterogeneity and infection patterns on epidemic transmission disclosed by a combined contact-dependent dynamics and compartmental model, PLoS ONE 18(6): e0286558 (2023). Doi: 10.1371/journal.pone.0286558

\bibitem{Gostiaux-etal2023}
L. Gostiaux, W. J. T. Bos , J.-P. Bertoglio, Periodic epidemic outbursts explained by local saturation of clusters, 
Phys. Rev. E 107, L012201 (2023).

\bibitem{Peyard2023} M. Peyrard, What can we learn from the dynamics of the Covid-19 epidemic? 
arXiv:2308.14090 

\bibitem{Soper1929} H. E. Soper, The interpretation of periodicity in disease prevalence, J. Royal Statistical Society 92, 34-61 (1929).

\bibitem{NohRieger2004} J.-D. Noh, H. Rieger, Random walks on complex networks, Phys. Rev. Lett. 92, No. 11 (2004). 

\bibitem{fractional_book2019} T. Michelitsch, A. P. Riascos, B. A. Collet, A. Nowakowski, and
F. Nicolleau, Fractional Dynamics on Networks and Lattices (ISTE-Wiley, London, 2019).

\bibitem{WidderSchilling2010} R. Schilling, R. Song, Z. Vondraček, Bernstein functions. Theory and Applications, Studies in Mathematics, 37, de Gruyter, Berlin (2010).

\bibitem{Simon2020} C. M. Simon, The SIR dynamic model of infectious disease transmission and its analogy with
chemical kinetics, PeerJ Physical Chemistry (2020).  DOI 10.7717/peerj-pchem.14 

\bibitem{supplementaries} Supplementary materials: PYTHON codes ($\copyright$ T\'eo Granger 2023) and animated films: \\ 
\href{https://sites.google.com/view/scirs-model-supplementaries/accueil}{https://sites.google.com/view/scirs-model-supplementaries/accueil}   

\end{thebibliography}
\end{document}